%% file: main.tex
\PassOptionsToPackage{table}{xcolor}
\documentclass[10pt,twocolumn,letterpaper]{article}

\usepackage[pagenumbers]{wacv} 

\input{preamble}

\definecolor{wacvblue}{rgb}{0.21,0.49,0.74}
\usepackage[pagebackref,breaklinks,colorlinks,allcolors=wacvblue]{hyperref}

\def\wacvPaperID{3137} 
\def\confName{WACV}
\def\confYear{2027}

\usepackage{inconsolata}

\usepackage{url}
\usepackage{graphicx}
\usepackage{caption}
\usepackage{subcaption}
\usepackage{booktabs}
\usepackage{siunitx}
\usepackage{multirow}
\usepackage{colortbl}
\usepackage{float}
\usepackage{listings}
\usepackage{xcolor}
\usepackage{comment}
\usepackage{algorithm}
\usepackage{algpseudocode}
\usepackage{fancyvrb}
\usepackage[accsupp]{axessibility}
\usepackage{tabularx}
\usepackage{makecell}
\usepackage{ragged2e}
\usepackage[table]{xcolor}
\usepackage{siunitx}
\usepackage{amsmath}

\usepackage[table]{xcolor} 
\definecolor{mutedblue}{RGB}{70,105,125} 
\definecolor{mutedred}{RGB}{145,85,80}
\definecolor{lightblue}{RGB}{70,130,180}

\definecolor{tblgray}{gray}{0.96}

\newcolumntype{L}[1]{>{\RaggedRight\arraybackslash}p{#1}}
\newcolumntype{Y}{>{\RaggedRight\arraybackslash}X}

\usepackage[most]{tcolorbox}
\usepackage{fvextra}

\title{Multimodal Graph RAG for Long-range\\Visually Rich Document Understanding}

\author{
Yi-Cheng Wang\textsuperscript{$\dagger,\ddagger$},
Chu-Song Chen\textsuperscript{$\dagger,\ddagger$}\thanks{Corresponding author}\\
\textsuperscript{$\dagger$}Department of Computer Science and Information Engineering, National Taiwan University\\
\textsuperscript{$\ddagger$}FinTech Center, National Taiwan University\\
{\tt\small \{d13922033, chusong\}@csie.ntu.edu.tw}
}

\begin{document}
\maketitle
\input{sec/0_abstract}    
\input{sec/1_intro}
\input{sec/2_related_work}
\input{sec/3_methodology}
\input{sec/4_experiments}
\input{sec/5_conclusion}

{
    \small
    \bibliographystyle{ieeenat_fullname}
    \bibliography{main}
}

\clearpage
\appendix
\input{appendix}

\end{document}

%% file: preamble.tex
\usepackage{adjustbox}
\usepackage{pifont}
\usepackage[most]{tcolorbox}
\usepackage{fvextra}
\tcbset{
  qapromptbox/.style={
    colback=gray!3,
    colframe=black!65,
    boxrule=0.5pt,
    arc=1.5mm,
    left=1mm,
    right=1mm,
    top=1mm,
    bottom=1mm,
    fonttitle=\bfseries
  }
}

\usepackage{tikz}
\definecolor{accent}{HTML}{1E3A8A}      
\definecolor{accentmid}{HTML}{2563EB}   
\definecolor{accentbg}{HTML}{EEF2FF}    
\definecolor{rule}{HTML}{D1D5DB}        
\definecolor{captiongray}{HTML}{4B5563}

%% file: sec/0_abstract.tex
\begin{abstract}
Multimodal large language models (MLLMs) are applicable to visual document understanding, yet the limited context window yields visual question answering (VQA) to be difficult for long documents. While recent multimodal retrieval-augmented generation (MMRAG) can address this issue by retrieving relevant pages for VQA, it still struggles with the questions requiring a holistic comprehension of the visual document. To tackle this issue, knowledge graph (KG) that summarizes global knowledge of a document can provide effective solutions. However, current evaluation of the LLM-based KG methods rely mainly on the indirect evidence such as comprehensiveness, diversity, and empowerment of the answers. The lack of annotated ground-truth datasets for comprehensive document-level VQA poses a significant issue to effective performance evaluation of KG-based RAG. To overcome this limitation, we introduce a new benchmark, DLVQA, for \textbf{d}ocument-\textbf{l}evel \textbf{VQA}, which provides reference summaries and corresponding supporting facts for long-term visual document understanding.  We also propose a new automatic method\footnote{Code and dataset: \url{https://github.com/AI-Application-and-Integration-Lab/KG4VD}.} for multimodal KGs (MMKGs) construction of visually rich documents. The conducted multimodal graph-based RAG approach can handle long document VQA better than previous methods. Experimental results show that our approach outperforms existing MMRAG or KG-based approaches on both DLVQA and existing multi-hop QA/VQA benchmarks.
\end{abstract}

%% file: sec/1_intro.tex
\section{Introduction}
\label{sec:intro}


Although MLLMs~\citep{openai2025gpt5systemcard, bai2025qwen2, comanici2025gemini, meta2025llama4, team2025gemma} are useful for VQA~\citep{kuang2025natural}, their performance remain limited on visually rich documents. Recent MMRAG methods~\citep{zhang2024gme, faysse2024colpali, yu2024visrag} mitigate this by retrieving relevant document pages directly; each page is an image, and the retrieved pages together with the questions are fed into an MLLM (which takes both images and text query as input) for answer generation. However, page-level retrieval alone remains limited for the queries requiring long-range, document-level understanding. These limitations leave two key gaps: reusable multimodal graphs for linking evidence across visually rich documents, and benchmarks for evaluating whether long-range answers are faithful with respect to dispersed textual and visual evidence. 



A promising strategy for the first gap is to construct KGs as global guidance, decomposing documents into entity-centric structures that integrate information across pages. Text-only KGs for RAG have been extensively studied~\citep{gutierrez2025rag, ma2025thinkongraph, guo2024lightrag, edge2024local}. However, automatically constructing MMKGs for visually rich documents are still less studied \cite{guo2025rag, hsiao2025megarag}, which introduces further challenges. First, visual document pages may have highly uneven information density: a sparse title page and a dense figure- or diagram-heavy page consume different KG construction budget. Second, grounding extracted entities and relations to concrete page regions is crucial for verifying whether they actually appear in the document or are hallucinated by the model.

\begin{figure*}
    \centering
    \includegraphics[width=1\linewidth]{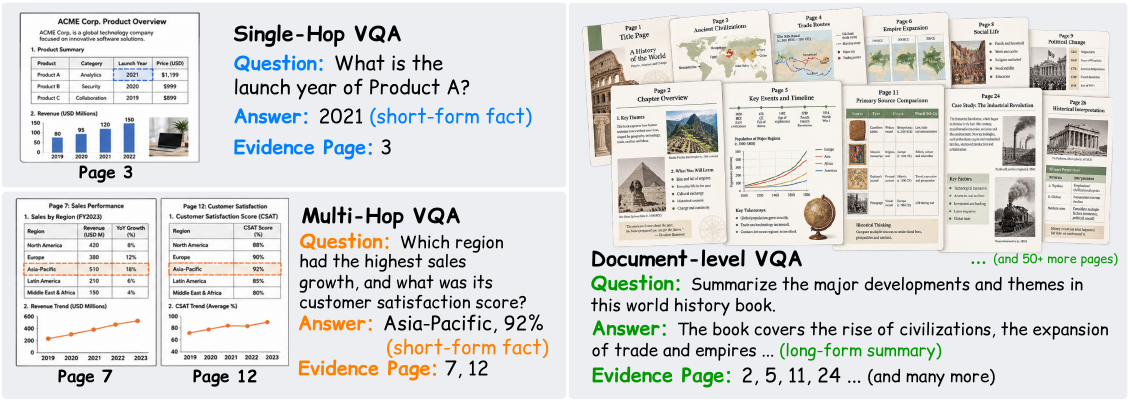}
    \vspace{-20pt}
    \caption{
        \textbf{Single-hop VQA} requires only a single page for question answering. \textbf{Multi-hop VQA} needs to reason over a few pages to produce answers. Both single- and multi-hop VQA tackle only multi-choice or short-answer questions. \textbf{Document-level VQA} requires document-wide understanding and collect dispersed multimodal evidence for answer generation; our \textbf{DLVQA} dataset provides a benchmark for evaluating document-level VQA. Our proposed \textbf{KG4VD} supports all these settings and performs better than previous approaches. 
    }
    \label{fig:kg4vd_teaser}
\end{figure*}



Several recent works have explored MMKG-augmented generation and multimodal graph-based RAG~\citep{lee-etal-2024-multimodal, liu2025aligning, wang2025taming, guo2025rag, hsiao2025megarag}, showing the value of structured multimodal knowledge. Earlier studies rely on manually constructed graphs~\citep{lee-etal-2024-multimodal, 10.1007/978-3-030-21348-0_30, wang2025taming}. Some approach focused social media images~\citep{liu2025aligning}. More recent systems start to target visually rich documents. RAGAnything~\citep{guo2025rag} first parses documents into components such as text, tables, and images, which are then processed separately. MegaRAG~\citep{hsiao2025megarag} incorporates page-level layout cues into MMKG construction. However, RagAnything employs a component-based pipeline which loses important page-level layout information. MegaRAG utilizes the same extraction strategy for all pages, regardless of their information density or structural complexity. 



We propose KG4VD, a zero-shot method for constructing and employing MM\textit{KG}s for \textbf{v}isual \textbf{d}ocuments. KG4VD grounds page-level graphs to layout components and uses adaptive extraction to handle uneven page complexity. It connects page-level graphs into a document-level MMKG for long-range reasoning. 
During inference, KG4VD avoids direct retrieval over the entire MMKG, which can introduce substantial noise. It first retrieves relevant pages as anchors and then applies query-adaptive personalized PageRank (PPR)~\citep{10.1145/775152.775191} to explore related evidence across pages. 
KG4VD 
introduces an MLLM-based query analyzer to adapt the PPR exploration budget to each query. The retrieved graph evidence is then combined with the retrieved page images to generate the final answer. 
These characteristics enable KG4VD to outperform representative automatic MMKG construction methods for RAG \cite{guo2025rag, hsiao2025megarag} on single and multi-hup VQA.


Regarding the second gap, KG (or MMKG) comprehend global information better for a document. However, existing QA/VQA benchmarks~\citep{cheng2024multihop, 10.5555/3737916.3740957, trivedi2022musique, yang2018hotpotqa} usually require evidence from only a small number of pages (eg., \textit{two} to \textit{seven}) with short-answer questions; therefore, they cannot reflect document-level understanding over dispersed evidence. To evaluate document-level KG-RAG performance, current studies employ LLM-as-a judge on only indirect metrics such as comprehensiveness, diversity, empowerment, and overall quality~\citep{guo2024lightrag, edge2024local} of the answers, rather than checking the factual correctness. To address this, we introduce DLVQA, a document-level benchmark with reference summaries and supporting facts for grounded evaluation.


Our DLVQA dataset targets grounded evaluation of document-level VQA with reference summaries and supporting facts.
Figure~\ref{fig:kg4vd_teaser} illustrates the traditional single- or multi-hop VQA settings, and the document-level setting. KG4VD supports all these settings with favorable performance.
Our contributions are summarized as follows.


\noindent $\bullet$~We propose KG4VD, a new zero-shot method that can handle variable page complexity for visually rich documents on automatic MMKG construction. 

\noindent $\bullet$~We introduce DLVQA, the first benchmark to our knowledge for evaluating document-level VQA performance in terms of faithfulness and completeness.


\noindent $\bullet$~KG4VD outperforms existing MMRAG and MMKG-augmented RAG methods on VQA tasks while remaining competitive in text-only settings.

%% file: sec/2_related_work.tex
\section{Related Work}
\label{related_work}


\noindent\textbf{MMRAG.} MMRAG methods address context-window limitations by accessing relevant visual document units instead of feeding the whole document to an MLLM. 
DSE~\citep{ma2024unifying} directly encodes document screenshots, preserving visual layout, text, and images in a unified representation. 
ColPaLi~\citep{faysse2024colpali} introduces multi-vector embeddings for document images to capture fine-grained visual cues, and ColQwen improves retrieval by replacing PaLI-Gemma~\citep{beyer2024paligemma} with Qwen2.5-VL~\citep{bai2025qwen2}. 
VisRAG~\citep{yu2024visrag} extends page-image retrieval to the full RAG pipeline by retrieving document images and reusing them for answer generation.
The above methods mainly focus on text-to-image retrieval and do not support a mixture of single-, cross-, and fused-modality retrieval. UniIR~\citep{wei2023uniir} introduces unified retrieval across these settings. GME~\citep{zhang2024gme} extends this direction by using an MLLM as the encoder, enabling more flexible modality combinations and stronger retrieval performance. CogniVerse~\citep{fang2026cogniverse} explores a complementary direction with geometry-aware retrieval, aligning multimodal representations in a Riemannian rather than Euclidean space.

\begin{figure*}
    \centering
    \includegraphics[width=1\linewidth]{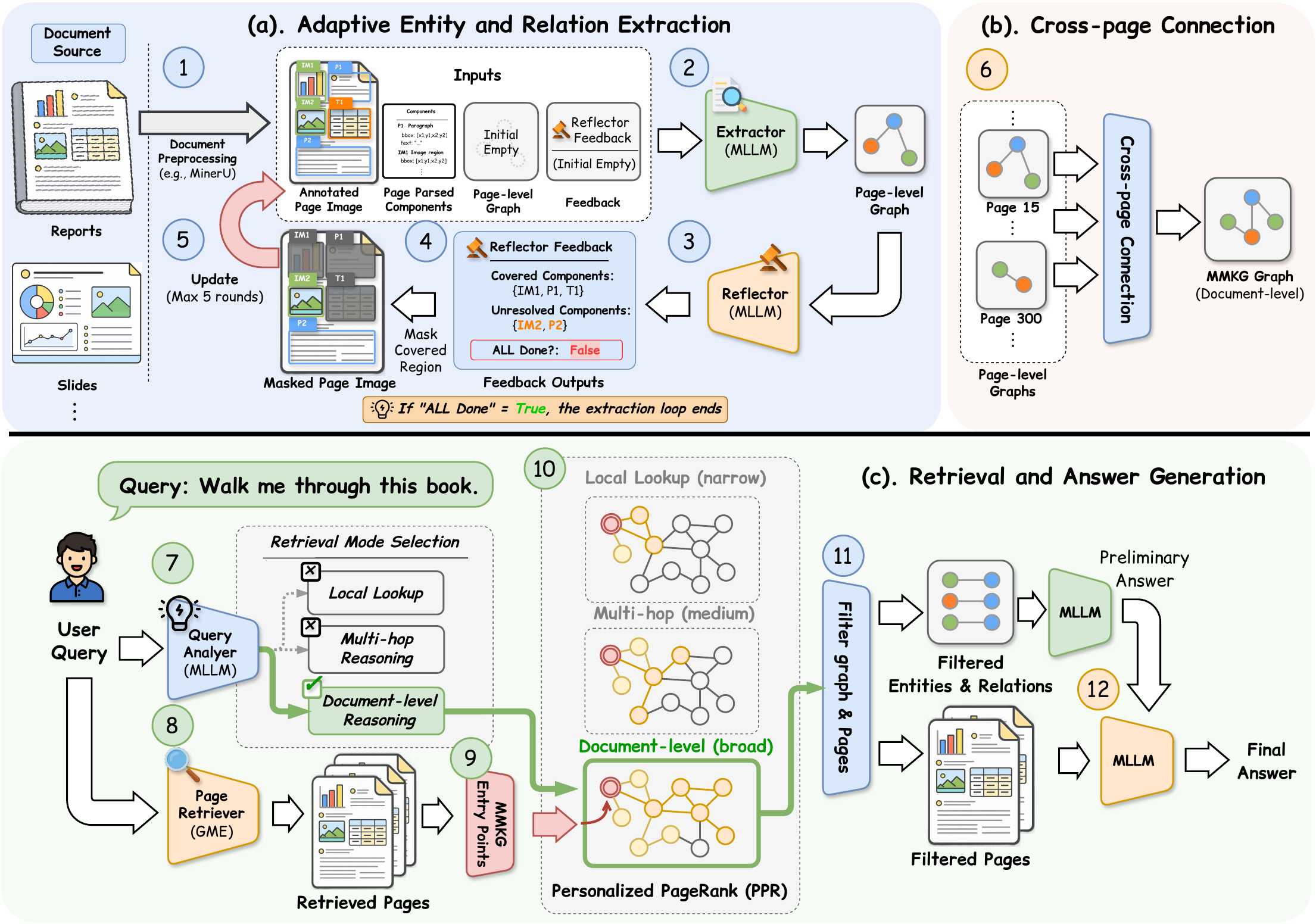}
    \caption{
        \textbf{Overall pipeline of KG4VD.}
        KG4VD constructs a reusable multimodal knowledge graph (MMKG) for visually rich documents and uses it to support retrieval and answer generation. Numbered circles indicate the major processing steps in the pipeline.
        }
    \label{fig:main_diagram_kg4vd}
    \vspace{-10pt}
\end{figure*}


\noindent\textbf{Graph-based RAG.}
Graph-based RAG improves long-range reasoning by pre-organizing documents into entity-relation structures. 
Earlier KG-RAG methods rely on existing KGs~\citep{li2025simple, he2024g, gao-etal-2022-graph, ma2025thinkongraph}, while recent LLM-based systems construct graphs directly from raw documents. 
GraphRAG~\citep{edge2024local} builds text KGs and organizes them with hierarchical community detection. LightRAG~\citep{guo2024lightrag} improves scalability through direct entity-relation retrieval. HippoRAG2~\citep{gutierrez2025rag} incorporates Personalized PageRank and deeper passage integration for multi-hop QA. However, these methods remain text-centric; they do not ground entities and relations to visual regions or figures, layouts, and other elements that are central to visual documents.

\noindent\textbf{Multimodal Graph-based RAG.}
Multimodal knowledge graphs extend conventional KGs by associating entities with visual information 
\citep{zhangmultimodal}, and have been applied to KG completion 
\citep{ xie2017image}, recommendation~\citep{sun2020multi}, captioning~\citep{zhao2023boosting}, and multimodal generation~\citep{bu2025query, lee-etal-2024-multimodal, wang2025taming, liu2025aligning}.
Recent MMKG-based RAG methods have also emerged across different multimodal domains. VideoRAG~\citep{10.1145/3770854.3783944} combines graph-based knowledge grounding with multimodal context retrieval for long-video understanding. M$^3$KG-RAG~\citep{Park_2026_CVPR} focuses on multi-hop audio-visual knowledge retrieval.


Our work focuses on multimodal Graph-RAG for visual documents. There are still few studies in this research track. RAGAnything~\citep{guo2025rag} parses documents into components such as text, tables, and images, and constructs an MMKG from these components. Because the visual components are processed individually, RAGAnything may lose page-level layout context due to its component-based processing. MegaRAG~\citep{hsiao2025megarag} adopts a two-stage construction pipeline. It first constructs an initial MMKG and then refines the graph to revise and complete extracted entities and relations. MegaRAG preserves page structure, but applies the same extraction and refinement procedure to every page regardless of the page complexity. Our KG4VD preserves layout clues too. It adapts graph extraction to each page with self-refinement mechanisms and adopts a refined PPR strategy for multimodal graph retrieval, resulting in higher accuracy, precision and recall than these methods. 

%% file: sec/3_methodology.tex
\section{Methodology}
\label{method}


KG4VD follows a page-centric strategy and further connects evidence across pages through a reusable MMKG for retrieval and answer generation.

\subsection{KG4VD Overview}
\label{sec:kg4vd}

Given a document containing \(N\) pages, KG4VD consists of two phases: 
(1) offline MMKG construction and 
(2) online retrieval and answer generation. 
Figure~\ref{fig:main_diagram_kg4vd} illustrates the overall pipeline.

\noindent\textbf{Offline MMKG construction.}
Graph-based RAG methods typically extract entities and relations from local units and merge them into a document-level graph~\citep{edge2024local, guo2024lightrag, guo2025rag, hsiao2025megarag}. For visually rich documents, this raises two challenges. First, fixed-budget extraction may miss entities and relations on dense pages containing tables, figures, or diagrams. Second, extracted graph elements may be weakly grounded or hallucinated, making their supporting regions hard to verify.

KG4VD addresses these issues with adaptive extraction and layout-component grounding. Adaptive extraction assigns more extraction rounds to complex pages while allowing simple pages to terminate early. Layout grounding links entities and relations to page-region bounding boxes for verification and refinement. The resulting page graphs are then connected into a document-level MMKG.

\noindent\textbf{Online retrieval and answer generation.}
Direct retrieval over the entire MMKG can introduce noise from semantically similar but contextually irrelevant graph elements~\citep{gutierrez2025rag}. KG4VD instead retrieves relevant page images as graph entry points. To support single-hop, multi-hop, and document-level VQA, KG4VD adapts the retrieval scope to each query. A query analyzer determines the required scope, from local page evidence to broader cross-page reasoning. PPR then expands from the retrieved pages to collect related entities and relations. The retrieved evidence is further filtered before answer generation.

\subsection{MMKG Construction}
\label{sec:mmkg_construction}

We define an MMKG as \(\mathcal{G}=(\mathcal{V},\mathcal{E})\). Each node \(v\in\mathcal{V}\) represents an entity and each edge \(e\in\mathcal{E}\) represents a relation between two entities. An entity contains a name, type, modality tag, and description. The modality tag indicates whether the entity is text-derived (such as a person or organization mentioned in a paragraph), or visual (such as a chart element, diagram region, illustrated character, or object in a figure). Each relation contains its source entity, target entity, relation label, and description.


KG4VD performs extraction at the page level to preserve visual, textual, and layout context. Each extracted item is grounded to concrete regions on the page, making the graph easier to inspect and verify.
For each page \(P_i\), we render a page image \(I_i\). A direct approach would ask the MLLM to extract entities and relations while simultaneously predicting their bounding boxes from \(I_i\). However, free-form coordinate prediction is unreliable on complex document pages. 

KG4VD therefore reframes localization as visual region selection. Following a Set-of-Marks (SoM)~\cite{yang2023set} strategy, we first use a document parser~\citep{wang2024mineru} to detect layout components \(\mathcal{C}_i=\{c_{i,1},c_{i,2},\ldots,c_{i,M_i}\}\). Each \(c_{i,j}\) corresponds to a parser-detected page region and contains a component identifier, type, bounding box, and parsed content when available. We draw each component's bounding box and identifier on \(I_i\) to obtain an annotated page image \(\hat{I}_i\). The MLLM then cites component identifiers for extracted entities and relations instead of generating coordinates directly. 

\begin{figure*}
    \centering
    \includegraphics[width=1\linewidth]{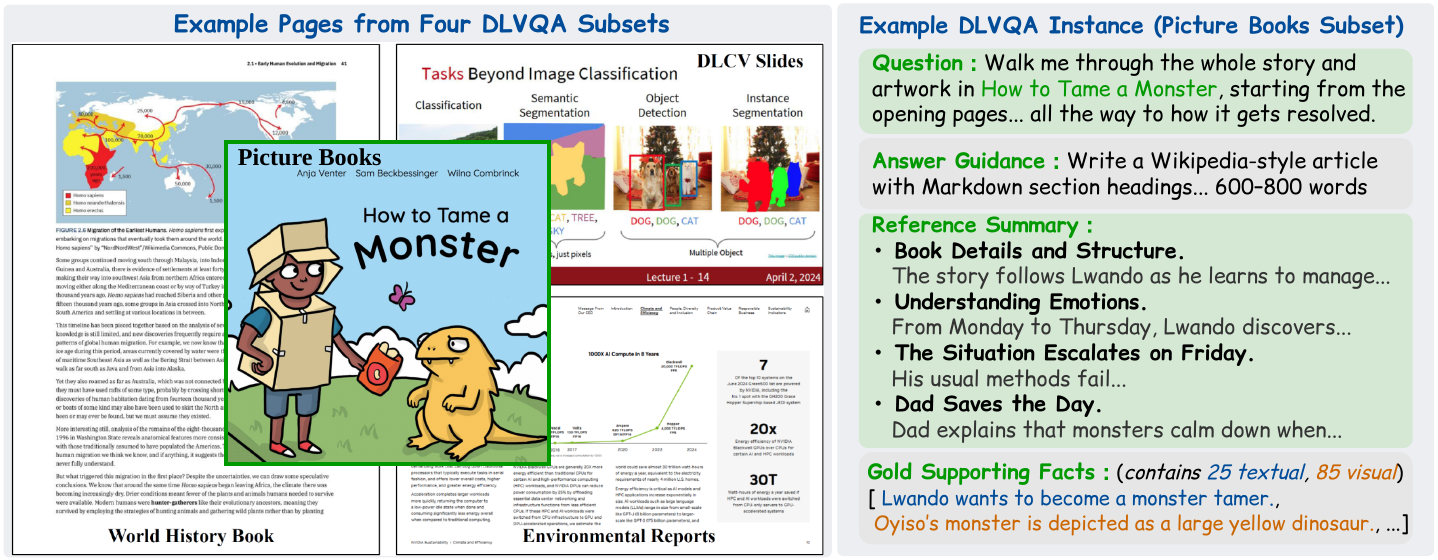}
    \caption{Example pages from the four DLVQA subsets are shown on the left. A representative DLVQA instance on the right includes the question, answer guidance, reference summary, and gold supporting facts. The full instance is provided in Appendix~\ref{sec:appendix_dlvqa}.}
    \label{fig:dlvqa_instance}
    \vspace{-10pt}
\end{figure*}

\subsubsection{Adaptive Entity and Relation Extraction}
\label{sec:adaptive_extraction}

Visual document pages vary greatly in information density. A fixed extraction budget may be inefficient for simple pages and insufficient for complex ones. KG4VD addresses this with an adaptive extract-reflect loop inspired by self-refinement mechanisms~\citep{madaan2023selfrefine, renze2024self}. The loop consists of an \textit{extractor} \(\mathrm{Ext}(\cdot)\), a \textit{reflector} \(\mathrm{Ref}(\cdot)\), and a deterministic controller \(\mathrm{Ctrl}(\cdot)\), as illustrated in Figure~\ref{fig:main_diagram_kg4vd}(a).

For each page \(P_i\), KG4VD maintains a page graph \(\mathcal{G}_i^{(t)}\) at extraction round \(t\), initialized as an empty graph \(\mathcal{G}_i^{(0)}\). Let \(\hat{I}_i^{(t)}\) denote the annotated page used at round \(t\), with \(\hat{I}_i^{(0)}=\hat{I}_i\). In later rounds, regions already judged as sufficiently covered are masked, so as to focus on unresolved content. Let \(B_i^{(t)}\) denote the reflector feedback from the previous round, summarizing covered components, unresolved regions, and graph elements that require refinement.

\noindent\textbf{Workflow.}
At each round, the extractor proposes graph edits \(\Delta_i^{(t)}=\mathrm{Ext}(\hat{I}_i^{(t)},\mathcal{C}_i,\mathcal{G}_i^{(t)},B_i^{(t)})\), which may add, replace, or delete nodes and edges. The controller checks these edits for schema validity, source-component references, and relation endpoints, and applies valid edits to obtain \(\mathcal{G}_i^{(t+1)}=\mathrm{Ctrl}(\mathcal{G}_i^{(t)},\Delta_i^{(t)},\mathcal{C}_i)\). The reflector then reviews the updated graph and returns \((B_i^{(t+1)},s_i^{(t+1)})=\mathrm{Ref}(\hat{I}_i^{(t)},\mathcal{C}_i,\mathcal{G}_i^{(t+1)})\), where \(B_i^{(t+1)}\) guides the next round and \(s_i^{(t+1)}\in\{0,1\}\) indicates whether extraction should stop. If another round is required, KG4VD masks regions judged covered according to \(B_i^{(t+1)}\) and constructs the next annotated page image \(\hat{I}_i^{(t+1)}\). 
We provide an example of reflector feedback \(B_i^{(t+1)}\) in Appendix~\ref{sec:appendix_kg4vd_impl} and the prompts in Appendix~\ref{sec:prompts}.

\noindent\textbf{Extractor.}
At round \(t\), the MLLM extractor takes the annotated page image, the current page graph, and reflector feedback \(B_i^{(t)}\) as input. It outputs a graph patch \(\Delta_i^{(t)}\), which contains a set of entity and relation updates to be applied to the current page graph. Entity updates include fields such as name, type, modality, description, and source components, while relation updates specify source and target entities, relation labels, descriptions, and source components. Each update specifies one of three operations: \textit{add}, \textit{replace}, or \textit{delete}. In the initial round, the extractor mainly adds entities and relations grounded to page components. In later rounds, it uses the current graph and reflector feedback to recover missing items, correct existing elements, refine grounding, or remove duplicates.

\noindent\textbf{Reflector.}
The MLLM reflector identifies coverage gaps in the current graph and returns feedback \(B_i^{(t+1)}\) together with a binary stop signal \(s_i^{(t+1)}\), where \(s_i^{(t+1)}=1\) indicates that extraction is complete. The feedback indicates which components are covered, which remain incomplete, and what the next extraction round should revisit or correct. It helps flag missing entities in a dense text component or request finer-grained entities for a complex visual region.

After reflection, components judged covered are masked while unresolved regions remain visible. This focuses later extraction rounds on content that still requires revision. The loop terminates when \(s_i^{(t+1)}=1\) or when the maximum number of rounds is reached.

\subsection{Cross-Page Connection}
After processing all pages, KG4VD connects the page graphs into a document-level MMKG \(\mathcal{G}=(\mathcal{V},\mathcal{E})\). Following common practice in MMKG construction~\citep{guo2025rag, hsiao2025megarag}, we canonicalize entity names and merge entities with matching canonical forms across pages. Relations incident to merged entities are redirected to the corresponding canonical nodes, and duplicate edges are removed.

\subsection{Retrieval and Answer Generation}
\label{sec:retrieval}

As illustrated in Figure~\ref{fig:main_diagram_kg4vd}(c), KG4VD performs online retrieval in three steps: query analysis, page-anchored graph expansion, and answer generation. It first retrieves relevant pages as visual anchors, and then expands from their grounded entities over the MMKG according to the query scope. The retrieved graph evidence and page images are finally combined for answer generation.

Direct retrieval over the full MMKG can introduce substantial noise. KG4VD uses the PPR-based graph retrieval strategy of HippoRAG2 \cite{gutierrez2025rag}. We extend this strategy with page-image anchoring and query-dependent graph expansion for visually rich documents.

\noindent\textbf{Query-dependent Graph Expansion.}
To support single-hop, multi-hop, and document-level VQA, KG4VD adapts the retrieval scope to each query. An MLLM-based Query Analyzer predicts one of three retrieval modes: \textit{local lookup}, \textit{multi-hop reasoning}, or \textit{document-level reasoning}. Local lookup uses limited graph expansion for fact-seeking questions. Multi-hop reasoning explores connected entities and relations for compositional questions. Document-level reasoning allows broader exploration across pages for questions requiring dispersed evidence.

\noindent\textbf{Page-Anchored PPR Retrieval.}
Given page images \(\{I_i\}_{i=1}^{N}\), KG4VD uses GME \cite{zhang2024gme} to encode each page as \(\mathbf{z}_{P_i}=\mathrm{GME}(I_i)\). Given a query \(q\), KG4VD retrieves a small set of relevant page images and uses them as visual anchors for subsequent graph exploration.

Entities grounded to the retrieved pages are used as seed nodes for PPR over the MMKG \(\mathcal{G}\). The query analyzer selects a retrieval mode and assigns its corresponding PPR exploration budget, defined by the maximum hop depth \(H_{\max}\) and candidate-node budget \(N_{\max}\). Local queries therefore explore a smaller neighborhood, while multi-hop and document-level queries allow progressively broader graph exploration. PPR then retrieves relevant entities, relations, and their grounded page images.

KG4VD retains the top-\(K\) entities and relations according to their PPR scores, together with the top-\(M\) retrieved page images. Answer generation is then performed in two stages. First, an MLLM organizes the retrieved entities and relations into a preliminary answer. Second, the MLLM combines this preliminary answer with the retrieved page images to produce the final grounded response. Details of the formulation and settings are provided in Appendix~\ref{sec:appendix_kg4vd_impl}.

\section{Document-Level VQA Benchmark}
\label{sec:data_construction}

Existing QA/VQA benchmarks mainly target short-answer questions that require evidence from only a small number of pages. They cannot fully capture document-level understanding over dispersed textual and visual evidence. In contrast, existing GraphRAG-style evaluations~\cite{edge2024local, guo2024lightrag} mainly focus on text-only documents and long-form question answering. However, they typically provide only questions without reference answers or annotated supporting facts. As a result, they rely on reference-free LLM judge metrics such as comprehensiveness and overall quality, which do not directly verify the faithfulness of generated answers.

\noindent\textbf{Task Definition.}
DLVQA targets document-level VQA that requires understanding and integrating dispersed textual and visual evidence across a document. Given a multi-page visual document and a question, the model must collect relevant evidence across pages and modalities and generate a complete and grounded response.

\noindent\textbf{DLVQA Instance.}
As illustrated in Figure~\ref{fig:dlvqa_instance}, each instance contains a question, answer guidance, a reference answer summary, and a list of supporting facts. The answer guidance specifies the expected response format and scope. The reference summary describes the expected answer content, while the supporting facts decompose this content into verifiable factual units for evaluation.

\noindent\textbf{Reference-Based Evaluation.}
We match factual claims in each generated answer against the gold supporting facts and compute Answer Fact Precision, Recall, and F1. Precision measures whether generated claims are faithful to the supporting evidence. Recall measures how completely the response covers the required facts. Together, these metrics directly evaluate factual correctness and evidence coverage.

\noindent\textbf{Dataset Construction.}
Due to space constraints, we summarize the construction pipeline here and provide full details, statistics, and examples in Appendix~\ref{sec:appendix_dlvqa}. We construct DLVQA through a human-in-the-loop pipeline assisted by \textit{Gemini-3.1-pro-preview}, with human review at every stage. The model first generates page- and document-level summaries and extracts atomic textual and visual facts from each page. Annotators revise these facts to remove hallucinations and align them with the source document. We then cluster related facts across pages to identify document-level topics. For each cluster, the model generates a question, answer guidance, reference summary, and supporting fact list. Annotators again verify all fields for correctness and evidence coverage. We further filter out instances whose supporting evidence spans fewer than 10 pages to ensure sufficient document-level coverage.

\noindent\textbf{Comparison with GraphRAG-Style Evaluations.}
The final DLVQA dataset contains 525 VQA instances over 3,441 pages from environmental reports, a world-history textbook, computer-vision lecture slides, and picture books. Detailed statistics on evidence distance, modality, and question types are provided in Appendix~\ref{sec:appendix_dlvqa}.

While modest in scale, DLVQA is comparable to recent document-level GraphRAG evaluations~\cite{edge2024local, guo2024lightrag, hsiao2025megarag}, which use 500 questions. DLVQA further provides reference summaries and gold supporting facts for direct evaluation of factual correctness and evidence coverage. We also report GraphRAG-style win-rate metrics in Appendix~\ref{sec:appendix_extra_exp}.

\input{table/main_mmlongbench2k}
\input{table/main_textonly}

\input{table/main_dlvqa_finesure_new}

%% file: table/main_mmlongbench2k.tex
\begin{table*}[t]
\centering
\scriptsize
\setlength{\tabcolsep}{9.5pt}
\renewcommand{\arraystretch}{0.82}

\begin{tabular}{lcccccccc}
\toprule
\multirow{2}{*}{\textbf{Method}} 
& \multicolumn{7}{c}{\textbf{MMLongBench-Doc~\citep{10.5555/3737916.3740957} (2K Packed)}} 
& \multirow{2}{*}{\textbf{Overall}} \\
\cmidrule(lr){2-8}
& \textbf{Res.} 
& \textbf{Tut.} 
& \textbf{Acad.} 
& \textbf{Guid.} 
& \textbf{Broch.} 
& \textbf{Admin.} 
& \textbf{Fin.} 
& \\
\midrule

\multicolumn{9}{c}{\textit{\textbf{Without Retrieval}}} \\
\midrule
No Documents & 27.36 & 17.95 & 23.58 & 17.14 & 20.69 & 26.83 & 10.34 & 20.56 \\

\midrule
\multicolumn{9}{c}{\textit{\textbf{Embedding RAG}}} \\
\midrule
NaiveRAG~\citep{guo2024lightrag} & 37.74 & 25.64 & 27.36 & 22.86 & \underline{41.38} & 31.71 & 37.93 & 32.09 \\
VisRAG~\citep{yu2024visrag} & 44.34 & \underline{48.72} & 23.58 & \underline{40.00} & 37.93 & 31.71 & 41.38 & 38.24 \\
ColQwen~\citep{faysse2024colpali} & \textbf{55.66} & 43.59 & 27.36 & \underline{40.00} & \underline{41.38} & \textbf{41.46} & 43.10 & \underline{41.79} \\
GME~\citep{zhang2024gme} & 48.11 & \textbf{51.28} & 31.13 & 31.43 & 37.93 & \underline{39.02} & \underline{45.30} & 40.60 \\

\midrule
\multicolumn{9}{c}{\textit{\textbf{Text-Only Graph-based RAG}}} \\
\midrule
GraphRAG~\citep{edge2024local} & 39.62 & 20.51 & 24.53 & 14.29 & 27.59 & 29.27 & 24.14 & 25.71 \\
LightRAG~\citep{guo2024lightrag} & 29.25 & 17.95 & 25.47 & 22.86 & 27.59 & 34.15 & 13.79 & 24.44 \\
HippoRAG2~\citep{gutierrez2025rag} & 43.40 & 23.08 & \textbf{33.02} & 25.71 & \underline{41.38} & 36.59 & 34.48 & 33.95 \\

\midrule
\multicolumn{9}{c}{\textit{\textbf{Multimodal Graph-based RAG}}} \\
\midrule
RAGAnything~\citep{guo2025rag} & 42.59 & 15.38 & 32.04 & 30.00 & 32.82 & \textbf{41.46} & 30.17 & 33.71 \\
MegaRAG~\citep{hsiao2025megarag} & \underline{51.42} & 41.03 & \underline{32.55} & 37.14 & 37.93 & 37.80 & 44.83 & 41.18 \\
KG4VD (ours) & \textbf{55.66} & \underline{48.72} & 31.13 & \textbf{45.71} & \textbf{44.83} & \underline{39.02} & \textbf{46.55} & \textbf{44.52} \\

\bottomrule
\end{tabular}

\caption{QA Accuracy (\%) on MMLongBench-Doc (2K).}
\label{tab:mmlongbench_doc_2k_packed_results}
\vspace{-8pt}
\end{table*}

%% file: table/main_textonly.tex
\begin{table}[t]
\centering
\resizebox{\columnwidth}{!}{%
\scriptsize
\begin{tabular}{lcccc}
\toprule
\multirow{2}{*}{\textbf{Method}} 
& \textbf{HotpotQA} 
& \textbf{MuSiQue} 
& \textbf{MultiHopQA} 
& \multirow{2}{*}{\textbf{Overall}} \\
& \citep{yang2018hotpotqa}
& \citep{trivedi2022musique}
& \citep{cheng2024multihop}
&  \\
\midrule

\multicolumn{5}{c}{\textit{\textbf{Without Retrieval}}} \\
\midrule
No Documents & 37.40 & 16.40 & 38.20 & 30.67 \\

\midrule
\multicolumn{5}{c}{\textit{\textbf{Embedding RAG}}} \\
\midrule
NaiveRAG~\citep{guo2024lightrag} & 69.50 & 43.30 & 48.10 & 53.63 \\

\midrule
\multicolumn{5}{c}{\textit{\textbf{Text-only Graph-based RAG}}} \\
\midrule
GraphRAG~\citep{edge2024local} & 44.50 & 33.80 & 46.30 & 41.53 \\
LightRAG~\citep{guo2024lightrag} & 60.00 & 29.70 & 60.50 & 50.07 \\
HippoRAG2~\citep{gutierrez2025rag} & \underline{80.60} & \underline{46.90} & \textbf{73.10} & \underline{66.87} \\

\midrule
KG4VD (ours) & \textbf{81.30} & \textbf{47.00} & \underline{72.50} & \textbf{66.93} \\

\bottomrule
\end{tabular}%
}
\caption{QA Accuracy (\%) on text-only QA setting.}
\label{tab:multihop_qa_results}
\end{table}

%% file: table/main_dlvqa_finesure_new.tex
\begin{table}[t]
\centering
\scriptsize
\setlength{\tabcolsep}{15pt}
\renewcommand{\arraystretch}{0.85}

\begin{tabular}{lccc}
\toprule
\multirow{2}{*}{\textbf{Method}} 
& \multicolumn{3}{c}{\textbf{DLVQA}} \\
\cmidrule(lr){2-4}
& \textbf{Precision} 
& \textbf{Recall} 
& \textbf{F1} \\
\midrule

\multicolumn{4}{c}{\textit{\textbf{Without Retrieval}}} \\
\midrule
No Documents & 55.48 & 19.33 & 28.67 \\

\midrule
\multicolumn{4}{c}{\textit{\textbf{Embedding RAG}}} \\
\midrule
NaiveRAG~\citep{guo2024lightrag} & 76.76 & 23.10 & 35.51 \\
VisRAG~\citep{yu2024visrag} & 78.38 & 27.89 & 41.14 \\
ColQwen~\citep{faysse2024colpali} & 82.26 & 28.05 & 41.83 \\
GME~\citep{zhang2024gme} & \textbf{83.93} & 28.33 & 42.36 \\

\midrule
\multicolumn{4}{c}{\textit{\textbf{Text-Only Graph-based RAG}}} \\
\midrule
GraphRAG~\citep{edge2024local} & 72.93 & 36.23 & 48.41 \\
LightRAG~\citep{guo2024lightrag} & 73.94 & 37.61 & 49.86 \\
HippoRAG2~\citep{gutierrez2025rag} & 76.60 & 37.84 & 50.66 \\

\midrule
\multicolumn{4}{c}{\textit{\textbf{Multimodal Graph-based RAG}}} \\
\midrule
RAGAnything~\citep{guo2025rag} & 72.64 & 40.38 & 51.91 \\
MegaRAG~\citep{hsiao2025megarag} & 74.20 & \underline{41.16} & \underline{52.95} \\
KG4VD (ours) & \underline{82.43} & \textbf{42.45} & \textbf{56.04} \\

\bottomrule
\end{tabular}

\caption{Main results on DLVQA dataset. We report Answer Fact Precision, Recall, and F1.}
\label{tab:dlvqa_fact_results}
\vspace{-8pt}
\end{table}

%% file: sec/4_experiments.tex
\section{Experiments}
\label{experiments}

We evaluate KG4VD on multimodal single and multi-hop VQA, text-only QA, and document-level VQA.

\noindent\textbf{Datasets.}
For multimodal single- and multi-hop VQA, we use MMLongBench-Doc~\citep{10.5555/3737916.3740957}, a long-document benchmark that combines documents from DUDE~\citep{van2023document}, SlideVQA~\citep{tanaka2023slidevqa}, ChartQA~\citep{masry2022chartqa}, and FinanceBench~\citep{islam2023financebench}, together with newly collected documents. It spans seven domains and includes both single-hop and multi-hop VQA requiring approximately 2$\sim$7 reasoning hops. To further stress long-range retrieval, we construct a packed-document setting by combining 48 documents into a 2,000-page input with 414 queries.
For text-only single and multi-hop QA, we follow HippoRAG2~\citep{gutierrez2025rag} and sample 1,000 queries each from HotpotQA~\citep{yang2018hotpotqa}, MuSiQue~\citep{trivedi2022musique}, and MultiHopQA~\citep{cheng2024multihop}.
For document-level VQA, we use DLVQA, which contains 525 QA pairs over 3,441 pages from environmental reports, a world history textbook, computer vision slides, and picture books. DLVQA targets long-form reasoning over dispersed evidence and provides reference summaries and supporting facts for grounded evaluation.


\noindent\textbf{Baselines.}
We compare KG4VD with MMRAGs and graph-based baselines, including \textit{No Documents}, NaiveRAG, VisRAG~\citep{yu2024visrag}, ColQwen~\citep{faysse2024colpali}, GME~\citep{zhang2024gme}, GraphRAG~\citep{edge2024local}, LightRAG~\citep{guo2024lightrag}, HippoRAG2~\citep{gutierrez2025rag}, RAGAnything~\citep{guo2025rag}, and MegaRAG~\citep{hsiao2025megarag}. The \textit{No Documents} setting provides no document input and tests whether questions can be answered from the model's internal knowledge alone. For methods without native multimodal support, we use MinerU~\citep{wang2024mineru} to extract text from visual documents. We do not include CogniVerse~\citep{fang2026cogniverse}, as it relies on a trained retrieval model whose implementation has not been released.


\noindent\textbf{Metrics.}
For short-answer QA, we use LLM judges to assess semantic equivalence and report accuracy. For DLVQA, we match generated answer sentences against gold supporting facts and report Answer Fact Precision, Recall, and F1. We use two judge models, \textit{GPT-5.4-mini} and \textit{Gemini-3.1-flash-lite}, to reduce model-specific bias and report their averaged scores. We provide detailed metric definitions in Appendix~\ref{sec:appendix_judge}.


\noindent\textbf{Implementation Details.}
Following GraphRAG~\citep{edge2024local}, LightRAG~\citep{guo2024lightrag}, and MegaRAG~\citep{hsiao2025megarag}, we use \textit{GPT-4o-mini} with temperature 0 for graph construction and answer generation. All methods share the same answer prompt. All page-retrieval methods use a top-10 page budget, and graph-based methods use the same predefined entity type schema when applicable, except HippoRAG2, which is schema-free.
For KG4VD, we use GME for multimodal indexing and \textit{text-embedding-3-small} for text-only settings. The adaptive extraction loop is capped at 5 rounds. We provide the code and dataset in the supplementary material.

\subsection{Main Results}

\noindent\textbf{Multimodal VQA.}
Table~\ref{tab:mmlongbench_doc_2k_packed_results} reports results on MMLongBench-doc. The low performance of \textit{No Documents} confirms the need for document grounding. MMRAGs such as ColQwen and GME outperform most text-only graph-based methods, highlighting the importance of visual information. Among multimodal graph baselines, RAGAnything reaches 33.71\% overall and MegaRAG improves to 41.18\%, showing the benefit of multimodal graph retrieval. KG4VD achieves the best overall accuracy of 44.52\%, outperforming ColQwen by 2.73 points and MegaRAG by 3.34 points. Its weaker performance on Academic documents suggests that dense pages with fine textual details may still favor text-based methods.

\noindent\textbf{Text-only QA.}
Table~\ref{tab:multihop_qa_results} reports results on HotpotQA, MuSiQue, and MultiHopQA. KG4VD obtains 66.93\% overall accuracy, comparable to HippoRAG2 at 66.87\%, indicating that KG4VD preserves text-only QA performance.

\noindent\textbf{Document-level VQA.}
Table~\ref{tab:dlvqa_fact_results} reports Answer Fact Precision, Recall, and F1 on DLVQA. The \textit{No Documents} baseline performs poorly, confirming the need for document evidence. Embedding-based RAG achieves high precision but low recall, often missing evidence distributed across pages. Graph-based RAG improves recall by aggregating broader evidence. KG4VD achieves the best recall and F1 while maintaining high precision, showing a better balance between evidence coverage and retrieval noise.

\input{table/ablation_study_v1}
\input{table/long_mllm}

\subsection{Detailed Analysis}

\noindent\textbf{Ablation and Query Analysis.}
Table~\ref{tab:retrieval_analysis} evaluates the main components of KG4VD. Replacing adaptive extraction with single-pass extraction mainly reduces evidence coverage, showing the benefit of allocating more extraction rounds to complex pages. Replacing query-adaptive PPR with direct MMKG retrieval introduces more irrelevant evidence and lowers performance. The query analyzer also achieves high accuracy, supporting reliable retrieval-mode selection.

\noindent\textbf{Comparison with Long-context MLLMs.}
We evaluate whether directly providing the full document to long-context MLLMs can replace retrieval. Gemini 2.5 Pro and GPT-5.1 achieve substantially lower precision and F1 than KG4VD, as shown in Table~\ref{tab:long_context_mllm}. This result is consistent with context rot, where performance degrades as relevant evidence becomes more dispersed within long input contexts.

\noindent\textbf{MMKG Construction Quality.}
We evaluate whether KG4VD constructs more reliable MMKGs. To our knowledge, existing visual-document benchmarks do not provide gold entity and relation annotations. We therefore construct human-verified references on sampled DLVQA pages with assistance from Gemini-3.1 Pro. Table~\ref{tab:mmkg_quality} shows that KG4VD achieves substantially higher entity extraction precision and F1 than RAGAnything and MegaRAG, together with the best relation F1. It also achieves 81.0\% layout-grounding accuracy, a capability not provided by prior methods. Due to space constraints, we report the inference cost comparison in Appendix~\ref{sec:appendix_extra_exp}.

\noindent\textbf{MMKG Construction Cost.}
Table~\ref{tab:computational_cost} compares the MMKG construction cost of KG4VD and MegaRAG. MegaRAG uses two document-wide stages. Its second-stage refinement can begin only after the initial MMKG has been fully constructed. KG4VD instead performs adaptive extraction independently for each page, so one page does not need to wait for the others.

KG4VD averages 3.22 and 3.82 extraction rounds per page on DLVQA World History and ESG Reports, respectively. Later rounds process only unresolved regions rather than repeating full-page extraction. Its token usage is therefore comparable to or moderately higher than MegaRAG. The lower recall of MegaRAG and RAGAnything in Table~\ref{tab:mmkg_quality} suggests under-extraction. KG4VD therefore uses additional extraction rounds to improve graph coverage, at the cost of moderately higher token usage.

\input{table/mmkg_quality}
\input{table/appendix_compute_overhead}



%% file: table/ablation_study_v1.tex





\begin{table}[t]
\centering
\scriptsize
\setlength{\tabcolsep}{4.5pt}
\renewcommand{\arraystretch}{1.05}

\newsavebox{\ablationbox}

\savebox{\ablationbox}{%
\begin{tabular}{lcccc}
\toprule
\multicolumn{5}{c}{\textbf{(a) Ablation Study on DLVQA}} \\
\midrule
\textbf{Variant}
& \textbf{Precision}
& \textbf{Recall}
& \textbf{F1}
& \(\boldsymbol{\Delta}\) \\
\midrule
KG4VD (ours)
& \textbf{82.43}
& \textbf{42.45}
& \textbf{56.04}
& -- \\

w/o Adaptive Extraction
& 80.35
& 38.20
& 51.78
& -4.26 \\

w/o Query-Adaptive PPR
& 77.90
& 39.85
& 52.73
& -3.31 \\
\bottomrule
\end{tabular}%
}

\usebox{\ablationbox}

\vspace{4pt}

\begin{tabularx}{\wd\ablationbox}{Xrrr}
\toprule
\multicolumn{4}{c}{\textbf{(b) Query Analyzer Accuracy}} \\
\midrule
\textbf{Subset}
& \textbf{Queries}
& \textbf{Correct}
& \textbf{Accuracy} \\
\midrule
Single-hop
& 187
& 164
& 87.70 \\

Multi-hop
& 130
& 108
& 83.08 \\

Document-level
& 525
& 505
& 96.19 \\
\midrule
Overall
& 842
& 797
& 94.66 \\
\bottomrule
\end{tabularx}

\caption{Analysis of KG4VD retrieval components.
(a) Ablation study on DLVQA, where \(\Delta\) denotes the Answer Fact F1 change relative to the full model.
(b) Accuracy of the query analyzer used for query-adaptive PPR retrieval in MMLongBench and DLVQA.}
\label{tab:retrieval_analysis}
\end{table}

%% file: table/long_mllm.tex
\begin{table}[t]
\centering
\scriptsize
\setlength{\tabcolsep}{4.5pt}
\renewcommand{\arraystretch}{1.12}
\begin{tabular}{lccc}
\toprule
\textbf{Method} & \textbf{Precision} & \textbf{Recall} & \textbf{F1} \\
\midrule
GPT-5.1 (full doc) & 30.18 & 30.09 & 30.13 \\
Gemini 2.5 Pro (full doc) & 39.53 & 31.72 & 35.21 \\
KG4VD & \textbf{80.03} & \textbf{34.96} & \textbf{48.66} \\
\bottomrule
\end{tabular}
\caption{Long-context MLLM results on DLVQA World History.}
\label{tab:long_context_mllm}
\end{table}

%% file: table/mmkg_quality.tex
\begin{table}[t]
\centering
\resizebox{\columnwidth}{!}{%
\scriptsize
\setlength{\tabcolsep}{3.5pt}
\renewcommand{\arraystretch}{1.08}
\begin{tabular}{llcccc}
\toprule
\textbf{Method} & \textbf{Metric} & \textbf{Precision} & \textbf{Recall} & \textbf{F1} & \textbf{Layout} \\
\midrule
\multirow{2}{*}{RAGAnything~\cite{guo2025rag}}
& Entity   & 14.21 & 32.05 & 19.69 & \multirow{2}{*}{\textsc{N/A}} \\
& Relation & 19.34 & 55.58 & 28.70 & \\

\midrule
\multirow{2}{*}{MegaRAG~\cite{hsiao2025megarag}}
& Entity   & 42.50 & 43.59 & 43.04 & \multirow{2}{*}{\textsc{N/A}} \\
& Relation & 80.81 & 57.44 & 67.15 & \\

\midrule
\multirow{2}{*}{KG4VD (ours)}
& Entity   & \textbf{86.18} & \textbf{46.15} & \textbf{60.11}
& \multirow{2}{*}{\textbf{81.00}} \\
& Relation & \textbf{84.45} & \textbf{67.67} & \textbf{75.13} & \\
\bottomrule
\end{tabular}
}
\caption{Evaluation of MMKG construction quality.}
\label{tab:mmkg_quality}
\end{table}

%% file: table/appendix_compute_overhead.tex
\begin{table}[t]
\centering
\scriptsize
\setlength{\tabcolsep}{4.5pt}
\renewcommand{\arraystretch}{1.08}

\begin{tabular}{llcc}
\toprule
\textbf{Dataset}
& \textbf{Metric}
& \textbf{MegaRAG~\cite{hsiao2025megarag}}
& \textbf{KG4VD} \\
\midrule

\multirow{3}{*}{World History (788)}
& Time (min)
& 31.0
& 24.5 \\

& Tokens (M)
& 38.2
& 40.7 \\

& Avg. Rounds/Page
& 2.0 
& 3.2 \\

\midrule

\multirow{3}{*}{Env. Report (422)}
& Time (min)
& 14.0
& 10.2 \\

& Tokens (M)
& 54.3
& 64.8 \\

& Avg. Rounds/Page
& 2.0 
& 3.8 \\

\bottomrule
\end{tabular}

\caption{MMKG construction cost. 
}
\label{tab:computational_cost}
\vspace{-10pt}
\end{table}

%% file: sec/5_conclusion.tex

\section{Conclusion}
In this paper, we presented KG4VD, a graph-based RAG method for automatically constructing and using MMKGs from visually rich documents. KG4VD addresses key challenges in document-level VQA through adaptive page-level extraction, which handles varying page complexity, and query-adaptive PPR retrieval, which balances visual grounding and long-range reasoning. We also introduced DLVQA, a document-level VQA benchmark with supporting facts for grounded evaluation. Experiments on open VQA benchmarks and DLVQA show that KG4VD consistently outperforms strong MMRAG and graph-based baselines across settings.
Future work will explore integrating KG4VD with advanced graph reasoning methods~\citep{ma2025thinkongraph}.

%% file: appendix.tex
\section*{Appendix}

This appendix is organized as follows.

\noindent $\bullet$~
\textbf{Appendix~\ref{sec:appendix_dlvqa}: DLVQA Dataset and Construction.}\\
\textit{Provides detailed dataset statistics, source-document descriptions, the human-in-the-loop construction pipeline, annotation and review procedures, and representative DLVQA examples.}\\
\noindent $\bullet$~
\textbf{Appendix~\ref{sec:appendix_kg4vd_impl}: KG4VD Implementation Details.}\\
\textit{Describes the entity schema, adaptive page-level MMKG construction, query analyzer, page-anchored PPR retrieval, mode-specific retrieval budgets, and two-stage answer generation.}\\
\noindent $\bullet$~
\textbf{Appendix~\ref{sec:appendix_extra_exp}: Additional Experimental Results.}\\
\textit{Provides comparisons with specialized document RAG systems, reference-free and domain-wise DLVQA results, online inference latency, and an analysis of fixed vs.\ query-adaptive PPR budgets.}\\
\noindent $\bullet$~
\textbf{Appendix~\ref{sec:appendix_qualitative}: Qualitative MMKG Analysis.}\\
\textit{Analyzes graph size and extraction behavior across DLVQA domains, shows how adaptive extraction evolves across rounds, and presents representative MMKG construction successes and errors.}\\
\noindent $\bullet$~
\textbf{Appendix~\ref{sec:appendix_judge}: Evaluation Metrics and LLM Judges.}\\
\textit{Defines Answer Fact Precision, Recall, and F1, explains the sentence-fact matching procedure, describes the judge models and score aggregation, reports inter-judge agreement, and provides evaluation examples.}\\
\noindent $\bullet$~
\textbf{Appendix~\ref{sec:llm_usage}: The Use of Large Language Models.}\\
\textit{Summarizes where LLMs and MLLMs are used in MMKG construction, DLVQA construction, query analysis, answer generation, and evaluation.}\\
\noindent $\bullet$~
\textbf{Appendix~\ref{sec:prompts}: MLLM Prompts.}\\
\textit{Provides the complete prompt templates used for adaptive page-level MMKG construction, query analysis, preliminary and final multimodal answer generation.}

\section{DLVQA Dataset and Construction}
\label{sec:appendix_dlvqa}

\subsection{Dataset Overview}
\label{sec:appendix_dlvqa_overview}

DLVQA evaluates document-level visual question answering over long, visually rich, multi-page documents.
Unlike localized VQA benchmarks, where the relevant evidence is typically confined to one page or a small page set, DLVQA requires models to aggregate dispersed evidence across long page ranges and, in many cases, across both textual and visual modalities.
The final dataset contains 525 QA pairs over 3,441 pages from four document collections: deep learning for computer vision slides, a world history textbook, environmental reports, and picture books.
Each QA instance contains a question, answer guidance, a reference summary, and gold supporting facts.

Table~\ref{tab:dlvqa_stats} summarizes corpus size, question characteristics, reference-summary length, grounding density, visual-fact ratio, and evidence span.
Several collections contain multiple coherent sub-documents, such as separate picture books, slide decks, or environmental reports from different companies and years.
Questions are generated from clusters of semantically related atomic facts rather than by forcing unrelated distant pages into one instance.
The reported spans therefore reflect coherent evidence groups, while the maximum spans show that DLVQA still contains extremely long-range questions.

\input{table/appendix_dlvqa_stats}
\input{table/dlvqa_construction_1}

\subsection{Source Documents}
\label{sec:appendix_dlvqa_sources}

\noindent\textbf{Deep Learning for Computer Vision Slides.}
This collection consists of 18 slide decks from a deep learning for computer vision course\footnote{\scriptsize{\url{https://cs231n.stanford.edu/slides/2024/}}}, totaling 1,984 pages, 2,018 figures, 75 tables, and 136,000 text tokens.
The slides cover image classification, object detection, visual recognition, generative models, and societal impacts of AI.
They provide a technical setting with dense slide layouts, equations, diagrams, visual examples, and concise explanatory text.

\noindent\textbf{Environmental Reports.}
This collection contains 5 corporate sustainability reports, totaling 422 pages, 416 figures, 122 tables, and 229,000 text tokens.
It includes reports from Google\footnote{\scriptsize{\url{https://sustainability.google/reports/google-2024-environmental-report/}}}, Apple\footnote{\scriptsize{\url{https://www.apple.com/environment/pdf/Apple_Environmental_Progress_Report_2024.pdf}}}, Microsoft\footnote{\scriptsize{\url{https://www.microsoft.com/en-us/corporate-responsibility/sustainability/report}}}, Meta\footnote{\scriptsize{\url{https://sustainability.atmeta.com/2024-sustainability-report/}}}, and NVIDIA\footnote{\scriptsize{\url{https://www.nvidia.com/en-us/sustainability/}}}.
These documents contain business narratives, tables, charts, and figures related to carbon reduction, energy use, supply chains, sustainability governance, and corporate environmental targets.

\noindent\textbf{World History Textbook.}
This textbook\footnote{\scriptsize{\url{https://open.umn.edu/opentextbooks/textbooks/1418}}} contains 788 pages, 468 figures, 5 tables, and 442,000 text tokens.
It traces global developments from prehistory to 1500 CE, covering early civilizations, empires, religious movements, trade, migration, and intercultural exchange.
This domain represents long-form narrative documents in which relevant evidence is distributed across chapters and often supported by maps, illustrations, and historical context.

\noindent\textbf{Picture Books.}
The picture-book collection is drawn from Free Kids Books\footnote{\scriptsize{\url{https://www.freekidsbooks.org/}}} and contains 10 books, totaling 247 pages, 240 figures, and fewer than 10,000 text tokens.
Most books contain sparse text and rely heavily on illustrations for narrative progression.
This domain tests whether models can aggregate visual evidence when textual cues are limited or absent.

\subsection{Human-in-the-loop Construction Pipeline}
\label{sec:appendix_dlvqa_pipeline}

DLVQA is constructed through a human-in-the-loop pipeline assisted by \textit{Gemini-3.1-pro-preview}.
The pipeline converts long documents into traceable intermediate artifacts, including page summaries, document summaries, grounded atomic facts, multi-page fact clusters, and document-level VQA instances.
All MLLM-assisted outputs are reviewed by two annotators with computer science backgrounds before being retained.
Questions are generated from fact clusters rather than document outlines alone, grounding each instance in concrete evidence distributed across the source document.

\noindent\textbf{Page and document summarization.}
For each document, the MLLM first produces a summary for every page image.
The page summaries are then aggregated into a document-level summary.
Annotators revise the document summaries for faithfulness, coverage, and clarity.
The revised summaries provide global context for subsequent fact extraction and question generation.

\noindent\textbf{Atomic fact extraction.}
The MLLM extracts self-contained textual and visual atomic facts from each page, conditioned on the page image, page summary, and document-level summary.
Each fact is stored with page-level provenance.
Textual facts describe information stated in paragraphs, tables, or other text regions, while visual facts describe information conveyed by figures, charts, maps, diagrams, or illustrations.
Annotators remove redundant, weakly grounded, or uninformative facts; revise unclear or incomplete facts; and add missing facts when necessary.

\noindent\textbf{Topic clustering and selection.}
We encode atomic facts using OpenAI's \texttt{text-embedding-3-large} model and cluster the resulting embeddings across pages with HDBSCAN~\citep{mcinnes2017hdbscan}.
This groups semantically related evidence without requiring a fixed number of clusters and filters sparse noise points.
The resulting clusters serve as candidate evidence groups for document-level questions.
Annotators remove clusters that are noisy, incoherent, or insufficiently useful for long-range reasoning.

\noindent\textbf{QA generation and validation.}
For each retained cluster, the MLLM generates a document-level VQA draft using the cluster facts and revised document summary.
Each draft contains a question, a reference summary following STORM~\citep{shao-etal-2024-assisting}, answer guidance, and gold supporting facts.
The answer guidance specifies only the expected response style, organization, and length range, such as a Wikipedia-style article with Markdown section headings and flowing paragraphs.
It does not reveal the reference summary, gold facts, or question-specific evidence.
The same guidance is provided to all evaluated methods to reduce variation caused by formatting choices or overly short responses.
Annotators revise unclear or inaccurate drafts, remove invalid drafts, and verify that the reference summary is supported by the associated gold facts.
Finally, we filter out instances whose supporting evidence spans fewer than 10 pages.
The retained QA pairs are used to evaluate factual precision and evidence coverage in long-form answers.

\subsection{Human Review and Agreement}
\label{sec:appendix_dlvqa_agreement}

Table~\ref{tab:dlvqa_pipeline_transparency} summarizes the construction and review statistics for DLVQA.
The left block reports the inputs and outputs of MLLM-assisted stages, while the right block reports human curation actions and the artifacts retained after review.

\noindent\textbf{Two-reviewer protocol.}
For stages requiring human judgment, two annotators independently inspect the MLLM-assisted outputs using stage-specific guidelines.
Before discussion, each annotator records a binary decision, such as keep/revise or valid/invalid, and suggests edits when needed.
We compute raw agreement and Cohen's kappa from these independent pre-adjudication decisions.

After agreement is measured, the annotators discuss disagreements and reach a consensus decision for each disputed item.
Artifacts accepted by both annotators are retained.
Artifacts with fixable issues, such as factual errors, unclear wording, incomplete evidence, or incorrect grounding, are revised according to the consensus decision.
Artifacts judged invalid or not reliably correctable are removed.
Thus, the agreement scores in Table~\ref{tab:dlvqa_pipeline_transparency} reflect reviewer consistency before adjudication, whereas the curation counts and retained artifacts reflect the final consensus-reviewed dataset.

\subsection{DLVQA Examples}
\label{sec:appendix_dlvqa_image_questions}

We present one DLVQA example from each subset.
The examples below illustrate how DLVQA requires models to aggregate textual and visual evidence across multiple pages while following a specified long-form answer format.

\input{table/dlvqa_examples_appendix}

\section{KG4VD Implementation Details}
\label{sec:appendix_kg4vd_impl}

This section provides the entity schema, adaptive page-level extraction algorithm, and query-adaptive retrieval settings used by KG4VD.

\subsection{Entity Type Schema}
\label{sec:appendix_entity_types}

Table~\ref{tab:entity_types} lists the predefined entity-type schema used by KG4VD and graph-based baselines when applicable.
We define the schema after inspecting documents from each DLVQA corpus.
For comparability, all graph-based methods share the same corpus-specific textual entity schema.
KG4VD additionally uses shared visual entity types for entities grounded to figures, charts, tables, and layout regions.
Text-only GraphRAG baselines use only textual entity types because they operate on parsed text rather than page images.
HippoRAG2 is excluded from this constraint because it is schema-free by design.

\input{table/appendix_entity_types}

\subsection{Adaptive MMKG Construction}
\label{sec:appendix_page_graph}

For each page $P_i$, KG4VD takes as input the annotated page image $\hat{I}_i$, layout components $\mathcal{C}_i$, the current page graph $\mathcal{G}_i^{(t)}$, and reflector feedback $B_i^{(t)}$.
The extractor proposes graph edits, the deterministic controller applies only schema-valid and grounded edits, and the reflector returns updated feedback together with a stop signal.
Algorithm~\ref{alg:extract_reflect} summarizes this adaptive loop.

\begin{algorithm}[t]
\small
\caption{Adaptive Page-level MMKG Construction}
\label{alg:extract_reflect}
\begin{algorithmic}[1]
\Require annotated page image $\hat{I}_i$, layout components $\mathcal{C}_i$, maximum rounds $R$
\State $\mathcal{G}_i^{(0)} \gets \emptyset$
\State $B_i^{(0)} \gets \emptyset$
\State $\hat{I}_i^{(0)} \gets \hat{I}_i$
\For{$t = 0, \dots, R-1$}
  \State $\Delta_i^{(t)} \gets
  \mathrm{Ext}(\hat{I}_i^{(t)}, \mathcal{C}_i,
  \mathcal{G}_i^{(t)}, B_i^{(t)})$
  \State $\tilde{\Delta}_i^{(t)} \gets
  \mathrm{Ctrl}(\mathcal{G}_i^{(t)}, \Delta_i^{(t)}, \mathcal{C}_i)$
  \State $\mathcal{G}_i^{(t+1)} \gets
  \mathcal{G}_i^{(t)} \oplus \tilde{\Delta}_i^{(t)}$
  \State $(B_i^{(t+1)}, s_i^{(t+1)}) \gets
  \mathrm{Ref}(\hat{I}_i^{(t)}, \mathcal{C}_i,
  \mathcal{G}_i^{(t+1)})$
  \If{$s_i^{(t+1)} = 1$}
    \State $\mathcal{G}_i^* \gets \mathcal{G}_i^{(t+1)}$
    \Return $\mathcal{G}_i^*$
  \EndIf
  \State update $\hat{I}_i^{(t+1)}$ by masking components marked
  \texttt{covered} in $B_i^{(t+1)}$
\EndFor
\State $\mathcal{G}_i^* \gets \mathcal{G}_i^{(R)}$
\Return $\mathcal{G}_i^*$
\end{algorithmic}
\end{algorithm}

\noindent\textbf{Reflector feedback and masking.}
The reflector produces a component-level coverage report in a structured format.
For each layout component, it predicts whether the component is \texttt{covered}, \texttt{partially\_covered}, or \texttt{uncovered}; summarizes remaining gaps; lists unresolved regions and revision targets; and returns a stop signal.
The feedback guides the next extractor call and determines which components remain visible in the next annotated page image.
Figure~\ref{fig:reflector_example} shows an example after the first extraction round on a page containing a paragraph component \texttt{P1}, a table component \texttt{T1}, and an image component \texttt{IM1}.

In this example, the next extractor round should add table-specific entities grounded to \texttt{T1}, add visual entities grounded to \texttt{IM1}, and revise relations connecting the table and figure evidence.
Components listed in \texttt{covered\_components}, such as \texttt{P1}, can be masked in the next round.
The loop terminates when the reflector returns \texttt{stop\_signal: 1} or the maximum of five rounds is reached.

\begin{figure*}[t]
\centering
\begin{tcolorbox}[
  width=\textwidth,
  colback=gray!3,
  colframe=black!65,
  boxrule=0.5pt,
  arc=1.5mm,
  left=1mm,
  right=1mm,
  top=1mm,
  bottom=1mm,
  fonttitle=\bfseries,
  title={Example output from the reflector}
]
\small
\texttt{\{}\\
\hspace*{1em}\texttt{"summary":}
\texttt{"The current page graph captures the main textual content in P1,
but the table evidence in T1 and the visual evidence in IM1 are still
under-specified.",}\\
\hspace*{1em}\texttt{"per\_component": [}\\
\hspace*{2em}\texttt{\{}\\
\hspace*{3em}\texttt{"component\_id": "P1",}\\
\hspace*{3em}\texttt{"status": "covered",}\\
\hspace*{3em}\texttt{"gap\_summary": "The main textual entities and
relations are grounded."}\\
\hspace*{2em}\texttt{\},}\\
\hspace*{2em}\texttt{\{}\\
\hspace*{3em}\texttt{"component\_id": "T1",}\\
\hspace*{3em}\texttt{"status": "partially\_covered",}\\
\hspace*{3em}\texttt{"gap\_summary": "Several key table entries and
comparisons are missing from the graph."}\\
\hspace*{2em}\texttt{\},}\\
\hspace*{2em}\texttt{\{}\\
\hspace*{3em}\texttt{"component\_id": "IM1",}\\
\hspace*{3em}\texttt{"status": "uncovered",}\\
\hspace*{3em}\texttt{"gap\_summary": "The figure contains visual
elements that are not represented as visual entities."}\\
\hspace*{2em}\texttt{\}}\\
\hspace*{1em}\texttt{],}\\
\hspace*{1em}\texttt{"unresolved\_regions": ["T1", "IM1"],}\\
\hspace*{1em}\texttt{"revision\_targets": [}\\
\hspace*{2em}\texttt{"Add entities for the key entries in T1.",}\\
\hspace*{2em}\texttt{"Add visual entities for the chart axes and salient
visual trends in IM1.",}\\
\hspace*{2em}\texttt{"Revise relations so that the table entries are
connected to the corresponding visual evidence."}\\
\hspace*{1em}\texttt{],}\\
\hspace*{1em}\texttt{"covered\_components": ["P1"],}\\
\hspace*{1em}\texttt{"stop\_signal": 0}\\
\texttt{\}}
\end{tcolorbox}
\caption{Example reflector output used to guide the next extraction round.}
\label{fig:reflector_example}
\end{figure*}

\subsection{Retrieval and Answer Generation}
\label{sec:appendix_retrieval_generation}

\noindent\textbf{Query analyzer.}
Given a query $q$, KG4VD predicts one of three retrieval modes:
\textit{local lookup}, \textit{multi-hop reasoning}, or
\textit{document-level reasoning}.
The analyzer uses only the query as input and determines the retrieval scope before accessing the document.
Its prediction selects the corresponding PPR configuration and final evidence budget.
Local lookup is intended for fact-seeking questions that can be answered from evidence near a small number of relevant pages.
Multi-hop reasoning allows broader graph exploration for questions that require combining several related facts.
Document-level reasoning uses the largest exploration budget for questions that require aggregating dispersed evidence across larger portions of the document.

\noindent\textbf{Page anchoring and seed construction.}
KG4VD first retrieves the top-10 page images from the page index
$\mathcal{I}_{\mathrm{page}}$.
The retrieved pages serve as visual anchors rather than as PPR nodes.
Entities grounded to these anchor pages form the initial seed set
$\mathcal{S}_q$.
To prioritize entities that are more relevant to the query, each seed entity is assigned a weight according to its similarity to the query in the entity index
$\mathcal{I}_{\mathrm{ent}}$.
The resulting weights are normalized to form the seed distribution
$\mathbf{s}$.
Using multiple retrieved pages as anchors allows evidence from different document regions to enter the graph exploration process simultaneously.

\noindent\textbf{PPR formulation.}
Given the seed set $\mathcal{S}_q$, KG4VD constructs a bounded query-specific subgraph from the page-grounded graph neighborhoods containing the seed entities.
For each retrieval mode, the neighborhood is expanded up to
$H_{\max}$ hops from the seeds, while the combined candidate set is capped at
$N_{\max}$ entity nodes.
Evidence associated with different anchor pages is therefore considered jointly within the same PPR computation.

The PPR update is
\begin{equation}
\mathbf{r}_{\tau+1}
=
(1-\alpha)\mathbf{P}^{\top}\mathbf{r}_{\tau}
+
\alpha\mathbf{p},
\label{eq:ppr}
\end{equation}
where $\alpha$ is the restart probability,
$\mathbf{P}$ is the column-normalized adjacency matrix of the bounded subgraph,
and $\mathbf{p}$ is the personalization vector.
We define
\begin{equation}
\mathbf{p}
=
(1-\beta)\mathbf{s}
+
\beta\mathbf{t}(q),
\label{eq:ppr_p}
\end{equation}
where $\beta$ controls the contribution of query relevance and
$\mathbf{t}(q)$ is proportional to the non-negative cosine similarity between the query embedding and candidate entity embeddings.
Dangling mass is redistributed through $\mathbf{p}$.

Tables~\ref{tab:ppr_params} and~\ref{tab:retrieval_budgets} list the mode-specific exploration settings and final evidence budgets.
The hyperparameters are chosen to instantiate progressively broader retrieval scopes rather than to optimize each query type independently.
From local lookup to document-level reasoning, we decrease the restart probability $\alpha$ so that PPR can propagate farther from the anchor-page seeds, while increasing the query-bias weight $\beta$ to preserve semantic relevance during broader exploration.
We simultaneously increase $H_{\max}$, $N_{\max}$, and the final evidence budgets to accommodate evidence distributed over larger portions of the document.
The exact values serve as practical operating points for narrow, medium, and broad graph exploration, and the same mode-specific configuration is used across all evaluated datasets rather than being tuned separately for each benchmark.

\begin{table}[t]
\centering
\small
\setlength{\tabcolsep}{6pt}
\renewcommand{\arraystretch}{1.15}
\begin{tabular}{lcccc}
\toprule
Mode & $\alpha$ & $\beta$ & $H_{\max}$ & $N_{\max}$ \\
\midrule
Local lookup              & 0.80 & 0.0 & 1 & 30  \\
Multi-hop reasoning       & 0.45 & 0.3 & 2 & 100 \\
Document-level reasoning  & 0.20 & 0.6 & 5 & 200 \\
\bottomrule
\end{tabular}
\caption{
Mode-specific PPR configurations for progressively broader graph exploration.
$N_{\max}$ caps the combined entity-candidate set used by PPR, and the page-retrieval budget is fixed to top-10 for all modes.
}
\label{tab:ppr_params}
\end{table}

\begin{table}[t]
\centering
\small
\setlength{\tabcolsep}{8pt}
\renewcommand{\arraystretch}{1.15}
\begin{tabular}{lcc}
\toprule
Mode & Entities $K_v$ & Relations $K_e$ \\
\midrule
Local lookup              & 12 & 12 \\
Multi-hop reasoning       & 30 & 40 \\
Document-level reasoning  & 60 & 80 \\
\bottomrule
\end{tabular}
\caption{
Mode-specific evidence budgets after PPR-based ranking.
The budgets increase with the predicted retrieval scope, while all modes retain up to 10 page images.
}
\label{tab:retrieval_budgets}
\end{table}

\noindent\textbf{PPR-based evidence selection.}
After PPR converges, KG4VD directly ranks candidate entities according to their PPR scores.
Relations are retained according to the scores of the connected high-ranking entities and the resulting ranked graph evidence.
For each retrieval mode, KG4VD retains the top-$K_v$ entities and top-$K_e$ relations according to the budgets in Table~\ref{tab:retrieval_budgets}.
The retrieved page images are retained separately in the order returned by the page retriever, with at most 10 pages provided to the answer-generation stage.
No additional learned reranker is used after PPR.

\noindent\textbf{Two-stage answer generation.}
KG4VD generates the final response in two stages.
In the first stage, an MLLM organizes the retained entities and relations into a preliminary graph-grounded answer.
This stage summarizes the structured evidence relevant to the query and provides an intermediate representation before consulting the original page images.
In the second stage, another MLLM call receives the preliminary answer together with the retrieved page images and graph evidence.
It verifies the preliminary response against the original visual document context and revises unsupported or incomplete content when necessary.

For DLVQA, the answer guidance is supplied to both generation stages only as an output-format constraint.
It specifies the desired organization, style, and approximate response length, but it does not contain question-specific evidence and is not treated as part of the retrieved evidence.

\section{Additional Experimental Results}
\label{sec:appendix_extra_exp}

\subsection{Comparison with Specialized Document RAG Systems}
\label{sec:appendix_recent_docrag}

We additionally compare KG4VD with VDocRAG~\cite{tanaka2025vdocrag} and M3DocRAG~\cite{cho2024m3docrag} using their official releases and default settings.
Both systems belong to the broad MMRAG family: they retrieve visual document pages and use an MLLM for answer generation.
Tables~\ref{tab:recent_docrag_mmlb} and~\ref{tab:recent_docrag_dlvqa} report end-to-end results on MMLongBench-Doc-2K and DLVQA.
Because the official systems use different generators, this comparison measures released system performance rather than a backbone-controlled retriever comparison.

\begin{table}[t]
\centering
\small
\setlength{\tabcolsep}{4pt}
\renewcommand{\arraystretch}{1.12}
\begin{tabular}{lcc}
\toprule
Method & Retriever / Generator & Accuracy \\
\midrule
VDocRAG & VDocRAG / Phi-3-Vision & 30.39 \\
M3DocRAG & ColPali / Qwen2-VL-7B & 37.54 \\
ColQwen & ColQwen / GPT-4o-mini & 41.79 \\
KG4VD & KG4VD / GPT-4o-mini & \textbf{44.52} \\
\bottomrule
\end{tabular}
\caption{Comparison with specialized document RAG systems on MMLongBench-Doc-2K.}
\label{tab:recent_docrag_mmlb}
\end{table}

\begin{table}[t]
\centering
\small
\setlength{\tabcolsep}{5pt}
\renewcommand{\arraystretch}{1.12}
\begin{tabular}{lccc}
\toprule
Method & Precision & Recall & F1 \\
\midrule
VDocRAG & 13.26 & 18.96 & 15.61 \\
M3DocRAG & 40.12 & 18.79 & 25.59 \\
ColQwen & 82.26 & 28.05 & 41.83 \\
KG4VD & \textbf{82.43} & \textbf{42.45} & \textbf{56.04} \\
\bottomrule
\end{tabular}
\caption{Answer Fact Precision, Recall, and F1 on DLVQA for specialized document RAG systems.}
\label{tab:recent_docrag_dlvqa}
\end{table}

\begin{table}[t]
\centering
\small
\setlength{\tabcolsep}{5pt}
\renewcommand{\arraystretch}{1.12}
\begin{tabular}{lr}
\toprule
\multicolumn{2}{c}{\textbf{KG4VD Stage-wise Breakdown}} \\
\midrule
Stage & Latency (s) \\
\midrule
Page retrieval & 0.5 \\
Entity-seed retrieval & 0.8 \\
Query-adaptive PPR and evidence selection & 2.0 \\
Preliminary graph-answer generation & 12.5 \\
Final multimodal answer generation & 26.0 \\
\midrule
KG4VD total & 41.8 \\
\midrule
\multicolumn{2}{c}{\textbf{End-to-end Comparison}} \\
\midrule
GME & 26.4 \\
MegaRAG & 42.0 \\
KG4VD & 41.8 \\
\bottomrule
\end{tabular}
\caption{
Average online inference latency on Environmental Reports after offline indexing.
KG4VD has similar end-to-end latency to MegaRAG, while GME is faster because it does not perform graph-based intermediate reasoning.
}
\label{tab:online_latency_breakdown}
\end{table}

KG4VD remains the strongest system on both benchmarks.
On MMLongBench-Doc-2K, it exceeds VDocRAG and M3DocRAG by 14.13 and 6.98 accuracy points, respectively.
On DLVQA, its main advantage is evidence coverage: KG4VD improves recall from 28.05 to 42.45 over ColQwen while retaining similarly high precision, producing a 14.21-point F1 gain.

\subsection{Fixed vs.\ Query-adaptive PPR Budgets}
\label{sec:appendix_fixed_ppr}

We further examine whether a single fixed PPR configuration can replace the query analyzer.
The fixed variants apply the same retrieval configuration and evidence budget to all queries, using the local, multi-hop, or document-level settings in Tables~\ref{tab:ppr_params} and~\ref{tab:retrieval_budgets}.
In contrast, query-adaptive PPR selects the configuration based on the estimated reasoning scope.

Table~\ref{tab:fixed_vs_adaptive_ppr} shows different preferences across the two benchmarks.
On MMLongBench-Doc-2K, the fixed local setting performs best.
Since the benchmark contains more single-hop questions, narrow retrieval is often sufficient and avoids noise from unnecessary graph expansion.
In contrast, DLVQA favors the fixed document-level setting, as its questions typically require evidence aggregation across many pages.

Query-adaptive PPR remains close to the best fixed setting on both benchmarks and achieves the best average performance.
This shows that no single retrieval budget is optimal across different query scopes, supporting adaptive retrieval for a unified system handling local, multi-hop, and document-level questions.

\begin{table}[t]
\centering
\small
\setlength{\tabcolsep}{5pt}
\renewcommand{\arraystretch}{1.12}
\begin{tabular}{lccc}
\toprule
Retrieval Policy
& MMLB-2K Acc.
& DLVQA F1
& Avg. \\
\midrule
Fixed Local
& \textbf{45.20}
& 49.00
& 47.10 \\

Fixed Multi-hop
& 43.30
& 53.70
& 48.50 \\

Fixed Document-level
& 41.90
& \textbf{56.70}
& 49.30 \\

Query-adaptive
& 44.52
& 56.04
& \textbf{50.28} \\
\bottomrule
\end{tabular}
\caption{
Comparison of fixed and query-adaptive PPR configurations.
}
\label{tab:fixed_vs_adaptive_ppr}
\end{table}

\subsection{Reference-free Pairwise Evaluation on DLVQA}
\label{sec:appendix_reference_free}

Table~\ref{tab:dlvqa_subset_reference_free_winrate} reports subset-wise pairwise evaluation, comparing KG4VD against each baseline in terms of comprehensiveness, diversity, and empowerment.
For each criterion, we report win, loss, and tie rates together with the margin $\Delta$ between KG4VD wins and losses.

\input{table/main_dlvqa_subsetwise}

KG4VD obtains positive average margins across all four DLVQA subsets and all three criteria.
The gains are strongest for diversity, indicating that KG4VD covers a broader range of relevant information.
Although the World History subset is more challenging, KG4VD maintains positive average margins across the reference-free criteria.

\subsection{Domain-wise DLVQA Results}
\label{sec:appendix_domainwise_dlvqa}

Table~\ref{tab:dlvqa_subset_finesure_results} reports Answer Fact Precision, Recall, and F1 across the four DLVQA subsets: DLCV Slides, Picture Books, Environmental Reports, and World History Textbook.
The domain-wise breakdown highlights different retrieval behaviors: high precision reflects better grounding of generated claims, while high recall indicates broader coverage of the required evidence.
KG4VD shows stronger gains on visually rich domains, where figures, layouts, and cross-page visual cues are important.
In contrast, the World History subset is predominantly text-based, which reduces the advantage of multimodal graph retrieval and helps explain why LightRAG performs better on this subset.

\input{table/appendix_dlvqa_domainwise}

\subsection{Online Inference Latency}
\label{sec:computational_overhead}

The main paper reports the one-time offline MMKG construction cost.
Here we compare the online inference latency after the reusable graph and indexes have been built.
All results are measured on the Environmental Reports subset.
Table~\ref{tab:online_latency_breakdown} first reports the stage-wise latency of KG4VD and then compares its end-to-end latency with GME and MegaRAG.

For KG4VD, page and entity retrieval are lightweight, while query-adaptive PPR and score-based evidence selection add about 2.0 seconds.
Most of the runtime is spent on the preliminary graph-grounded answer and final multimodal answer generation.
Overall, KG4VD requires approximately 41.8 seconds per question.

For comparison, GME requires approximately 26.4 seconds per question, as it performs page retrieval followed by a single answer-generation stage.
MegaRAG requires approximately 42.0 seconds per question using its two-stage generation pipeline, which separately reasons over graph and page evidence before fusion.
KG4VD therefore has a similar end-to-end latency to MegaRAG while outperforming it on DLVQA, and adds approximately 15.4 seconds over the page-retrieval-only GME baseline.
The additional latency comes mainly from graph-based reasoning and the intermediate answer stage rather than retrieval itself.

\section{Qualitative MMKG Analysis}
\label{sec:appendix_qualitative}

\subsection{Graph Statistics Across DLVQA Subsets}
\label{sec:appendix_graph_statistics}

\begin{figure*}[t]
    \centering
    \includegraphics[width=\textwidth]{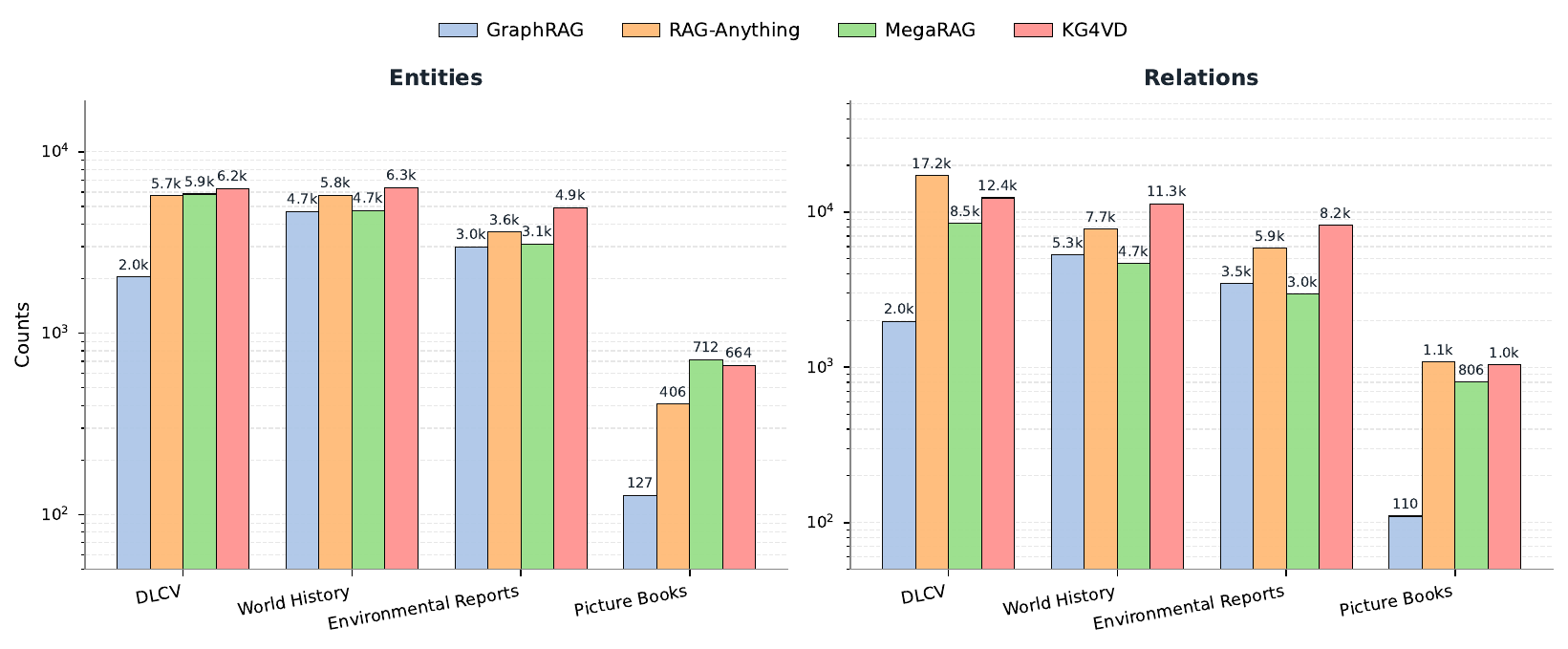}
    \caption{Entity and relation counts of graphs constructed by GraphRAG, RAGAnything, MegaRAG, and KG4VD across the four DLVQA subsets.}
    \label{fig:graph_statistics_dlvqa}
\end{figure*}

Figure~\ref{fig:graph_statistics_dlvqa} compares the total entity and relation counts produced by GraphRAG, RAGAnything, MegaRAG, and KG4VD.
GraphRAG consistently produces smaller graphs than the multimodal methods, especially on visually rich subsets such as DLCV Slides and Picture Books.
KG4VD yields the largest entity counts on DLCV Slides, World History, and Environmental Reports, while MegaRAG produces slightly more entities on Picture Books.
For relations, KG4VD is largest on World History and Environmental Reports, whereas RAGAnything is largest on DLCV Slides and Picture Books.
These statistics describe coverage and extraction behavior, but graph size alone is not a quality metric; correctness is evaluated separately by the human-verified MMKG audit in the main paper.

\subsection{Adaptive Extraction Across Rounds}
\label{sec:appendix_adaptive_extraction_rounds}

\begin{figure*}[t]
    \centering
    \includegraphics[width=\textwidth]{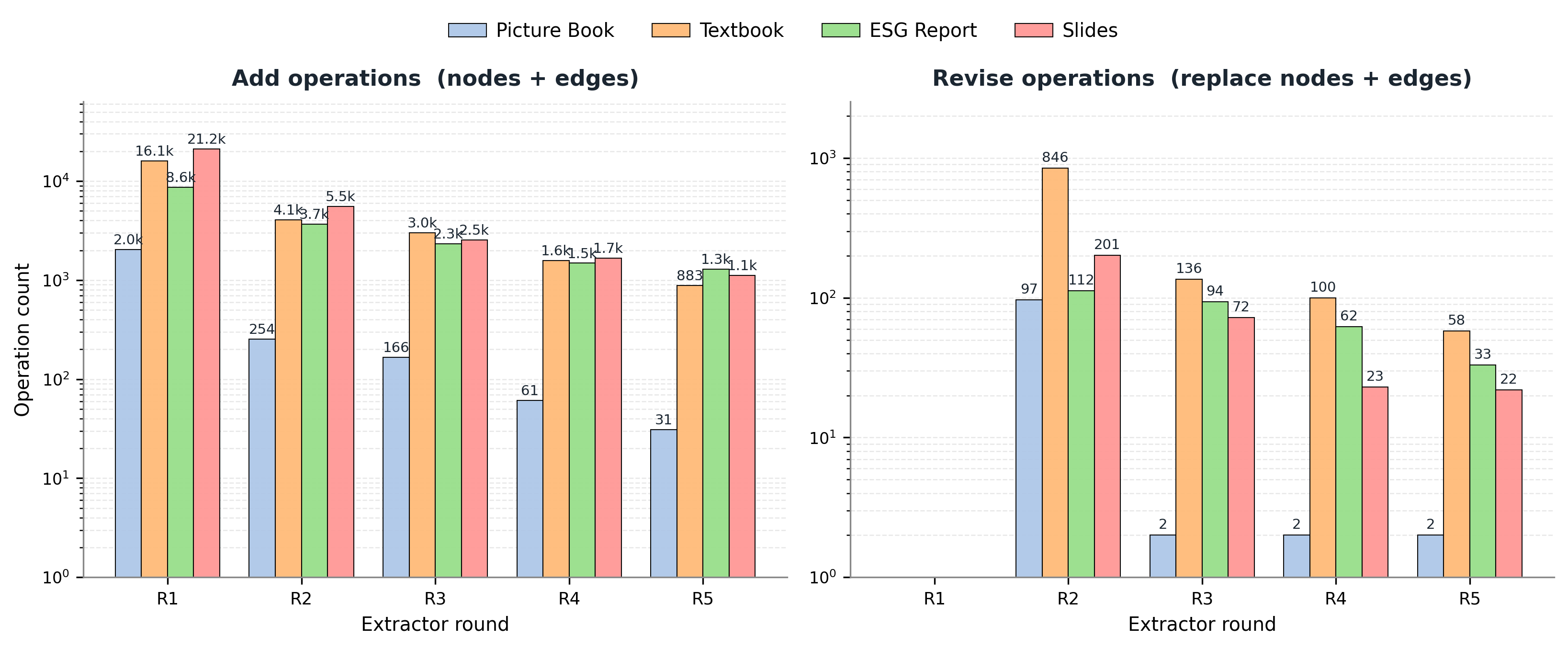}
    \caption{Add and revise operations across adaptive extraction rounds on four DLVQA subsets. Add operations dominate the first round, while later rounds recover missing evidence and refine earlier graph elements.}
    \label{fig:adaptive_extraction_rounds}
\end{figure*}

\begin{figure}[t]
    \centering
    \includegraphics[width=\linewidth]{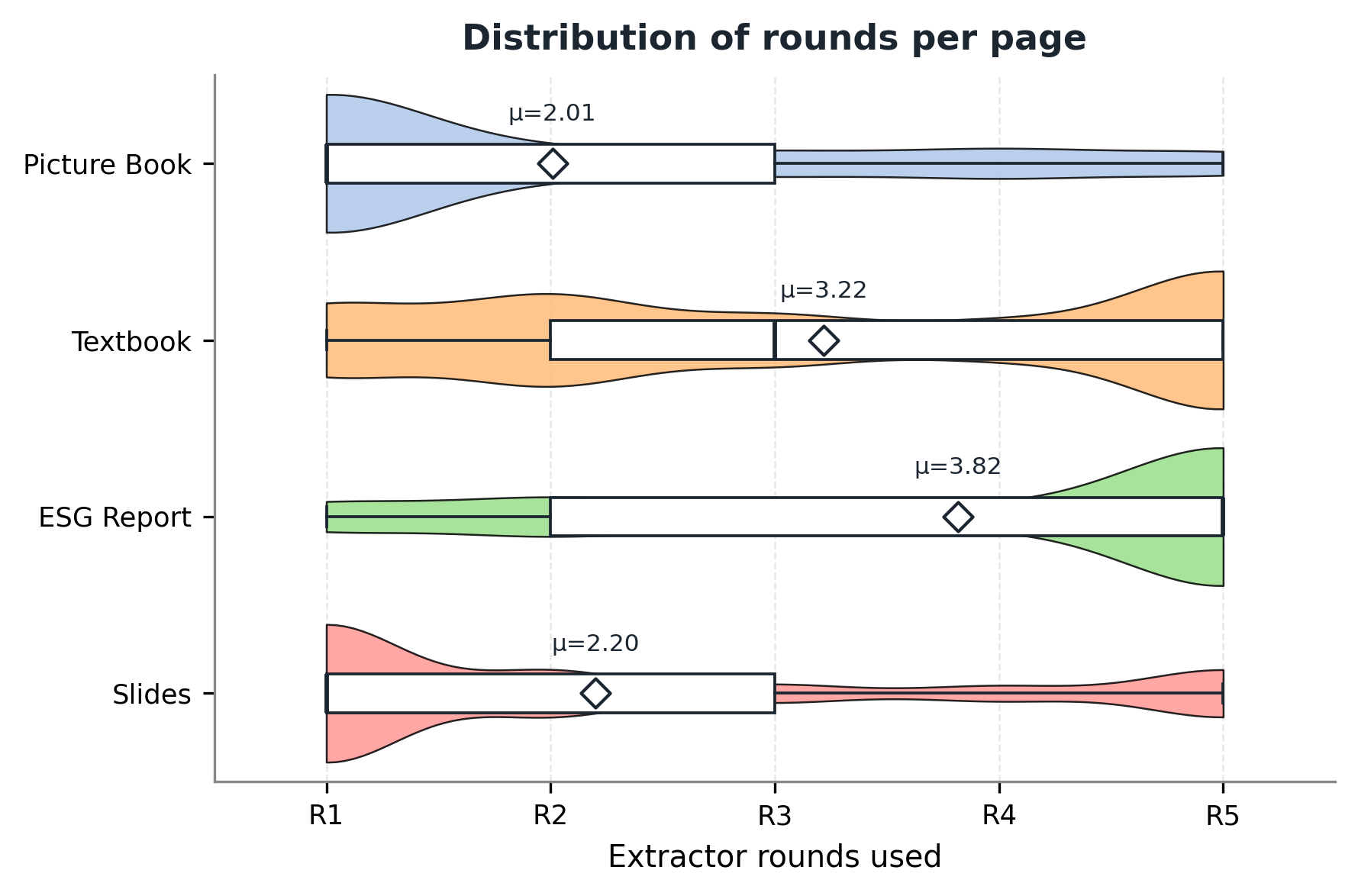}
    \caption{Per-page extraction-round distribution across the four DLVQA subsets. Violin plots show the round distribution, boxplots show the median and interquartile range, and diamonds mark the mean.}
    \label{fig:round_distribution}
\end{figure}

\begin{figure*}[t]
    \centering
    \includegraphics[width=0.9\textwidth]{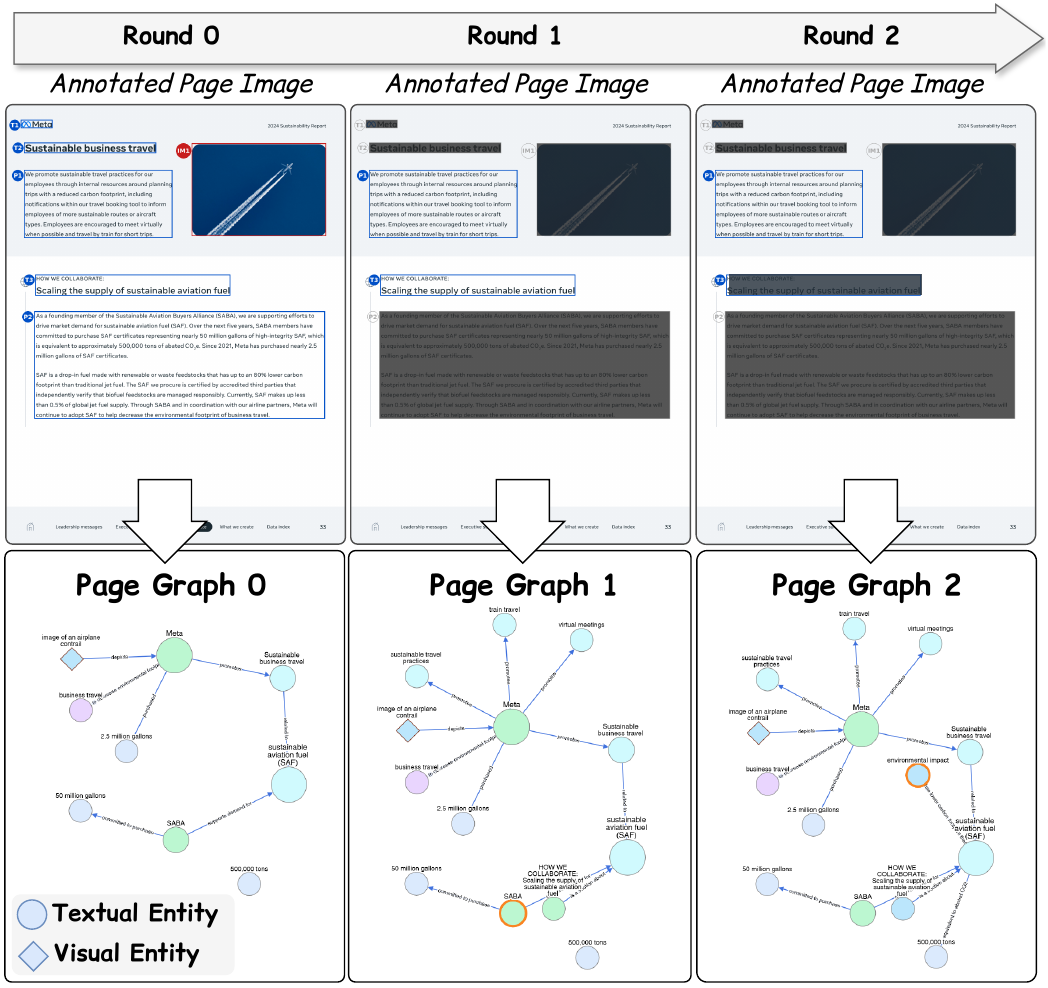}
    \caption{Qualitative example of adaptive extraction across rounds. The top row shows annotated page images, and the bottom row shows the corresponding page graphs. Covered regions are masked in later rounds to focus extraction on unresolved components. Orange outlines indicate revised nodes or edges.}
    \label{fig:adaptive_extraction_case}
\end{figure*}

Figure~\ref{fig:adaptive_extraction_rounds} shows add and revise operations across extraction rounds.
Because every page graph is initialized as empty, the first round is dominated by add operations and contains no revisions.
Most graph construction occurs in the first round, but later rounds still add entities and relations, especially for structurally dense slides, textbooks, and environmental reports.
This indicates that the initial extractor captures major page content, while reflector-guided rounds recover evidence from unresolved regions.

Revise operations are concentrated in the second round and decrease thereafter, suggesting that the reflector mainly corrects coarse or incomplete early graph elements before the page graph stabilizes.
World History pages show the most revisions, consistent with their longer narrative context and dense textual evidence.

Figure~\ref{fig:round_distribution} summarizes the number of rounds used per page before the reflector returns the stop signal or the maximum budget is reached, with $1 \leq r \leq 5$.
Picture Books and DLCV Slides converge relatively quickly, with mean rounds of $\mu=2.01$ and $\mu=2.20$, whereas World History and Environmental Reports require deeper iteration, with $\mu=3.22$ and $\mu=3.82$.
The Environmental Reports distribution is strongly skewed toward the fifth round, suggesting that dense report pages may benefit from a larger extraction budget.

Figure~\ref{fig:adaptive_extraction_case} presents an environmental-report example.
As extraction proceeds, covered regions are masked so that later rounds focus on unresolved components, while the page graph is expanded and refined.
Textual entities are shown as circles, visual entities as diamonds, and orange outlines indicate revised graph elements.
The example illustrates how KG4VD improves page-graph coverage through iterative extraction and revision rather than a single pass.

\subsection{Representative MMKG Construction Errors}
\label{sec:appendix_mmkg_error}

\begin{table*}[t]
\centering
\small
\setlength{\tabcolsep}{4pt}
\renewcommand{\arraystretch}{1.12}
\begin{tabularx}{\textwidth}{>{\raggedright\arraybackslash}p{0.8cm}>{\raggedright\arraybackslash}p{2.6cm}>{\raggedright\arraybackslash}X}
\toprule
Type & Example & Observation \\
\midrule
TP & World History, p.127 & KG4VD correctly constructs \textit{Hattusas} $\rightarrow$ \textit{capital\_of} $\rightarrow$ \textit{Hittites}. \\
TP & Picture Books, p.170 & KG4VD correctly extracts the multimodal relation \textit{Gecko} $\rightarrow$ \textit{claims} $\rightarrow$ \textit{bus}. \\
FP & World History, p.279 & \textit{Italy} appears only as a multiple-choice option, but KG4VD incorrectly treats it as an asserted fact: \textit{Alexander the Great} $\rightarrow$ \textit{conquered} $\rightarrow$ \textit{Italy}. \\
FP & Environmental Reports, p.276 & KG4VD constructs \textit{Shipping Containers} $\rightarrow$ \textit{depicts} $\rightarrow$ \textit{Meta}, although the image does not support this relation. \\
FN & DLCV Slides, p.1694 & KG4VD misses an arrow relation originating from \textit{Simple Gaussian prior}. \\
FN & Picture Books, p.79 & KG4VD misses fine-grained visual relations, such as a woman wearing a blue dress and carrying a black bag. \\
\bottomrule
\end{tabularx}
\caption{Representative true-positive, false-positive, and false-negative MMKG construction cases.}
\label{tab:mmkg_error_examples}
\end{table*}

KG4VD performs well when textual or multimodal evidence is explicit, but still exhibits three recurring errors: treating non-asserted content such as answer options as facts, introducing relations that are not visually supported, and missing fine-grained visual relations.
These cases are consistent with the quantitative audit in the main paper: precision is strong, while moderate entity recall shows that graph construction remains incomplete.

\section{Evaluation Metrics and LLM Judges}
\label{sec:appendix_judge}

\subsection{Answer Fact Precision, Recall, and F1}
\label{sec:appendix_answer_fact_metrics}

DLVQA evaluates long-form answers by matching generated answer sentences against gold supporting facts.
Our implementation adapts the released FineSurE evaluation codebase~\citep{song-etal-2024-finesure} for long-form output preprocessing, sentence segmentation, construction of sentence-fact comparison inputs, and aggregation of the judge decisions.
We modify this workflow for DLVQA by using the gold supporting facts as the reference units and by reporting Answer Fact Precision, Recall, and F1.

\noindent\textbf{Output segmentation.}
Each generated response is first processed and split into a list of evaluable sentences using the adapted FineSurE sentence-segmentation pipeline.
Let $A=\{a_i\}_{i=1}^{m}$ denote these generated-answer sentences and $F=\{f_j\}_{j=1}^{n}$ denote the gold supporting facts.
Using the same segmentation procedure for all methods keeps the denominator of sentence-level precision consistent across systems with different output styles and lengths.

\noindent\textbf{Sentence--fact matching.}
For each generated sentence $a_i$, an LLM judge determines whether its factual content is fully supported by one or more gold facts in $F$.
Separately, for each gold fact $f_j$, the judge determines whether its content is covered by at least one generated sentence in $A$.
The first direction yields one binary support decision per generated sentence, while the reverse direction yields one binary coverage decision per gold fact.
A generated sentence may be supported by multiple facts, and a gold fact may be expressed across multiple sentences; the final decision remains binary for each evaluated sentence or fact.
The reference summary is not treated as a single matching unit in these calculations; the gold supporting facts provide the explicit factual units.

We compute
\begin{equation}
\mathrm{Precision}
=
\frac{\sum_{i=1}^{m}\mathbb{1}[a_i\ \mathrm{is\ supported}]}{m},
\end{equation}
\begin{equation}
\mathrm{Recall}
=
\frac{\sum_{j=1}^{n}\mathbb{1}[f_j\ \mathrm{is\ covered}]}{n},
\end{equation}
and
\begin{equation}
\mathrm{F1}
=
\frac{2\,\mathrm{Precision}\,\mathrm{Recall}}
{\mathrm{Precision}+\mathrm{Recall}}.
\end{equation}
Precision is therefore the fraction of generated sentences supported by the annotated facts, while Recall is the fraction of gold facts covered by the generated answer.
Unsupported or irrelevant sentences reduce Precision, and omitted gold facts reduce Recall.

\subsection{Judge Models and Aggregation}
\label{sec:appendix_judge_models}

\noindent\textbf{Judge models.}
We use two LLM judges, \textit{gpt-5.4-mini} and \textit{gemini-3.1-flash-lite}.
Using two model families reduces reliance on a single judge and helps mitigate model-specific bias.

\noindent\textbf{DLVQA aggregation.}
Each judge independently produces the generated-sentence support decisions and gold-fact coverage decisions for every method.
We compute Precision, Recall, and F1 per query for each judge and report the unweighted average of the two judges.
No oracle normalization is applied.

\noindent\textbf{Short-answer VQA accuracy.}
For MMLongBench-Doc-2K, each judge gives a binary semantic-correctness verdict.
We average the two verdicts directly.

\noindent\textbf{Reference-free pairwise evaluation.}
Each judge evaluates every answer pair in both answer orders to reduce positional bias.
A win is counted only when the judge selects the same winner in both orders; order-dependent verdicts are treated as ties.
The outcomes are then averaged across the two judges.

\subsection{Inter-judge Agreement}
\label{sec:appendix_judge_agreement}

For DLVQA, we measure the per-query Pearson correlation between the two judges across evaluated methods.
The correlations are moderate for Answer Fact Precision ($r=0.47$--$0.70$) and lower for Recall ($r=0.22$--$0.48$), reflecting the greater ambiguity of deciding whether a long-form answer fully covers each required fact.
For MMLongBench-Doc-2K, binary VQA judgments are highly consistent across major methods, with agreement ranging from $97.6\%$ to $99.7\%$ and Cohen's $\kappa$ ranging from $0.948$ to $0.994$.

\subsection{Judge Output Examples}
\label{sec:appendix_judge_examples}

Table~\ref{tab:judge_case_winrate} shows a reference-free pairwise example in which both judges prefer KG4VD across the reported criteria.
Table~\ref{tab:judge_case_finesure} shows a reference-based sentence--fact matching example in which KG4VD produces the strongest grounded coverage.

\input{table/appendix_judge_example}

\section{The Use of Large Language Models}
\label{sec:llm_usage}

We use LLMs and MLLMs at several stages of the pipeline.

\noindent\textbf{MMKG construction and query-time reasoning.}
Unless otherwise noted, \textit{GPT-4o-mini} is used for page-level extraction, reflection, query analysis, preliminary graph-answer generation, and final multimodal answer generation.
Candidate entities and relations are selected deterministically from their PPR scores.

\noindent\textbf{Dataset construction.}
\textit{Gemini-3.1-pro-preview} assists page and document summarization, atomic-fact extraction, topic-based QA drafting, and reference-summary generation in the DLVQA human-in-the-loop pipeline.
No MLLM-assisted artifact is retained without human review.

\noindent\textbf{Evaluation.}
\textit{GPT-5.4-mini} and \textit{Gemini-3.1-flash-lite} are used as independent LLM judges.

\noindent\textbf{Fine-tuning.}
All LLM and MLLM usage is zero-shot; no model is fine-tuned on DLVQA or the evaluated QA benchmarks.

\section{MLLM Prompts}
\label{sec:prompts}

%
\tcbset{
  qapromptbox/.style={
    colback=gray!3,
    colframe=black!65,
    boxrule=0.5pt,
    arc=1.5mm,
    left=1mm,
    right=1mm,
    top=1mm,
    bottom=1mm,
    fonttitle=\bfseries
  }
}

For reproducibility, we provide the prompt templates used for the LLM-mediated
steps in KG4VD. Two prompts are used during offline MMKG construction:
the page-level extractor and the page-level reflector. Three prompts are used during online retrieval and answer generation:
the query analyzer, the preliminary answer organizer, and the final multimodal
answer generator.

\begin{figure*}[t]
\centering
\begin{subfigure}{0.98\textwidth}
\centering
\begin{tcolorbox}[qapromptbox,title={KG4VD Page-level Extractor Prompt}]
\begin{Verbatim}[fontsize=\footnotesize, breaklines=true, breakanywhere=true]
You are a multimodal knowledge-graph extractor for page {page_id} of a visually rich document. The page image is annotated with bounding boxes and component identifiers, such as P1, T1, or IM1. These identifiers correspond to parser-detected layout components.

INPUTS
1. Annotated page image I_hat_i^(t).
2. Component manifest C_i: {component_manifest}
3. Current page graph G_i^(t): {current_graph}
4. Previous reflector feedback B_i^(t), if available: {reflector_feedback}

TASK. Propose graph edits Delta_i^(t) for the current page. On the initial round, extract candidate entities and relations from the annotated page image and component manifest. On later rounds, use the current page graph and reflector feedback to revise the previous extraction. Allowed operation types are:
- add_node, replace_node, delete_node
- add_edge, replace_edge, delete_edge

SCHEMA. Every node must contain name, type, modality_tag (text or visual), description, and a non-empty source_components list of component IDs. Every edge must contain source_entity, target_entity, relation_label, description, and a non-empty source_components list.

RULES.
1. Cite only component IDs that appear in the component manifest.
2. Do not generate free-form coordinates. Ground each node or edge by citing source layout components.
3. Use add operations for new entities or relations, replace operations to correct attributes, descriptions, or grounding, and delete operations to remove unsupported or duplicate items.
4. Avoid duplicating entities or relations that are already present in the current page graph unless the operation explicitly replaces them.

OUTPUT a JSON object {"ops": [...]}, no markdown.
\end{Verbatim}
\end{tcolorbox}
\caption{Page-level extractor \(\mathrm{Ext}\).}
\label{fig:kg4vd_extractor_prompt}
\end{subfigure}
\caption{Build-time prompt for adaptive page-level entity and relation extraction.}
\label{fig:kg4vd_extractor_prompt_group}
\end{figure*}

\begin{figure*}[t]
\centering
\begin{subfigure}{0.98\textwidth}
\centering
\begin{tcolorbox}[qapromptbox,title={KG4VD Page-level Reflector Prompt}]
\begin{Verbatim}[fontsize=\footnotesize, breaklines=true, breakanywhere=true]
You are a reflector for a multimodal knowledge-graph extraction pipeline. Your job is to diagnose coverage gaps in the current page graph and decide whether the extract-reflect loop should stop for page {page_id}. You do not propose graph edits directly; instead, you return feedback for the next extractor round.

INPUTS
1. Annotated page image I_hat_i^(t).
2. Component manifest C_i: {component_manifest}
3. Updated page graph G_i^(t+1): {updated_page_graph}

TASK. Produce a component-level coverage report. For each component, classify whether it is covered, partially_covered, or uncovered. Identify unresolved regions, missing entities, missing relations, duplicate items, unsupported items, and nodes or edges that should be revised in the next extractor round.

MASKING GUIDANCE. Components judged sufficiently covered may be listed as covered components so that they can be masked in the next annotated page image. Components that remain partially covered or uncovered should remain visible for the next extraction round.

STOPPING. Return stop_signal: 1 only when the current page graph sufficiently covers the page or when no useful revision target remains. Return stop_signal: 0 when another extractor round is needed.

OUTPUT a JSON object:
{
  "per_component": [{"component_id": ..., "status": ..., "gap_summary": ...}, ...],
  "unresolved_regions": [component_id, ...],
  "revision_targets": [string, ...],
  "covered_components": [component_id, ...],
  "stop_signal": 0 | 1
}
\end{Verbatim}
\end{tcolorbox}
\caption{Page-level reflector \(\mathrm{Ref}\).}
\label{fig:kg4vd_reflector_prompt}
\end{subfigure}
\caption{Build-time prompt for reflection, coverage diagnosis, masking guidance, and stop-signal prediction.}
\label{fig:kg4vd_reflector_prompt_group}
\end{figure*}

\begin{figure*}[t]
\centering
\begin{subfigure}{0.98\textwidth}
\centering
\begin{tcolorbox}[qapromptbox,title={KG4VD Query Analyzer Prompt}]
\begin{Verbatim}[fontsize=\footnotesize, breaklines=true, breakanywhere=true]
Classify the user query into ONE of three PPR expansion modes, then output a JSON object {"mode": ..., "rationale": ...}.

MODES
local_lookup: The query is fact-seeking and can usually be answered using evidence close to the relevant anchor page. PPR should stay close to the anchor-page entities.
multi_hop_reasoning: The query is compositional and requires combining entities and relations from a small set of relevant page-level graphs.
document_level_reasoning: The query requires broader evidence collection across many relevant pages or document-level reasoning.

JUDGMENT CUES
- Questions asking for a single fact, value, location, name, or count usually indicate local_lookup.
- Questions involving comparison, explanation of relationships, or combining several facts usually indicate multi_hop_reasoning.
- Questions asking for an overview, summary, trend, narrative, or document-level interpretation usually indicate document_level_reasoning.

QUERY: {query_text}
\end{Verbatim}
\end{tcolorbox}
\caption{Query analyzer.}
\label{fig:kg4vd_query_analyzer_prompt}
\end{subfigure}
\caption{Query-time prompt for selecting the query-adaptive PPR expansion mode.}
\label{fig:kg4vd_query_analyzer_prompt_group}
\end{figure*}

\begin{figure*}[t]
\centering
\begin{subfigure}{0.98\textwidth}
\centering
\begin{tcolorbox}[qapromptbox,title={KG4VD Preliminary Answer Organizer Prompt}]
\begin{Verbatim}[fontsize=\footnotesize, breaklines=true, breakanywhere=true]
You organize retrieved graph evidence from a visually rich document into a preliminary answer. Use only the retrieved entities and relations in the EVIDENCE block. Do not introduce external knowledge or facts that are not supported by the retrieved graph evidence. When answer guidance is provided, follow its requested scope and organization without treating it as evidence.

QUESTION: {query_text}

ANSWER GUIDANCE (optional):
{answer_guidance}

EVIDENCE:
{evidence_block: retrieved entity list, and retrieved relation list}

TASK. Write a preliminary answer that summarizes the evidence relevant to the question. Identify which entities and relations support the answer, and note when the retrieved graph evidence is incomplete or insufficient.

OUTPUT JSON:
{
  "preliminary_answer": <answer grounded in retrieved entities and relations>,
  "supporting_entities": [<entity id>, ...],
  "supporting_relations": [<relation id>, ...],
  "source_pages": [<int>, ...],
  "confidence": "high" | "medium" | "low",
  "insufficient_evidence": true | false
}
\end{Verbatim}
\end{tcolorbox}
\caption{Preliminary answer organizer.}
\label{fig:kg4vd_preliminary_answer_prompt}
\end{subfigure}
\caption{Query-time prompt for organizing retrieved entities and relations into a preliminary answer.}
\label{fig:kg4vd_preliminary_answer_prompt_group}
\end{figure*}

\begin{figure*}[t]
\centering
\begin{subfigure}{0.98\textwidth}
\centering
\begin{tcolorbox}[qapromptbox,title={KG4VD Final Multimodal Answer Generator Prompt}]
\begin{Verbatim}[fontsize=\footnotesize, breaklines=true, breakanywhere=true]
You produce the final answer to a question about a visually rich document. Use the preliminary answer, the retrieved graph evidence, and the retrieved page images. Ground the final response in the original document context. Do not invent information that is not supported by the retrieved evidence or page images. When answer guidance is provided, follow its requested scope, organization, and length without treating it as evidence.

QUESTION: {query_text}

ANSWER GUIDANCE (optional):
{answer_guidance}

PRELIMINARY ANSWER:
{preliminary_answer}

RETRIEVED GRAPH EVIDENCE:
{retrieved_entities_and_relations}

RETRIEVED PAGE IMAGES:
{retrieved_page_images}

TASK. Verify the preliminary answer against the retrieved page images and produce the final response. Follow the answer guidance when it is provided. If the visual evidence contradicts or weakens the preliminary answer, revise the answer accordingly. If the retrieved evidence is insufficient, state that the question is not answerable from the available evidence.

OUTPUT JSON:
{
  "answer": <final grounded answer>,
  "used_pages": [<int>, ...],
  "confidence": "high" | "medium" | "low",
  "failure_reason": null | "insufficient_evidence"
}
\end{Verbatim}
\end{tcolorbox}
\caption{Final multimodal answer generator.}
\label{fig:kg4vd_final_answer_prompt}
\end{subfigure}
\caption{Query-time prompt for combining the preliminary answer with retrieved page images.}
\label{fig:kg4vd_final_answer_prompt_group}
\end{figure*}

%% file: table/appendix_dlvqa_stats.tex
\begin{table*}[t]
\centering
\small
\renewcommand{\arraystretch}{1.18}
\setlength{\tabcolsep}{6pt}
\begin{tabular}{lrrrrr} 
\toprule
\multicolumn{1}{c}{\textbf{Metric}} & \multicolumn{1}{c}{\textbf{DLCV Slides}} & \multicolumn{1}{c}{\textbf{World History}} & \multicolumn{1}{c}{\textbf{Env. Reports}} & \multicolumn{1}{c}{\textbf{Picture Books}} & \multicolumn{1}{c}{\textbf{All}} \\ 
\hline
\multicolumn{6}{l}{{\cellcolor[rgb]{0.957,0.957,0.957}}\textbf{\textit{Corpus size}}} \\
Pages & 1,984 & 788 & 422 & 247 & 3,441 \\
Atomic facts & 12,527 & 7,692 & 3,171 & 1,022 & 24,412 \\
QA pairs & 306 & 120 & 62 & 37 & \textbf{525} \\ 
\hline
\multicolumn{6}{l}{{\cellcolor[rgb]{0.957,0.957,0.957}}\textbf{\textit{Question characteristics}}} \\
Question type (T / V / M) & 162 / 71 / 73 & 93 / 0 / 27 & 33 / 0 / 29 & 7 / 12 / 18 & 295 / 83 / 147 \\
Mean question length & 33.1 & 47.8 & 43.7 & 29.4 & 37.5 \\ 
\hline
\multicolumn{6}{l}{{\cellcolor[rgb]{0.957,0.957,0.957}}\textbf{\textit{Reference summary}}} \\
Mean summary length & 250.3 & 710.6 & 662.7 & 241.7 & 403.6 \\ 
\hline
\multicolumn{6}{l}{{\cellcolor[rgb]{0.957,0.957,0.957}}\textbf{\textit{Grounding}}} \\
Gold facts per QA, mean / median & 38.1 / 28 & 90.5 / 79 & 82.7 / 56 & 43.4 / 33 & 55.7 / 36 \\
Visual gold-fact ratio (\%) & 37.8 & 8.3 & 17.6 & 60.7 & 24.6 \\ 
\hline
\multicolumn{6}{l}{{\cellcolor[rgb]{0.957,0.957,0.957}}\textbf{\textit{Page span}}} \\
Distinct pages, min / mean / max & 10 / 30.5 / 639 & 10 / 36.4 / 165 & 10 / 28.1 / 80 & 10 / 17.2 / 35 & 10 / 30.6 / 639 \\
\bottomrule
\end{tabular}

\caption{
DLVQA dataset statistics across four document collections.
T / V / M denotes textual-heavy, visual-heavy, and mixed questions.
Question and reference-summary lengths are measured in words.
Visual gold-fact ratio is the percentage of cited atomic facts grounded in visual elements such as figures, charts, diagrams, tables, photos, or maps.
Page span reports the number of distinct pages touched by each QA instance's gold supporting facts.
}
\label{tab:dlvqa_stats}
\end{table*}

%% file: table/dlvqa_construction_1.tex
\begin{table*}[t]
\centering
\footnotesize
\setlength{\tabcolsep}{3.8pt}
\renewcommand{\arraystretch}{1.22}

\begin{tabularx}{\textwidth}{
    @{}
    L{0.135\textwidth}
    L{0.115\textwidth}
    L{0.185\textwidth}
    Y
    L{0.105\textwidth}
    S[table-format=2.1]
    S[table-format=1.2]
    @{}
}
\toprule
\multicolumn{3}{c}{\textbf{MLLM-Assisted Construction}}
&
\multicolumn{4}{c}{\textbf{Human Review}} \\
\cmidrule(lr){1-3}
\cmidrule(l){4-7}
\textbf{Stage}
&
\textbf{Input}
&
\textbf{Output}
&
\textbf{Human Curation}
&
\textbf{Retained}
&
\multicolumn{1}{c}{\textbf{Agr. (\%)}}
&
\multicolumn{1}{c}{\(\boldsymbol{\kappa}\)} \\
\midrule

\rowcolor{tblgray}
Summarization
&
\makecell[l]{4 collections;\\3,441 pages}
&
\makecell[l]{3,441 page summaries;\\4 document summaries}
&
\makecell[l]{4 document summaries\\revised}
&
\makecell[l]{All}
& \multicolumn{1}{c}{\textsc{N/A}}
& \multicolumn{1}{c}{\textsc{N/A}} \\

\addlinespace[2pt]

Fact extraction
&
3,441 pages
&
24,450 atomic facts
&
\makecell[l]{99 facts removed;\\4,014 facts edited;\\61 facts added}
&
24,412 atomic facts
& 94.6
& 0.82 \\

\addlinespace[2pt]

\rowcolor{tblgray}
\makecell[l]{Topic clustering\\and selection}
&
24,412 atomic facts
&
\makecell[l]{649 candidate clusters\\by HDBSCAN}
&
35 clusters removed
&
614 selected clusters
& 91.2
& 0.76 \\

\addlinespace[2pt]

\makecell[l]{QA generation\\and validation}
&
614 selected clusters
&
614 QA drafts
&
\makecell[l]{312 QA drafts revised;\\89 invalid drafts removed}
&
525 QA pairs
& \multicolumn{1}{c}{\makecell[c]{88.4}}
& \multicolumn{1}{c}{\makecell[c]{0.72}} \\

\bottomrule
\end{tabularx}

\caption{
Construction statistics for DLVQA.
The left block reports inputs and outputs from MLLM-assisted stages, while the right block summarizes human curation and retained artifacts.
Agr. denotes raw pre-adjudication agreement, and \(\kappa\) denotes Cohen's kappa. \textsc{N/A} indicates stages without annotator-agreement measurement.
}
\label{tab:dlvqa_pipeline_transparency}
\end{table*}

%% file: table/dlvqa_examples_appendix.tex
\onecolumn


\clearpage
\begin{tcolorbox}[enhanced, breakable, colback=white, colframe=accent!30, boxrule=0.4pt, sharp corners, left=8pt, right=8pt, top=6pt, bottom=8pt, before skip=10pt, after skip=12pt]
\sffamily\small\textbf{\color{accent}DLVQA Example - world\_history\_q143}\hfill\textit{\color{captiongray} Material Culture in World History}\\[-0.2em]
\par\vspace{0.5em}{\color{accent!30}\hrule height 0.4pt}\vspace{0.5em}

\rmfamily\normalsize
\paragraph{Question} How does this textbook trace the development of material culture, from early Paleolithic fossils and artifacts to the architecture and art of Ming China and early modern Europe?

\noindent\begin{minipage}[t]{0.34\linewidth}
  {\sffamily\footnotesize\textbf{\color{accent}Fact composition}\hfill{\scriptsize\color{captiongray}296 gold facts}}\\[2pt]
  \begin{tikzpicture}
    \draw[rounded corners=1pt, fill=rule!40, draw=none] (0,0) rectangle (\linewidth, 0.22cm);
    \fill[blue!60!black, opacity=0.8] (0,0) rectangle (0.5473\linewidth, 0.22cm);
    \fill[orange!85!black, opacity=0.85] (0.5473\linewidth,0) rectangle (1.0000\linewidth, 0.22cm);
  \end{tikzpicture}\\[1pt]
  {\scriptsize\sffamily\color{blue!60!black}\textbf{162}\,Textual \color{captiongray}(55\%)\hfill\color{captiongray}(45\%)\,\color{orange!85!black}\textbf{134}\,Visual}
\end{minipage}\hfill\begin{minipage}[t]{0.62\linewidth}
  {\sffamily\footnotesize\textbf{\color{accent}Page coverage}\hfill{\scriptsize\color{captiongray}165 / 788 pages}}\\[2pt]
  \begin{tikzpicture}
    \fill[rule!40] (0,0) rectangle (\linewidth,0.32cm);
    \fill[accentmid, opacity=0.339] (0.0250\linewidth,0) rectangle (0.0292\linewidth,0.32cm);
    \fill[accentmid, opacity=0.339] (0.0542\linewidth,0) rectangle (0.0583\linewidth,0.32cm);
    \fill[accentmid, opacity=0.417] (0.0583\linewidth,0) rectangle (0.0625\linewidth,0.32cm);
    \fill[accentmid, opacity=0.378] (0.0625\linewidth,0) rectangle (0.0667\linewidth,0.32cm);
    \fill[accentmid, opacity=1.000] (0.0667\linewidth,0) rectangle (0.0708\linewidth,0.32cm);
    \fill[accentmid, opacity=0.689] (0.0708\linewidth,0) rectangle (0.0750\linewidth,0.32cm);
    \fill[accentmid, opacity=0.378] (0.0792\linewidth,0) rectangle (0.0833\linewidth,0.32cm);
    \fill[accentmid, opacity=0.417] (0.0875\linewidth,0) rectangle (0.0917\linewidth,0.32cm);
    \fill[accentmid, opacity=0.494] (0.0958\linewidth,0) rectangle (0.1000\linewidth,0.32cm);
    \fill[accentmid, opacity=0.456] (0.1000\linewidth,0) rectangle (0.1042\linewidth,0.32cm);
    \fill[accentmid, opacity=0.494] (0.1042\linewidth,0) rectangle (0.1083\linewidth,0.32cm);
    \fill[accentmid, opacity=0.689] (0.1083\linewidth,0) rectangle (0.1125\linewidth,0.32cm);
    \fill[accentmid, opacity=0.456] (0.1125\linewidth,0) rectangle (0.1167\linewidth,0.32cm);
    \fill[accentmid, opacity=0.378] (0.1167\linewidth,0) rectangle (0.1208\linewidth,0.32cm);
    \fill[accentmid, opacity=0.417] (0.1208\linewidth,0) rectangle (0.1250\linewidth,0.32cm);
    \fill[accentmid, opacity=0.456] (0.1292\linewidth,0) rectangle (0.1333\linewidth,0.32cm);
    \fill[accentmid, opacity=0.339] (0.1333\linewidth,0) rectangle (0.1375\linewidth,0.32cm);
    \fill[accentmid, opacity=0.456] (0.1375\linewidth,0) rectangle (0.1417\linewidth,0.32cm);
    \fill[accentmid, opacity=0.533] (0.1417\linewidth,0) rectangle (0.1458\linewidth,0.32cm);
    \fill[accentmid, opacity=0.339] (0.1458\linewidth,0) rectangle (0.1500\linewidth,0.32cm);
    \fill[accentmid, opacity=0.378] (0.1583\linewidth,0) rectangle (0.1625\linewidth,0.32cm);
    \fill[accentmid, opacity=0.378] (0.1667\linewidth,0) rectangle (0.1708\linewidth,0.32cm);
    \fill[accentmid, opacity=0.456] (0.1917\linewidth,0) rectangle (0.1958\linewidth,0.32cm);
    \fill[accentmid, opacity=0.456] (0.1958\linewidth,0) rectangle (0.2000\linewidth,0.32cm);
    \fill[accentmid, opacity=0.339] (0.2000\linewidth,0) rectangle (0.2042\linewidth,0.32cm);
    \fill[accentmid, opacity=0.494] (0.2042\linewidth,0) rectangle (0.2083\linewidth,0.32cm);
    \fill[accentmid, opacity=0.417] (0.2083\linewidth,0) rectangle (0.2125\linewidth,0.32cm);
    \fill[accentmid, opacity=0.339] (0.2125\linewidth,0) rectangle (0.2167\linewidth,0.32cm);
    \fill[accentmid, opacity=0.339] (0.2167\linewidth,0) rectangle (0.2208\linewidth,0.32cm);
    \fill[accentmid, opacity=0.339] (0.2208\linewidth,0) rectangle (0.2250\linewidth,0.32cm);
    \fill[accentmid, opacity=0.417] (0.2250\linewidth,0) rectangle (0.2292\linewidth,0.32cm);
    \fill[accentmid, opacity=0.339] (0.2333\linewidth,0) rectangle (0.2375\linewidth,0.32cm);
    \fill[accentmid, opacity=0.378] (0.2375\linewidth,0) rectangle (0.2417\linewidth,0.32cm);
    \fill[accentmid, opacity=0.572] (0.2500\linewidth,0) rectangle (0.2542\linewidth,0.32cm);
    \fill[accentmid, opacity=0.494] (0.2542\linewidth,0) rectangle (0.2583\linewidth,0.32cm);
    \fill[accentmid, opacity=0.533] (0.2625\linewidth,0) rectangle (0.2667\linewidth,0.32cm);
    \fill[accentmid, opacity=0.456] (0.2667\linewidth,0) rectangle (0.2708\linewidth,0.32cm);
    \fill[accentmid, opacity=0.456] (0.2750\linewidth,0) rectangle (0.2792\linewidth,0.32cm);
    \fill[accentmid, opacity=0.378] (0.2792\linewidth,0) rectangle (0.2833\linewidth,0.32cm);
    \fill[accentmid, opacity=0.378] (0.2875\linewidth,0) rectangle (0.2917\linewidth,0.32cm);
    \fill[accentmid, opacity=0.494] (0.2917\linewidth,0) rectangle (0.2958\linewidth,0.32cm);
    \fill[accentmid, opacity=0.339] (0.2958\linewidth,0) rectangle (0.3000\linewidth,0.32cm);
    \fill[accentmid, opacity=0.339] (0.3000\linewidth,0) rectangle (0.3042\linewidth,0.32cm);
    \fill[accentmid, opacity=0.417] (0.3125\linewidth,0) rectangle (0.3167\linewidth,0.32cm);
    \fill[accentmid, opacity=0.417] (0.3167\linewidth,0) rectangle (0.3208\linewidth,0.32cm);
    \fill[accentmid, opacity=0.339] (0.3208\linewidth,0) rectangle (0.3250\linewidth,0.32cm);
    \fill[accentmid, opacity=0.339] (0.3250\linewidth,0) rectangle (0.3292\linewidth,0.32cm);
    \fill[accentmid, opacity=0.339] (0.3375\linewidth,0) rectangle (0.3417\linewidth,0.32cm);
    \fill[accentmid, opacity=0.339] (0.3417\linewidth,0) rectangle (0.3458\linewidth,0.32cm);
    \fill[accentmid, opacity=0.456] (0.3708\linewidth,0) rectangle (0.3750\linewidth,0.32cm);
    \fill[accentmid, opacity=0.339] (0.3750\linewidth,0) rectangle (0.3792\linewidth,0.32cm);
    \fill[accentmid, opacity=0.494] (0.3792\linewidth,0) rectangle (0.3833\linewidth,0.32cm);
    \fill[accentmid, opacity=0.339] (0.4250\linewidth,0) rectangle (0.4292\linewidth,0.32cm);
    \fill[accentmid, opacity=0.456] (0.4292\linewidth,0) rectangle (0.4333\linewidth,0.32cm);
    \fill[accentmid, opacity=0.339] (0.4333\linewidth,0) rectangle (0.4375\linewidth,0.32cm);
    \fill[accentmid, opacity=0.339] (0.4375\linewidth,0) rectangle (0.4417\linewidth,0.32cm);
    \fill[accentmid, opacity=0.339] (0.4417\linewidth,0) rectangle (0.4458\linewidth,0.32cm);
    \fill[accentmid, opacity=0.339] (0.4500\linewidth,0) rectangle (0.4542\linewidth,0.32cm);
    \fill[accentmid, opacity=0.339] (0.4583\linewidth,0) rectangle (0.4625\linewidth,0.32cm);
    \fill[accentmid, opacity=0.378] (0.4667\linewidth,0) rectangle (0.4708\linewidth,0.32cm);
    \fill[accentmid, opacity=0.378] (0.4750\linewidth,0) rectangle (0.4792\linewidth,0.32cm);
    \fill[accentmid, opacity=0.378] (0.5000\linewidth,0) rectangle (0.5042\linewidth,0.32cm);
    \fill[accentmid, opacity=0.339] (0.5042\linewidth,0) rectangle (0.5083\linewidth,0.32cm);
    \fill[accentmid, opacity=0.378] (0.5125\linewidth,0) rectangle (0.5167\linewidth,0.32cm);
    \fill[accentmid, opacity=0.339] (0.5250\linewidth,0) rectangle (0.5292\linewidth,0.32cm);
    \fill[accentmid, opacity=0.339] (0.5292\linewidth,0) rectangle (0.5333\linewidth,0.32cm);
    \fill[accentmid, opacity=0.339] (0.5333\linewidth,0) rectangle (0.5375\linewidth,0.32cm);
    \fill[accentmid, opacity=0.378] (0.5417\linewidth,0) rectangle (0.5458\linewidth,0.32cm);
    \fill[accentmid, opacity=0.339] (0.5500\linewidth,0) rectangle (0.5542\linewidth,0.32cm);
    \fill[accentmid, opacity=0.378] (0.5583\linewidth,0) rectangle (0.5625\linewidth,0.32cm);
    \fill[accentmid, opacity=0.456] (0.5667\linewidth,0) rectangle (0.5708\linewidth,0.32cm);
    \fill[accentmid, opacity=0.339] (0.5833\linewidth,0) rectangle (0.5875\linewidth,0.32cm);
    \fill[accentmid, opacity=0.417] (0.5875\linewidth,0) rectangle (0.5917\linewidth,0.32cm);
    \fill[accentmid, opacity=0.378] (0.5917\linewidth,0) rectangle (0.5958\linewidth,0.32cm);
    \fill[accentmid, opacity=0.378] (0.6000\linewidth,0) rectangle (0.6042\linewidth,0.32cm);
    \fill[accentmid, opacity=0.378] (0.6167\linewidth,0) rectangle (0.6208\linewidth,0.32cm);
    \fill[accentmid, opacity=0.339] (0.6292\linewidth,0) rectangle (0.6333\linewidth,0.32cm);
    \fill[accentmid, opacity=0.378] (0.6333\linewidth,0) rectangle (0.6375\linewidth,0.32cm);
    \fill[accentmid, opacity=0.417] (0.6417\linewidth,0) rectangle (0.6458\linewidth,0.32cm);
    \fill[accentmid, opacity=0.378] (0.6458\linewidth,0) rectangle (0.6500\linewidth,0.32cm);
    \fill[accentmid, opacity=0.378] (0.6500\linewidth,0) rectangle (0.6542\linewidth,0.32cm);
    \fill[accentmid, opacity=0.417] (0.6542\linewidth,0) rectangle (0.6583\linewidth,0.32cm);
    \fill[accentmid, opacity=0.533] (0.6583\linewidth,0) rectangle (0.6625\linewidth,0.32cm);
    \fill[accentmid, opacity=0.417] (0.6667\linewidth,0) rectangle (0.6708\linewidth,0.32cm);
    \fill[accentmid, opacity=0.883] (0.6750\linewidth,0) rectangle (0.6792\linewidth,0.32cm);
    \fill[accentmid, opacity=0.339] (0.6792\linewidth,0) rectangle (0.6833\linewidth,0.32cm);
    \fill[accentmid, opacity=0.339] (0.6917\linewidth,0) rectangle (0.6958\linewidth,0.32cm);
    \fill[accentmid, opacity=0.378] (0.6958\linewidth,0) rectangle (0.7000\linewidth,0.32cm);
    \fill[accentmid, opacity=0.378] (0.7000\linewidth,0) rectangle (0.7042\linewidth,0.32cm);
    \fill[accentmid, opacity=0.417] (0.7167\linewidth,0) rectangle (0.7208\linewidth,0.32cm);
    \fill[accentmid, opacity=0.339] (0.7333\linewidth,0) rectangle (0.7375\linewidth,0.32cm);
    \fill[accentmid, opacity=0.339] (0.7458\linewidth,0) rectangle (0.7500\linewidth,0.32cm);
    \fill[accentmid, opacity=0.339] (0.7542\linewidth,0) rectangle (0.7583\linewidth,0.32cm);
    \fill[accentmid, opacity=0.339] (0.8125\linewidth,0) rectangle (0.8167\linewidth,0.32cm);
    \fill[accentmid, opacity=0.378] (0.8292\linewidth,0) rectangle (0.8333\linewidth,0.32cm);
    \fill[accentmid, opacity=0.339] (0.8417\linewidth,0) rectangle (0.8458\linewidth,0.32cm);
    \fill[accentmid, opacity=0.378] (0.8500\linewidth,0) rectangle (0.8542\linewidth,0.32cm);
    \fill[accentmid, opacity=0.417] (0.8583\linewidth,0) rectangle (0.8625\linewidth,0.32cm);
    \fill[accentmid, opacity=0.339] (0.8667\linewidth,0) rectangle (0.8708\linewidth,0.32cm);
    \fill[accentmid, opacity=0.339] (0.8708\linewidth,0) rectangle (0.8750\linewidth,0.32cm);
    \fill[accentmid, opacity=0.339] (0.8958\linewidth,0) rectangle (0.9000\linewidth,0.32cm);
    \fill[accentmid, opacity=0.417] (0.9167\linewidth,0) rectangle (0.9208\linewidth,0.32cm);
    \fill[accentmid, opacity=0.378] (0.9208\linewidth,0) rectangle (0.9250\linewidth,0.32cm);
    \fill[accentmid, opacity=0.339] (0.9333\linewidth,0) rectangle (0.9375\linewidth,0.32cm);
    \fill[accentmid, opacity=0.572] (0.9375\linewidth,0) rectangle (0.9417\linewidth,0.32cm);
    \fill[accentmid, opacity=0.417] (0.9417\linewidth,0) rectangle (0.9458\linewidth,0.32cm);
    \fill[accentmid, opacity=0.378] (0.9583\linewidth,0) rectangle (0.9625\linewidth,0.32cm);
  \end{tikzpicture}\\[1pt]
  {\scriptsize\sffamily\color{captiongray}p.1\hfill{}p.22--p.759 \textbullet{} span 738\hfill{}p.788}
\end{minipage}\par\vspace{0.5em}

\begin{tcolorbox}[enhanced, breakable, colback=accentbg!40, colframe=rule, boxrule=0.4pt, sharp corners, left=8pt, right=8pt, top=5pt, bottom=5pt, before skip=4pt, after skip=6pt]
{\sffamily\footnotesize\textbf{\color{accent}Answer Summary Guidelines}\hfill{\scriptsize\color{captiongray}\texttt{wiki} \textbullet\ 1150--1570 words}}\par\vspace{0.3em}
{\rmfamily\footnotesize Before you start writing, take a good look through the document(s) provided. Your goal is to piece everything together into a natural, Wikipedia-style article that's strictly based on the evidence in document(s).

To keep things organized, you'll need to break your answer down into logical sections using Markdown headings (for example, \#\# Historical Background). Since we're going for a clean, encyclopedic feel, please stick entirely to flowing paragraphs, this means no bullet points, numbered lists, tables, or code blocks. Your response should be between 1150 and 1570 words long; this ensures your narrative is thorough enough to cover the whole story from start to finish without including unnecessary filler.
}
\end{tcolorbox}

\paragraph{Reference summary} \paragraph{Prehistory and Early Human Settlements}

Early human material culture encompasses Acheulean and Mousterian stone tools alongside physical remnants ranging from marsupial lions to hominin fossils like Lucy. Early human creativity is preserved in Venus statuettes carved to represent fertility. Prominent examples include the forty-thousand-year-old Venus of Hohle Fels, a suspended two-inch figure carved from mammoth ivory, and the corpulent Venus of Willendorf, theorized by some to be a self-portrait. Prehistoric environments were captured by forty-thousand-year-old Spanish cave paintings depicting prey, utilizing natural rock protrusions to create three-dimensional depth within restricted spaces likely meant for ritual exhibition rather than public viewing. As the transition into the Neolithic era continued, monumental architecture and permanent settlements emerged at foundational archaeological sites like Göbekli Tepe and the walls of Jericho. Within the complex layout of Çatalhöyük, inhabitants venerated bulls through elaborate interior murals describing animal sacrifices directed toward a mother-deity, while genuine physical bull horns were explicitly mounted onto domestic furniture. Growing social organization eventually yielded massive European megalithic alignments such as Stonehenge, constructed around 2500 BCE.

\paragraph{Ancient Civilizations and the Dawn of Writing}

Urban settlement coalesced in Sumer at places like Uruk, leading to the invention of a phonetic cuneiform script around 3000 BCE. This system evolved from clay tokens encapsulated within bullae representing agricultural exchange, eventually growing into sophisticated stylus imprints detailing civic transactions. The vast wealth of this society is illustrated by the wooden Standard of Ur, an intricate mosaic-covered artifact depicting a king directing his soldiers. Monumental architecture shaped the landscape, most notably the hundred-foot-tall Ziggurat of Ur completed in the twenty-first century BCE. Sumerian spiritual life is reflected through worship statues crafted for perpetual devotion, while elite status remains visible in lapis lazuli and gold leaf headdresses. Overarching legal frameworks were documented on the Babylonian black stele detailing the Code of Hammurabi. In Egypt, intricate hierarchical symbols were inscribed upon stone steles and dry papyrus sheets to record physical sounds alongside abstract concepts. Egyptian building mastery is demonstrated by the stepped Pyramid of Djoser and the vast Giza complex. Within South Asia, Mohenjo-Daro established vast urban layouts encompassing a Citadel Mound and a stupa. Indus Valley artisans manufactured distinctive seals containing an undeciphered script, featuring scenes like a seated three-faced figure paralleling the later Hindu deity Shiva. Ceremonial performance within early Indian contexts is encapsulated by the Mohenjo-Daro Dancing Girl, a heavily bangled bronze and copper figure spanning back to 2500 BCE.

\paragraph{Classical Antiquity in the Near East and Mediterranean}

Imperial consolidation and ancient record-keeping expanded across Persia, maintained on the buried clay Cyrus Cylinder dating to 539 BCE. Darius I subsequently commissioned the monumental Behistun Relief, offering a massive trilingual justification of his rule carved directly into stone. Within Persepolis, administrative operations were tracked using diverse clay tablets. Sustaining these expansions, advanced infrastructure relied upon long-distance qanat irrigation tunnels necessary for massive agricultural yields. Concurrently in ancient Israelite history, defining architectural structures like Solomon's gates at Gezer maintain vital records of early state sovereignty, alongside profound historic expansions of major religious centers like Jerusalem’s Second Temple. Across the Mediterranean, Minoan society centered upon palatial complexes dominated by labrys symbols, female snake goddesses, and Knossos frescoes showcasing acrobatic bull leaping. Following the Minoan collapse, Greek classical culture visually immortalized athletic marvels wrapped within the physical forms captured in the Discobolus statue, accompanied by marble statues detailing kouroi youths and deities like Zeus. Continuing this classical tradition, Roman society utilized material culture for widespread utility, producing vast quantities of amphorae optimized for shipping liquids securely. Spiritual practices required miniature domestic shrines, displaying household penates alongside protective genii symbolized as a serpent, thus securing piety within local households. Regional religious integration is structurally preserved by sanctuaries portraying the Persian god Mithras sacrificing bulls upon Italian soil. Public spectacles and physical conquests were continually immortalized in stone reliefs capturing female gladiators during the first or second century CE, and through bronze statues of a winged Victory honoring complex milestones.

\paragraph{The Religious Art and Architecture of Medieval Asia}

Early Chinese written languages transitioned away from ritual divination on turtle shell oracle bones into structured characters mapping natural concepts. Imperial unification under Qin Shi Huang generated the formidable achievements of the renowned Terracotta Army, featuring thousands of individualized clay soldiers defending an elite estate. Advancing belief systems left structural footprints like the staggering cliffside Yungang Buddha statues carved into rock under the Northern Wei dynasty. Buddhism was disseminated utilizing breakthrough reproductions like the Diamond Sutra, operating as the world's first block-printed text featuring a central illustration honoring the Buddha. Medieval Chinese visual storytelling encompasses immense Dunhuang murals emphasizing religious tales of the enlightened Buddha repelling the hostile forces of Mara. These specific murals concurrently captured groundbreaking battlefield technology, showing figures holding incendiary bombs alongside deployed flame-emitting tubes. In neighboring Japan, evolving traditions moved from foundational Jōmon food containers and dogu goddess figurines to sophisticated illustrations preserving refined courtly lore surrounding The Tale of Genji, including paintings showing hierarchical ranks of priests attending the Funeral of Lady Aoi. Within Korea, exquisite ceremonial murals detailed vibrant noble couples and horse-mounted soldiers inside Goguryeo tombs, while grand devotions were embodied by the monumental Seokguram Grotto. In South Asia, multiethnic Gandhara blended cultural styles, generating unique Buddha statues wearing Greek-inspired togas. Further profound achievements extended across temple structures like the sprawling Ajanta caves, immense stupas decorated with narrative relief panels at Borobudur, and expansive spaces in Cambodia where Angkor Wat displayed overwhelming scenes drawn intimately from the Mahabharata.

\paragraph{Global Material Records of the Americas, Africa, and the Islamic World}

In the Americas, deciphered texts illustrate Maya elites recording detailed battle histories and genealogies upon varied stone monuments. Archaeological evidence confirming extensive cultural activities includes depictions of a sprawling Maya calendar system alongside ball courts and massive ceremonial architectures populated with intricate stucco characters adorning centralized structures in Palenque. By tracking complex administrative records using elaborate knotted strings known as a quipu, active Inca management safely bypassed stone tablets to maintain imperial control. Significant developmental strides across African landscapes featured the Western Deffufa, a massive mud-brick religious complex within Nubia rising nearly sixty feet tall. Other monumental architectural milestones developed throughout arid regions included the magnificent lines of the Djinguereber Mosque safely stationed in Timbuktu. Intact drywall construction functioning entirely without mortar formed reliable protective barriers entirely enclosing the open foundations of the central Great Enclosure resting upon the expansive Zimbabwean plateau. Advancing across vast Middle Eastern geographies into Central Asia, flourishing religious spheres focused intensively upon a cubic structure known as the Kaaba, shrouded in black to serve as passive centralized anchoring mapping vast networks of religious pilgrimages. These spiritual environments developed alongside incredible technological engineering, exploring mechanisms encompassing complex automatons like al-Jazari's elaborate elephant clock. This advanced artifact operated via integrated structural water mechanics causing layered chain reactions connecting an automated goad-wielding driver dropping sequential spheres.

\paragraph{The Late Middle Ages to the Early Modern Landscape}

In medieval Europe, Christian piety influenced visual materials by developing opulent manuscript illuminations, masterfully represented through the tenth-century Paris Psalter showcasing King David playing a harp. Spiritual architectural campaigns spanned deep interior aesthetics mapped by vibrant historical mosaics built capturing religious scenes inside the Italian Basilica of San Vitale and immense towering strongholds guiding foundational monastic traditions such as the third Abbey located in Cluny. Imposing military ventures eventually established massive sprawling protective fortresses across shifting territories, projecting dominance through defensive structures operating as fortified regional strongpoints such as the Krak des Chevaliers. Transitioning into the expanding early modern landscape, artistic masterworks developed layered technical depth, yielding precise representations mapping detailed Netherlandish merchants within the Arnolfini Portrait. This highly scrutinized creation accurately presented fashionable silhouettes defining prevalent gathered gown designs, rather than expressing physical representations of pregnancy. Contemporaneously framing moving early modern horizons, vast imperial authority restructuring the prosperous Ming dynasty launched centralized projects operating under overarching control, establishing lasting historical aesthetics highlighted by detailed silk hangings rendering figures displaying Emperor Zhu Yuanzhang. Consolidating immense expanding realms structurally required integrated administrations functioning within fortified parameters, framing massive boundaries comprising Halls of Military Eminence anchoring the intricate administrative network constructing the secure Forbidden City.

\paragraph{Cited pages.} \mbox{}\\[0.2em]
\noindent\centering
\includegraphics[width=\linewidth, keepaspectratio]{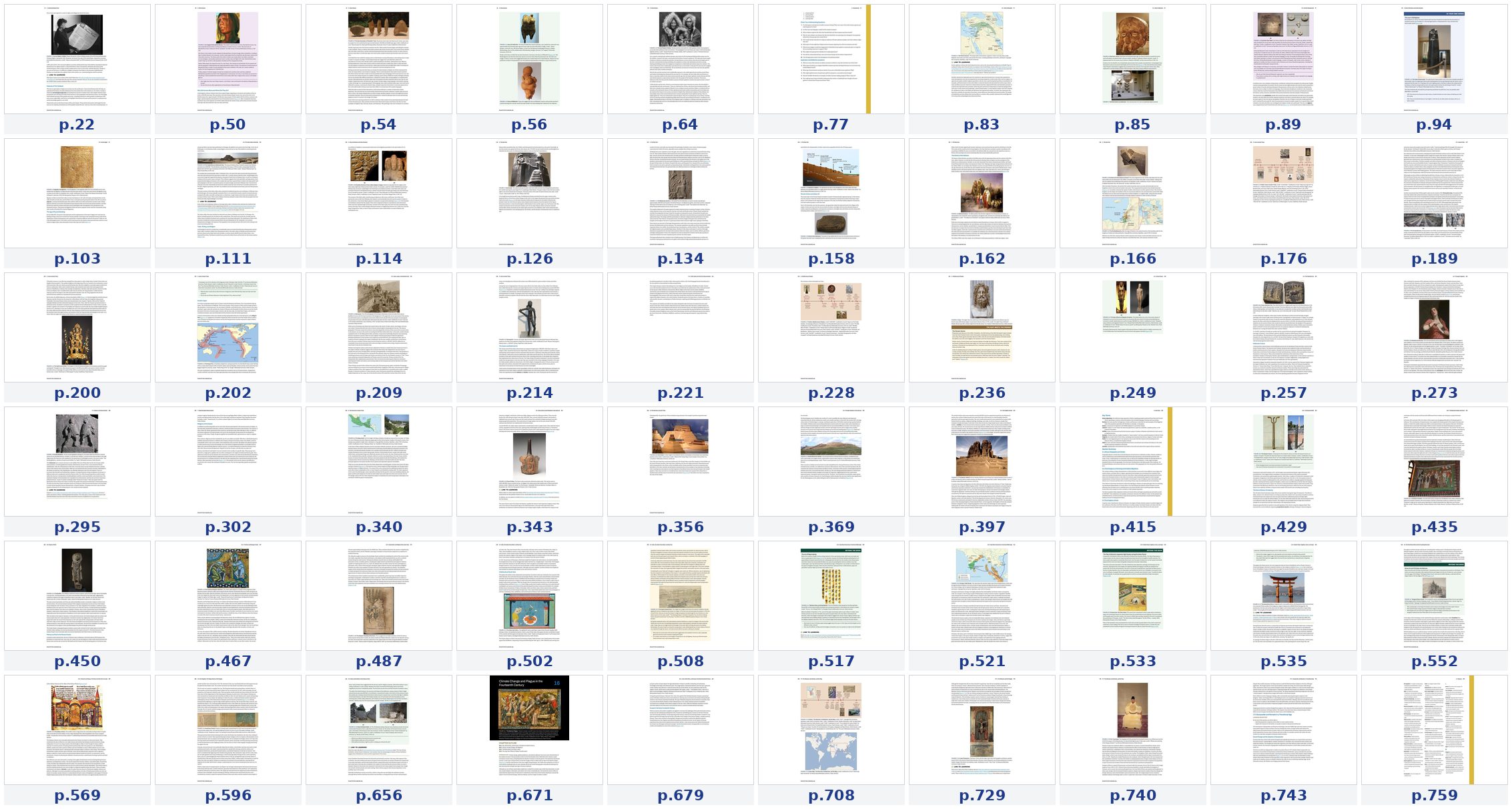}
\\[0.2em]{\footnotesize\itshape showing 60 of 165 (evenly sampled)}
\par\raggedright

\end{tcolorbox}

\clearpage
\begin{tcolorbox}[enhanced, breakable, colback=white, colframe=accent!30, boxrule=0.4pt, sharp corners, left=8pt, right=8pt, top=6pt, bottom=8pt, before skip=10pt, after skip=12pt]
\sffamily\small\textbf{\color{accent}DLVQA Example - environmental\_reports\_q056}\hfill\textit{\color{captiongray} Corporate Environmental Reporting and Strategy}\\[-0.2em]
\par\vspace{0.5em}{\color{accent!30}\hrule height 0.4pt}\vspace{0.5em}

\rmfamily\normalsize
\paragraph{Question} In Google's 2024 Environmental Report, walk me through all of their sustainability initiatives, from their AI-supported climate mitigation and adaptation solutions, through their operations and net-zero carbon energy goals, all the way to their external partnerships and policy advocacy.

\noindent\begin{minipage}[t]{0.34\linewidth}
  {\sffamily\footnotesize\textbf{\color{accent}Fact composition}\hfill{\scriptsize\color{captiongray}219 gold facts}}\\[2pt]
  \begin{tikzpicture}
    \draw[rounded corners=1pt, fill=rule!40, draw=none] (0,0) rectangle (\linewidth, 0.22cm);
    \fill[blue!60!black, opacity=0.8] (0,0) rectangle (0.8356\linewidth, 0.22cm);
    \fill[orange!85!black, opacity=0.85] (0.8356\linewidth,0) rectangle (1.0000\linewidth, 0.22cm);
  \end{tikzpicture}\\[1pt]
  {\scriptsize\sffamily\color{blue!60!black}\textbf{183}\,Textual \color{captiongray}(84\%)\hfill\color{captiongray}(16\%)\,\color{orange!85!black}\textbf{36}\,Visual}
\end{minipage}\hfill\begin{minipage}[t]{0.62\linewidth}
  {\sffamily\footnotesize\textbf{\color{accent}Page coverage}\hfill{\scriptsize\color{captiongray}68 / 422 pages}}\\[2pt]
  \begin{tikzpicture}
    \fill[rule!40] (0,0) rectangle (\linewidth,0.32cm);
    \fill[accentmid, opacity=0.533] (0.3625\linewidth,0) rectangle (0.3667\linewidth,0.32cm);
    \fill[accentmid, opacity=0.650] (0.3667\linewidth,0) rectangle (0.3708\linewidth,0.32cm);
    \fill[accentmid, opacity=0.825] (0.3708\linewidth,0) rectangle (0.3750\linewidth,0.32cm);
    \fill[accentmid, opacity=0.767] (0.3750\linewidth,0) rectangle (0.3792\linewidth,0.32cm);
    \fill[accentmid, opacity=0.417] (0.3792\linewidth,0) rectangle (0.3833\linewidth,0.32cm);
    \fill[accentmid, opacity=0.825] (0.3833\linewidth,0) rectangle (0.3875\linewidth,0.32cm);
    \fill[accentmid, opacity=0.708] (0.3875\linewidth,0) rectangle (0.3917\linewidth,0.32cm);
    \fill[accentmid, opacity=0.533] (0.3917\linewidth,0) rectangle (0.3958\linewidth,0.32cm);
    \fill[accentmid, opacity=0.358] (0.3958\linewidth,0) rectangle (0.4000\linewidth,0.32cm);
    \fill[accentmid, opacity=0.650] (0.4000\linewidth,0) rectangle (0.4042\linewidth,0.32cm);
    \fill[accentmid, opacity=0.592] (0.4042\linewidth,0) rectangle (0.4083\linewidth,0.32cm);
    \fill[accentmid, opacity=0.592] (0.4083\linewidth,0) rectangle (0.4125\linewidth,0.32cm);
    \fill[accentmid, opacity=0.475] (0.4125\linewidth,0) rectangle (0.4167\linewidth,0.32cm);
    \fill[accentmid, opacity=0.708] (0.4167\linewidth,0) rectangle (0.4208\linewidth,0.32cm);
    \fill[accentmid, opacity=0.942] (0.4208\linewidth,0) rectangle (0.4250\linewidth,0.32cm);
    \fill[accentmid, opacity=0.767] (0.4250\linewidth,0) rectangle (0.4292\linewidth,0.32cm);
    \fill[accentmid, opacity=0.533] (0.4292\linewidth,0) rectangle (0.4333\linewidth,0.32cm);
    \fill[accentmid, opacity=1.000] (0.4333\linewidth,0) rectangle (0.4375\linewidth,0.32cm);
    \fill[accentmid, opacity=0.417] (0.4375\linewidth,0) rectangle (0.4417\linewidth,0.32cm);
    \fill[accentmid, opacity=0.592] (0.4417\linewidth,0) rectangle (0.4458\linewidth,0.32cm);
    \fill[accentmid, opacity=0.650] (0.4458\linewidth,0) rectangle (0.4500\linewidth,0.32cm);
    \fill[accentmid, opacity=0.883] (0.4500\linewidth,0) rectangle (0.4542\linewidth,0.32cm);
    \fill[accentmid, opacity=0.825] (0.4542\linewidth,0) rectangle (0.4583\linewidth,0.32cm);
    \fill[accentmid, opacity=0.592] (0.4583\linewidth,0) rectangle (0.4625\linewidth,0.32cm);
    \fill[accentmid, opacity=0.358] (0.4708\linewidth,0) rectangle (0.4750\linewidth,0.32cm);
    \fill[accentmid, opacity=0.417] (0.4750\linewidth,0) rectangle (0.4792\linewidth,0.32cm);
    \fill[accentmid, opacity=0.417] (0.4792\linewidth,0) rectangle (0.4833\linewidth,0.32cm);
    \fill[accentmid, opacity=0.475] (0.4833\linewidth,0) rectangle (0.4875\linewidth,0.32cm);
    \fill[accentmid, opacity=0.358] (0.4917\linewidth,0) rectangle (0.4958\linewidth,0.32cm);
    \fill[accentmid, opacity=0.417] (0.4958\linewidth,0) rectangle (0.5000\linewidth,0.32cm);
    \fill[accentmid, opacity=0.358] (0.5000\linewidth,0) rectangle (0.5042\linewidth,0.32cm);
    \fill[accentmid, opacity=0.533] (0.5042\linewidth,0) rectangle (0.5083\linewidth,0.32cm);
    \fill[accentmid, opacity=0.708] (0.5083\linewidth,0) rectangle (0.5125\linewidth,0.32cm);
    \fill[accentmid, opacity=0.767] (0.5125\linewidth,0) rectangle (0.5167\linewidth,0.32cm);
    \fill[accentmid, opacity=0.358] (0.5167\linewidth,0) rectangle (0.5208\linewidth,0.32cm);
    \fill[accentmid, opacity=0.825] (0.5208\linewidth,0) rectangle (0.5250\linewidth,0.32cm);
    \fill[accentmid, opacity=0.942] (0.5250\linewidth,0) rectangle (0.5292\linewidth,0.32cm);
    \fill[accentmid, opacity=0.358] (0.5292\linewidth,0) rectangle (0.5333\linewidth,0.32cm);
    \fill[accentmid, opacity=0.417] (0.5375\linewidth,0) rectangle (0.5417\linewidth,0.32cm);
    \fill[accentmid, opacity=0.358] (0.5417\linewidth,0) rectangle (0.5458\linewidth,0.32cm);
    \fill[accentmid, opacity=0.533] (0.5458\linewidth,0) rectangle (0.5500\linewidth,0.32cm);
    \fill[accentmid, opacity=0.650] (0.5542\linewidth,0) rectangle (0.5583\linewidth,0.32cm);
    \fill[accentmid, opacity=0.358] (0.5583\linewidth,0) rectangle (0.5625\linewidth,0.32cm);
    \fill[accentmid, opacity=0.533] (0.5625\linewidth,0) rectangle (0.5667\linewidth,0.32cm);
  \end{tikzpicture}\\[1pt]
  {\scriptsize\sffamily\color{captiongray}p.1\hfill{}p.155--p.240 \textbullet{} span 86\hfill{}p.422}
\end{minipage}\par\vspace{0.5em}

\begin{tcolorbox}[enhanced, breakable, colback=accentbg!40, colframe=rule, boxrule=0.4pt, sharp corners, left=8pt, right=8pt, top=5pt, bottom=5pt, before skip=4pt, after skip=6pt]
{\sffamily\footnotesize\textbf{\color{accent}Answer Summary Guidelines}\hfill{\scriptsize\color{captiongray}\texttt{wiki} \textbullet\ 880--1190 words}}\par\vspace{0.3em}
{\rmfamily\footnotesize Before you start writing, take a good look through the document(s) provided. Your goal is to piece everything together into a natural, Wikipedia-style article that's strictly based on the evidence in document(s).

To keep things organized, you'll need to break your answer down into logical sections using Markdown headings (for example, \#\# Historical Background). Since we're going for a clean, encyclopedic feel, please stick entirely to flowing paragraphs, this means no bullet points, numbered lists, tables, or code blocks. Your response should be between 880 and 1190 words long; this ensures your narrative is thorough enough to cover the whole story from start to finish without including unnecessary filler.
}
\end{tcolorbox}

\paragraph{Reference summary} \paragraph{Report Structure and Governance}

The Google 2024 Environmental Report outlines the company's sustainability strategy for the 2023 fiscal year. The document chapters cover artificial intelligence, product architecture, global operations, and detailed data appendices. The introduction features target summaries and a leadership overview of goals running to 2050, accompanied by links to overarching corporate value missions. Oversight for environmental risk according to TCFD parameters is managed by the Board of Directors' Audit and Compliance Committee.

Operations span six continents and nearly 60 countries, aiming to accelerate net-zero timelines, water stewardship, circular economies, and biodiversity restoration. Extensive tables highlight progress against diverse metrics as of December 31, 2023. Reinforcing these benchmarks, Google earned positions as a CDP Supplier Engagement Leader and secured an A score on the CDP Climate Change list. Independent assessments provided top evaluations in TIME's World's Best Companies array and a 5/5 score inside the Gartner Sustainability assessment.

\paragraph{AI for Climate Mitigation}

Integrating artificial intelligence toward environmental goals is a core strategy supported by estimates that machine learning models could mitigate 10\% of global emissions by 2030. As an information hub, Google predicts and optimizes climate impacts for individuals and enterprise environments. To combat misleading environmental narratives, accurate United Nations resources are presented directly across video platforms.

Reimagining everyday commuting, navigation algorithms guide users toward mass-transit layouts, energy-saving driving paths, and electric vehicle corridors. AI additionally predicts and optimizes flight paths to reduce contrails by 54\%. For heavy commercial arteries, Project Green Light synchronizes urban intersections and suppresses stop-and-go emission cycles.

Broader analytical utilities rely on aerial mapping models identifying local buildings suitable for clean energy architecture, while Cloud capabilities analyze MethaneSAT arrays to pinpoint global leaks and continuous energy transmission behaviors. Residential energy reductions focus on smart appliances, like Nest Learning Thermostats, analyzing usage to automate efficient internal climates.

\paragraph{Extreme Weather Adaptation}

Google's mapping tools act as an alert system against extreme environmental variations. High-resolution monitoring leverages Earth-observation satellites and databases visualizing intense heat fluctuations, wildfire risks, and hazardous pollution indicators. Advanced hydrological neural models anticipate severe river dynamics to generate localized flood timelines, and specific warning displays track shifting vulnerabilities to alert nearby populations ahead of hazard boundaries. Systems also issue notifications to residents situated near moving pollution plumes and historic dangerous heat sinks.

Modern meteorology incorporates DeepMind GraphCast and MetNet-3 frameworks, computing 10-day forecasts in under one minute and boosting global weather accuracy by 15\%. Concurrent AI deployments detect live wildfires, establishing instantaneous geographic outlines mapping active hazard edges. Guarding sensitive biological habitats utilizes outside ecosystem alliances such as the Forest Data Partnership preventing unauthorized remote logging, and the Global Fishing Watch tracking mass industrial fishery vessels via satellite optics.

\paragraph{Operations and Ecosystem Resilience}

Sustaining computational networks worldwide requires balancing infrastructure across numerous geographical regions. As intensive algorithms demand elevated hardware outputs, efficiency optimizations maximize distributions uniformly across operating server networks. Hardware facilities utilize integrated physical aesthetics covering heavy mass timber architectures and broad solar canopies to build carbon-reduction templates, mirroring natural architectural interiors that support human environmental interaction formats comfortably.

Maintaining corporate material integrity restricts toxic chemical exposures inside third-party manufacturing environments. Google drives circular consumer product layouts ensuring core technical items guarantee maximized secondary lifespans and avoid destructive disposal methodologies. Environmental restitution protocols mandate external habitat verifications documenting discrete bio-anomalies highlighting distinct land recovery obstacles. Complementing these measures, long-term groundwater conservation programs work closely to restore widespread natural soil drainage patterns.

\paragraph{The Net-Zero Carbon Commitment}

Achieving global reductions mandates a net-zero footprint across the entire supply chain by 2030. Rigorous procedures evaluate performance data confirming standard gross corporate calculations against accepted international accounting criteria. Curbing external hardware impacts relies on structural treaties requiring partner manufacturers to execute verifiable clean power utilization paths achieving 100\% matches by 2029. To promote improved results, interconnected hardware organizations formalize specialized collaborations testing diverse tracking architectures supporting operational alignment.

Additional infrastructural investments support the expansion of localized electric vehicle chargers alongside backing specialized advancements around commercial aviation fuel models. Addressing dense atmospheric gases lacking rapid conversion replacements involves drafting long-term funding contracts targeting early theoretical geological carbon removal technologies.

\paragraph{Clean Energy Progress and 24/7 CFE}

Google consistently achieves total equivalence matching cumulative global electricity consumption with pure clean energy. Attaining neutrality demands immense generation capacities relying heavily upon extensive international energy acquisitions. Financial operations feed those combined architectures, utilizing over \$3.1 billion assigned toward advancing 4.5 GW of upcoming capabilities.

Accelerating advanced algorithmic loads significantly raises absolute data center electricity consumption metrics globally. Executing viable climate mitigation revolves around targeted localized alignment matching physical energy demand by utilizing zero-carbon sourcing on an hourly basis. Despite technical generation issues, operational average integration parameters remain steady. Global facility mappings visualize discrete localized hourly frameworks, determining 10 out of 44 worldwide internal systems crossed 90\% performance parameters over 2023. Extensive structural breakdowns emphasize prominent continental American zones achieving well above typical temporal marks regularly.

Substantial logistical gridlock restricts clean structural development heavily across widespread interconnected Asian segments. Overcoming immense regional inertia relies exclusively on bypassing local structural barriers like complex supply chain disruptions, difficult regulatory market tariffs, and outdated generation boundaries.

\paragraph{Collaborations, Philanthropy, and Policy Advocacy}

Systemic climate adaptation depends upon integrating open-source visualizations allowing researchers to map complex macro variations. Further centralized visualization layers track individual business resource workflows, dissecting complicated interactive corporate financing mechanics clearly.

Grants issue resources directing core ecological intelligence programs that record large physical biosystems directly. Philanthropic measures expand startup incubators, data security networks defending high-risk communities, and tech organizations shaping ecological commercial strategies. Addressing residential inequities formally channels distinct facility power revenues to specific adaptations aiding marginalized, low-income segments.

Macro planning aligns broad structural directives sharing mutual objectives supported across major international hubs including the C40 Cities program towards precise IEA analytical targets. Influencing local transitions emphasizes advocating transparent clean utility guidelines alongside accurate operational environmental labeling mechanics universally. Oversight parameters ensure active external trade alignments reflect stated Paris Agreement thresholds faithfully, positioning Google representatives to back commitments pushing to triple total international renewable capacities throughout COP-28 procedures. External verification utilizes university alignments framing institutional sustainability communities, matched against broad organizational allocations awarding engineering personnel a 20\% timing release probing original ecological engagements internally.

\paragraph{Cited pages.} \mbox{}\\[0.2em]
\noindent\centering
\includegraphics[width=\linewidth, keepaspectratio]{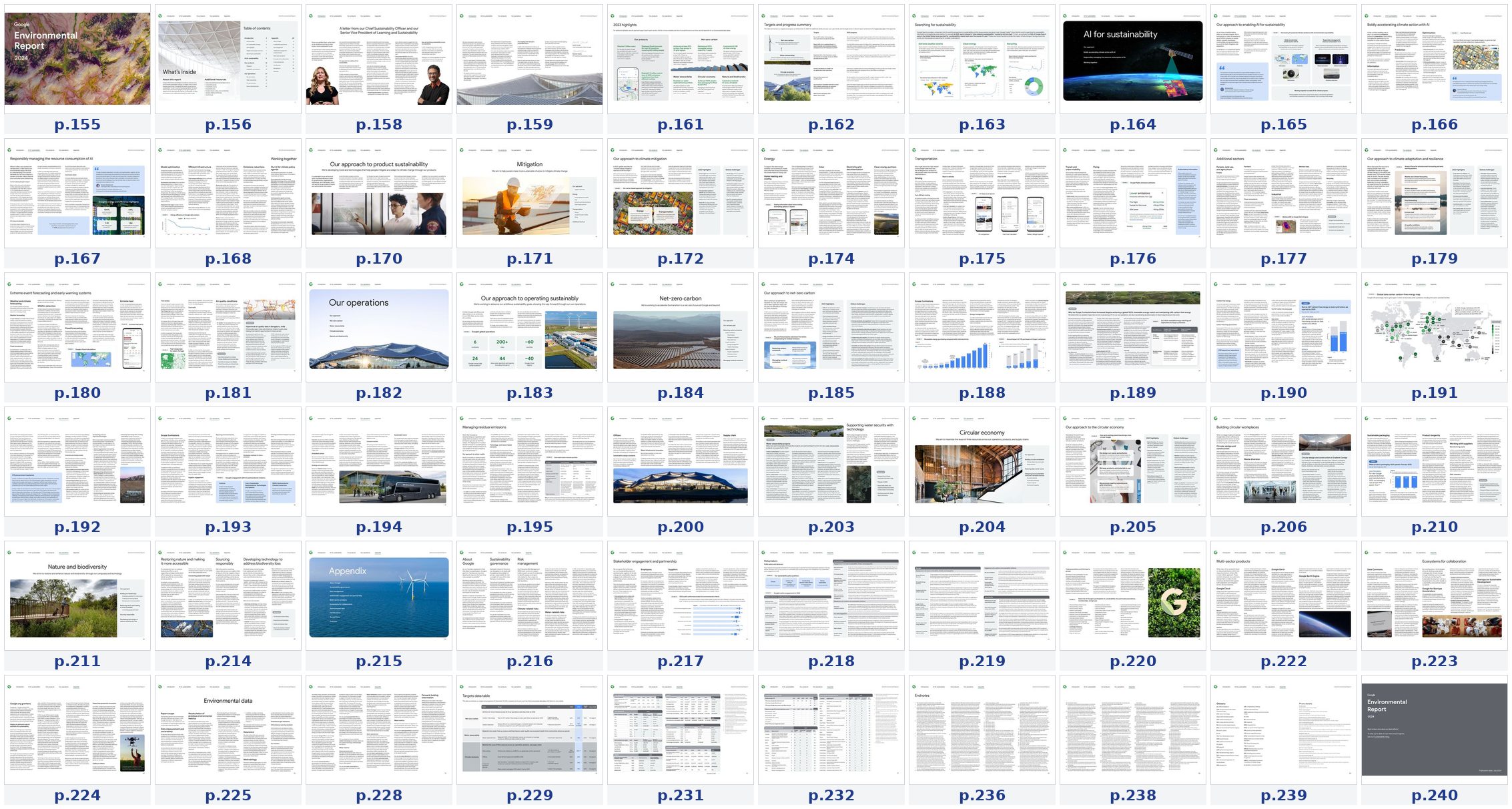}
\\[0.2em]{\footnotesize\itshape showing 60 of 68 (evenly sampled)}
\par\raggedright

\end{tcolorbox}

\clearpage
\begin{tcolorbox}[enhanced, breakable, colback=white, colframe=accent!30, boxrule=0.4pt, sharp corners, left=8pt, right=8pt, top=6pt, bottom=8pt, before skip=10pt, after skip=12pt]
\sffamily\small\textbf{\color{accent}Example DLCV\_q300}\hfill\textit{\color{captiongray} Foundations of Attention Mechanisms}\\[-0.2em]{\scriptsize\color{captiongray}Pages spanned: 219--798 (71 pages)}
\par\vspace{0.5em}{\color{accent!30}\hrule height 0.4pt}\vspace{0.5em}

\rmfamily\normalsize
\paragraph{Question} Walk me through how the math and architecture of attention mechanisms evolved, starting from basic alignment intuition and context vectors through to self-attention, Multi-head Transformers, and specialized vision applications.

\noindent\begin{minipage}[t]{0.34\linewidth}
  {\sffamily\footnotesize\textbf{\color{accent}Fact composition}\hfill{\scriptsize\color{captiongray}154 gold facts}}\\[2pt]
  \begin{tikzpicture}
    \draw[rounded corners=1pt, fill=rule!40, draw=none] (0,0) rectangle (\linewidth, 0.22cm);
    \fill[blue!60!black, opacity=0.8] (0,0) rectangle (0.5000\linewidth, 0.22cm);
    \fill[orange!85!black, opacity=0.85] (0.5000\linewidth,0) rectangle (1.0000\linewidth, 0.22cm);
  \end{tikzpicture}\\[1pt]
  {\scriptsize\sffamily\color{blue!60!black}\textbf{77}\,Textual \color{captiongray}(50\%)\hfill\color{captiongray}(50\%)\,\color{orange!85!black}\textbf{77}\,Visual}
\end{minipage}\hfill\begin{minipage}[t]{0.62\linewidth}
  {\sffamily\footnotesize\textbf{\color{accent}Page coverage}\hfill{\scriptsize\color{captiongray}71 / 1984 pages}}\\[2pt]
  \begin{tikzpicture}
    \fill[rule!40] (0,0) rectangle (\linewidth,0.32cm);
    \fill[accentmid, opacity=0.745] (0.1083\linewidth,0) rectangle (0.1125\linewidth,0.32cm);
    \fill[accentmid, opacity=0.650] (0.1125\linewidth,0) rectangle (0.1167\linewidth,0.32cm);
    \fill[accentmid, opacity=0.395] (0.1167\linewidth,0) rectangle (0.1208\linewidth,0.32cm);
    \fill[accentmid, opacity=0.650] (0.1208\linewidth,0) rectangle (0.1250\linewidth,0.32cm);
    \fill[accentmid, opacity=0.555] (0.1250\linewidth,0) rectangle (0.1292\linewidth,0.32cm);
    \fill[accentmid, opacity=1.000] (0.1292\linewidth,0) rectangle (0.1333\linewidth,0.32cm);
    \fill[accentmid, opacity=0.873] (0.1333\linewidth,0) rectangle (0.1375\linewidth,0.32cm);
    \fill[accentmid, opacity=0.841] (0.1375\linewidth,0) rectangle (0.1417\linewidth,0.32cm);
    \fill[accentmid, opacity=0.586] (0.1417\linewidth,0) rectangle (0.1458\linewidth,0.32cm);
    \fill[accentmid, opacity=0.491] (0.1458\linewidth,0) rectangle (0.1500\linewidth,0.32cm);
    \fill[accentmid, opacity=0.491] (0.1500\linewidth,0) rectangle (0.1542\linewidth,0.32cm);
    \fill[accentmid, opacity=0.332] (0.1542\linewidth,0) rectangle (0.1583\linewidth,0.32cm);
    \fill[accentmid, opacity=0.364] (0.1583\linewidth,0) rectangle (0.1625\linewidth,0.32cm);
    \fill[accentmid, opacity=0.555] (0.1625\linewidth,0) rectangle (0.1667\linewidth,0.32cm);
    \fill[accentmid, opacity=0.427] (0.1667\linewidth,0) rectangle (0.1708\linewidth,0.32cm);
    \fill[accentmid, opacity=0.555] (0.2625\linewidth,0) rectangle (0.2667\linewidth,0.32cm);
    \fill[accentmid, opacity=0.332] (0.2792\linewidth,0) rectangle (0.2833\linewidth,0.32cm);
    \fill[accentmid, opacity=0.395] (0.3667\linewidth,0) rectangle (0.3708\linewidth,0.32cm);
    \fill[accentmid, opacity=0.332] (0.3708\linewidth,0) rectangle (0.3750\linewidth,0.32cm);
    \fill[accentmid, opacity=0.332] (0.4000\linewidth,0) rectangle (0.4042\linewidth,0.32cm);
  \end{tikzpicture}\\[1pt]
  {\scriptsize\sffamily\color{captiongray}p.1\hfill{}p.219--p.798 \textbullet{} span 580\hfill{}p.1984}
\end{minipage}\par\vspace{0.5em}

\begin{tcolorbox}[enhanced, breakable, colback=accentbg!40, colframe=rule, boxrule=0.4pt, sharp corners, left=8pt, right=8pt, top=5pt, bottom=5pt, before skip=4pt, after skip=6pt]
{\sffamily\footnotesize\textbf{\color{accent}Answer Summary Guidelines}\hfill{\scriptsize\color{captiongray}\texttt{wiki} \textbullet\ 200--270 words}}\par\vspace{0.3em}
{\rmfamily\footnotesize Before you start writing, take a good look through the document(s) provided. Your goal is to piece everything together into a natural, Wikipedia-style article that's strictly based on the evidence in document(s).

To keep things organized, you'll need to break your answer down into logical sections using Markdown headings (for example, \#\# Historical Background). Since we're going for a clean, encyclopedic feel, please stick entirely to flowing paragraphs, this means no bullet points, numbered lists, tables, or code blocks. Your response should be between 200 and 270 words long; this ensures your narrative is thorough enough to cover the whole story from start to finish without including unnecessary filler.
}
\end{tcolorbox}

\paragraph{Reference summary} \paragraph{Foundational Intuition and Alignment Computation}

Attention mechanisms contextually weight inputs to form block alignments, handling word orders unsupervised. Passing hidden states and inputs through a linear layer or MLP produces scalar alignment scores, which a softmax function normalizes into probabilities summing to one.

\paragraph{Architecting the Context Vector}

Normalized weights guide the summation of sequence elements into a single context vector, mapping queries onto base features. Since processing relies purely on distributed weighting, operations are permutation equivariant, ignoring sequence order unless explicit positional encodings are used.

\paragraph{Query, Key, and Value Operations}

Self-attention uses projection layers to form parallel Query, Key, and Value matrices from inputs. Alignment scores equal the inner products between queries and keys, divided by an adjustment factor. Softmax provides mapping probabilities for the weighted combination. Autoregressive masking sets specific parameters to negative infinity, preventing future token glimpses.

\paragraph{Transformers and Multi-Head Networks}

Models scale this process by dividing streams across parallel heads to calculate distinct alignments simultaneously. Inside Transformer encoders, self-attention provides direct feature communication alongside integrated residual connections. Decoding pipelines pair masked teacher forcing with cross-attention that parses encoder outputs.

\paragraph{Spatiotemporal and Visual Implementation}

These concepts extend to vision by transposing queried features against key layouts on spatial grids. Normalized values feed 1x1 convolutions to map dimensions while preserving scales. Video frameworks expand this via spatiotemporal cross-multiplications. Unsupervised Vision Transformers can also generate segmentations around bottleneck CLS tokens.

\paragraph{Cited pages.} \mbox{}\\[0.2em]
\noindent\centering
\includegraphics[width=\linewidth, keepaspectratio]{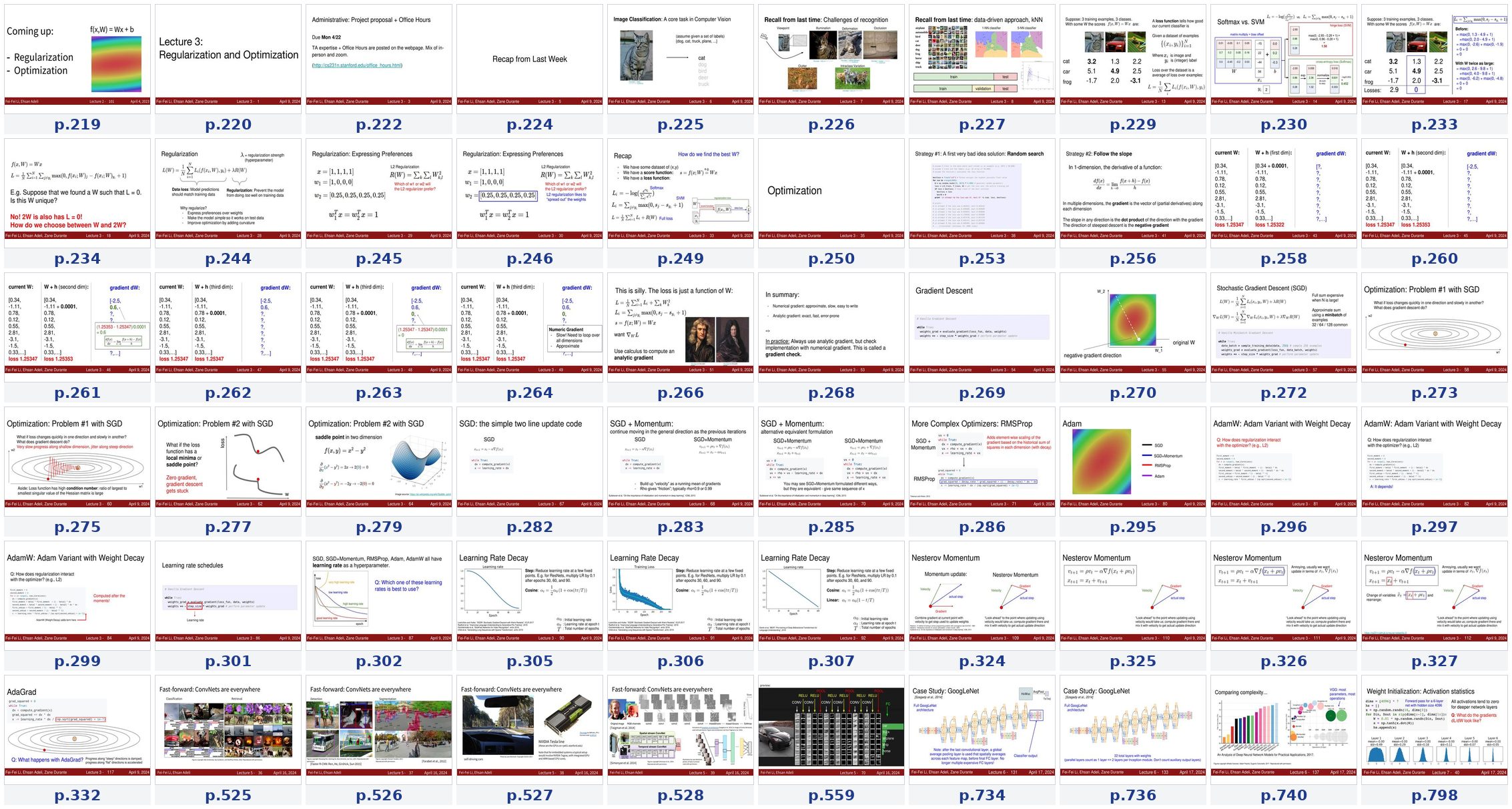}
\\[0.2em]{\footnotesize\itshape showing 60 of 71 (evenly sampled)}
\par\raggedright

\end{tcolorbox}

\clearpage
\begin{tcolorbox}[enhanced, breakable, colback=white, colframe=accent!30, boxrule=0.4pt, sharp corners, left=8pt, right=8pt, top=6pt, bottom=8pt, before skip=10pt, after skip=12pt]
\sffamily\small\textbf{\color{accent}DLVQA Example - picture\_books\_q033}\hfill\textit{\color{captiongray} 'How to Tame a Monster' Analysis}\\[-0.2em]
\par\vspace{0.5em}{\color{accent!30}\hrule height 0.4pt}\vspace{0.5em}

\rmfamily\normalsize
\paragraph{Question} Walk me through the whole story and the artwork in 'How to Tame a Monster', starting from the opening pages and character intros, tracking how the emotions build up every day, all the way to how it gets resolved.

\noindent\begin{minipage}[t]{0.34\linewidth}
  {\sffamily\footnotesize\textbf{\color{accent}Fact composition}\hfill{\scriptsize\color{captiongray}114 gold facts}}\\[2pt]
  \begin{tikzpicture}
    \draw[rounded corners=1pt, fill=rule!40, draw=none] (0,0) rectangle (\linewidth, 0.22cm);
    \fill[blue!60!black, opacity=0.8] (0,0) rectangle (0.2544\linewidth, 0.22cm);
    \fill[orange!85!black, opacity=0.85] (0.2544\linewidth,0) rectangle (1.0000\linewidth, 0.22cm);
  \end{tikzpicture}\\[1pt]
  {\scriptsize\sffamily\color{blue!60!black}\textbf{29}\,Textual \color{captiongray}(25\%)\hfill\color{captiongray}(75\%)\,\color{orange!85!black}\textbf{85}\,Visual}
\end{minipage}\hfill\begin{minipage}[t]{0.62\linewidth}
  {\sffamily\footnotesize\textbf{\color{accent}Page coverage}\hfill{\scriptsize\color{captiongray}30 / 247 pages\%}}\\[2pt]
  \begin{tikzpicture}
    \fill[rule!40] (0,0) rectangle (\linewidth,0.32cm);
    \fill[accentmid, opacity=0.738] (0.0708\linewidth,0) rectangle (0.0750\linewidth,0.32cm);
    \fill[accentmid, opacity=0.562] (0.0750\linewidth,0) rectangle (0.0792\linewidth,0.32cm);
    \fill[accentmid, opacity=0.650] (0.0833\linewidth,0) rectangle (0.0875\linewidth,0.32cm);
    \fill[accentmid, opacity=0.475] (0.0875\linewidth,0) rectangle (0.0917\linewidth,0.32cm);
    \fill[accentmid, opacity=0.562] (0.0917\linewidth,0) rectangle (0.0958\linewidth,0.32cm);
    \fill[accentmid, opacity=0.650] (0.0958\linewidth,0) rectangle (0.1000\linewidth,0.32cm);
    \fill[accentmid, opacity=0.738] (0.1000\linewidth,0) rectangle (0.1042\linewidth,0.32cm);
    \fill[accentmid, opacity=0.738] (0.1042\linewidth,0) rectangle (0.1083\linewidth,0.32cm);
    \fill[accentmid, opacity=0.738] (0.1083\linewidth,0) rectangle (0.1125\linewidth,0.32cm);
    \fill[accentmid, opacity=0.562] (0.1125\linewidth,0) rectangle (0.1167\linewidth,0.32cm);
    \fill[accentmid, opacity=0.562] (0.1167\linewidth,0) rectangle (0.1208\linewidth,0.32cm);
    \fill[accentmid, opacity=0.738] (0.1208\linewidth,0) rectangle (0.1250\linewidth,0.32cm);
    \fill[accentmid, opacity=0.650] (0.1250\linewidth,0) rectangle (0.1292\linewidth,0.32cm);
    \fill[accentmid, opacity=0.562] (0.1292\linewidth,0) rectangle (0.1333\linewidth,0.32cm);
    \fill[accentmid, opacity=0.562] (0.1333\linewidth,0) rectangle (0.1375\linewidth,0.32cm);
    \fill[accentmid, opacity=0.738] (0.1375\linewidth,0) rectangle (0.1417\linewidth,0.32cm);
    \fill[accentmid, opacity=0.825] (0.1417\linewidth,0) rectangle (0.1458\linewidth,0.32cm);
    \fill[accentmid, opacity=0.562] (0.1458\linewidth,0) rectangle (0.1500\linewidth,0.32cm);
    \fill[accentmid, opacity=0.562] (0.1500\linewidth,0) rectangle (0.1542\linewidth,0.32cm);
    \fill[accentmid, opacity=0.650] (0.1542\linewidth,0) rectangle (0.1583\linewidth,0.32cm);
    \fill[accentmid, opacity=0.562] (0.1583\linewidth,0) rectangle (0.1625\linewidth,0.32cm);
    \fill[accentmid, opacity=0.562] (0.1625\linewidth,0) rectangle (0.1667\linewidth,0.32cm);
    \fill[accentmid, opacity=0.738] (0.1667\linewidth,0) rectangle (0.1708\linewidth,0.32cm);
    \fill[accentmid, opacity=0.738] (0.1708\linewidth,0) rectangle (0.1750\linewidth,0.32cm);
    \fill[accentmid, opacity=0.650] (0.1750\linewidth,0) rectangle (0.1792\linewidth,0.32cm);
    \fill[accentmid, opacity=0.650] (0.1792\linewidth,0) rectangle (0.1833\linewidth,0.32cm);
    \fill[accentmid, opacity=0.475] (0.1833\linewidth,0) rectangle (0.1875\linewidth,0.32cm);
    \fill[accentmid, opacity=1.000] (0.1875\linewidth,0) rectangle (0.1917\linewidth,0.32cm);
    \fill[accentmid, opacity=0.475] (0.1917\linewidth,0) rectangle (0.1958\linewidth,0.32cm);
  \end{tikzpicture}\\[1pt]
  {\scriptsize\sffamily\color{captiongray}p.1\hfill{}p.19--p.49 \textbullet{} span 31\hfill{}p.247}
\end{minipage}\par\vspace{0.5em}

\begin{tcolorbox}[enhanced, breakable, colback=accentbg!40, colframe=rule, boxrule=0.4pt, sharp corners, left=8pt, right=8pt, top=5pt, bottom=5pt, before skip=4pt, after skip=6pt]
{\sffamily\footnotesize\textbf{\color{accent}Answer Summary Guidelines}\hfill{\scriptsize\color{captiongray}\texttt{wiki} \textbullet\ 620--850 words}}\par\vspace{0.3em}
{\rmfamily\footnotesize Before you start writing, take a good look through the document(s) provided. Your goal is to piece everything together into a natural, Wikipedia-style article that's strictly based on the evidence in document(s).

To keep things organized, you'll need to break your answer down into logical sections using Markdown headings (for example, \#\# Historical Background). Since we're going for a clean, encyclopedic feel, please stick entirely to flowing paragraphs, this means no bullet points, numbered lists, tables, or code blocks. Your response should be between 620 and 850 words long; this ensures your narrative is thorough enough to cover the whole story from start to finish without including unnecessary filler.
}
\end{tcolorbox}

\paragraph{Reference summary} \paragraph{Book Details and Structure}

The children's book 'How to Tame a Monster' appears on page 22 of the digital anthology, with its large, sans-serif title featured prominently early on. The cover credits illustrator Anja Venter, writer Sam Beckbessinger, and designer Wilna Combrinck, alongside editor Carla Lever. The cover uses a simple color-blocked background with a white top and a light green lower section. Early pages feature a standard "This book belongs to" line and a monochrome illustration of an adult reading to a child alongside a lizard-like monster. The main narrative follows a young boy named Lwando as he learns to manage the emotional outbursts of his sister, Oyiso.

\paragraph{Unveiling the Monster}

Lwando, a brown-skinned boy wearing a blue-green shirt, dark shorts, and red shoes, is on a mission to be a monster tamer. His sister Oyiso is usually cute but sometimes turns into a monster. Visually represented early on as unhappy girls in yellow dresses, her monstrous form manifests as a large, roaring yellow creature crying blue tears. To combat these transformations, Lwando outfits himself in a homemade cardboard box helmet and armor, carrying an orange bag with an 'O' symbol and a yellow bag with a red circle. During one early chaotic encounter, the grassy monster glares at Lwando while generating jagged red lightning shapes to signify loud noise.

\paragraph{Understanding Emotions (Monday to Thursday)}

Lwando learns to identify the root causes of the emotional outbursts over the course of the week. On Monday, the golden-yellow monster appears angry with downward-sloping eyes. A pink butterfly floats closely by as Lwando deduces she is just hungry, and the scene resolves with Oyiso eating. On Tuesday, the monster's anger is depicted with jagged spikes as she roars over a blue bed. Realizing she is tired, Lwando checks on the bed to find Oyiso sleeping peacefully beside a pale beige animal. Wednesday brings an even angrier monster clawing at the air next to three violent red lightning symbols. Near a beige school building with a red sign, Lwando recognizes that she is acting out because she is sad and comforts her with a supportive hug. By Thursday, the monster is furious inside a jagged red starburst amidst a noisy, crowded street. Sensing she simply needs quiet, Lwando takes her to a park bench near a sleeping dog to retreat from the overwhelming environment.

\paragraph{The Situation Escalates on Friday}

On Friday, Lwando's taming methods fail, leaving him watching with concerned, wide eyes. The pages show a grumpy yellow monster holding a bullseye toy alongside a roaring counterpart draped in an orange blanket. When Lwando reaches toward the shouting creature amidst sharp red lightning, his frustration inadvertently creates a second monster. A grumpy Lwando in his square helmet now faces new blue monsters, culminating in a confrontation between a large blue dinosaur-like creature and a smaller alarmed yellow lizard. The text notes that the monsters continue to grow bigger, evolving into massive reptilian beasts dominating a background of swirling flames and dark smoke. The booming volume is emphasized with the text "RAWWWWWWR! GRRRAWWWR!" inside a grey speech bubble. These manifestations escalate into a Godzilla-like beast aggressively attacking the yellow monster against a burning city skyline.

\paragraph{Dad Saves the Day}

As the beasts grow even larger, bright flames threaten to engulf the blue monster, a dog, and a man. This man, identifiable in a green sweater with a serious expression, is revealed to be Dad, the best monster tamer of all. He steps in to mediate between the giant blue monster and the yellow creatures, imparting the core lesson that monsters cannot be tamed by being bigger or angrier. Holding a hot pink mug, Dad explains that calming them requires making them feel safe. A dark-skinned boy points to a cozy orange tent decorated with yellow stars. Inside, Lwando sits safely and calmly near a white puppy resting on a yellow mat. Oyiso is finally seen at peace nearby, eating a cookie from a white plate. A retrospective grid of illustrations summarizes the emotional week, revisiting the monster's prior distresses like crying blue tears, screaming with an exposed red tongue, or roaring. The overarching narrative resolves with a tranquil final image of the yellow monster resting against a comforting pale blue background.

\paragraph{Cited pages.} \mbox{}\\[0.2em]
\noindent\centering
\includegraphics[width=\linewidth, keepaspectratio]{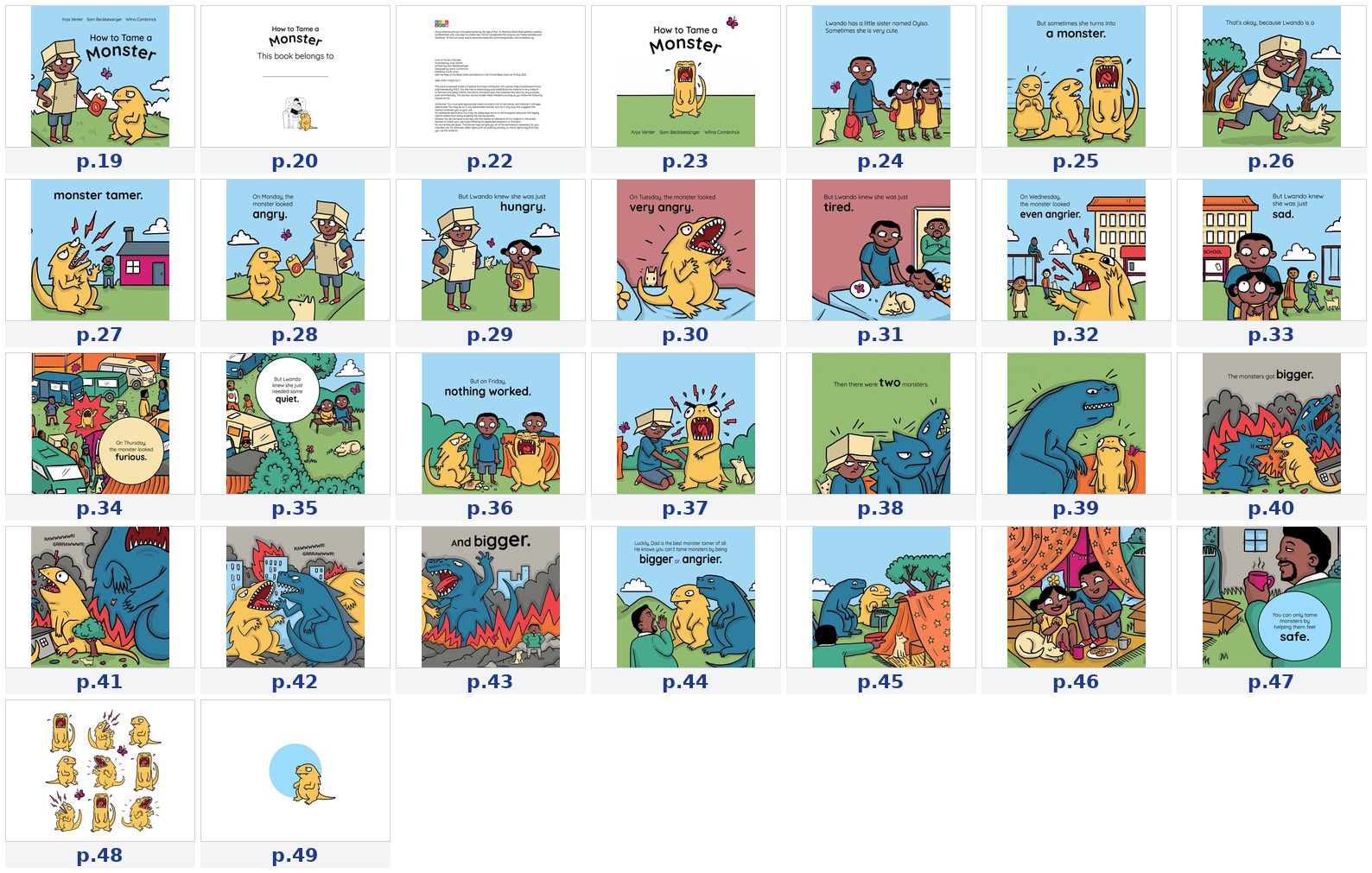}
\par\raggedright

\end{tcolorbox}

\twocolumn

%% file: table/appendix_entity_types.tex
\begin{table*}[t]
\centering
\scriptsize
\renewcommand{\arraystretch}{1.18}
\setlength{\tabcolsep}{5pt}
\begin{tabular}{@{}p{0.16\linewidth}c >{\raggedright\arraybackslash}p{0.76\linewidth}@{}}
\toprule
\textbf{DLVQA corpus} & \textbf{\#} & \textbf{Entity types} \\
\midrule

\textbf{Picture book}
& 12
&
\textit{Characters and story:}
\texttt{character}, \texttt{creature}, \texttt{setting},
\texttt{action\_event}, \texttt{emotion}, \texttt{narrative\_arc},
\texttt{dialogue\_speaker}.\\
&&
\textit{General:}
\texttt{person}, \texttt{location}, \texttt{date},
\texttt{object}, \texttt{scientific\_concept}. \\

\addlinespace[0.3em]

\textbf{World History}
& 13
&
\textit{General:}
\texttt{person}, \texttt{organization}, \texttt{group},
\texttt{location}, \texttt{event}, \texttt{date}.\\
&&
\textit{Content and structure:}
\texttt{work\_of\_art}, \texttt{product},
\texttt{scientific\_concept}, \texttt{chapter},
\texttt{section}, \texttt{unit}, \texttt{assessment\_type}. \\

\addlinespace[0.3em]

\textbf{Env. Reports}
& 14
&
\textit{General:}
\texttt{person}, \texttt{organization}, \texttt{location},
\texttt{event}, \texttt{date}.\\
&&
\textit{Sustainability and business:}
\texttt{kpi}, \texttt{target}, \texttt{commitment},
\texttt{material\_topic}, \texttt{stakeholder\_group},
\texttt{regulatory\_framework}, \texttt{business\_segment},
\texttt{geographic\_region}, \texttt{fiscal\_period}. \\

\addlinespace[0.3em]

\textbf{DLCV Slides}
& 24
&
\textit{General:}
\texttt{person}, \texttt{organization}, \texttt{location},
\texttt{event}, \texttt{date}, \texttt{work\_of\_art},
\texttt{product}, \texttt{scientific\_concept}.\\
&&
\textit{Lecture and ML-specific:}
\texttt{concept}, \texttt{term}, \texttt{slide\_title},
\texttt{agenda\_item}, \texttt{code\_snippet},
\texttt{citation}, \texttt{course\_code}, \texttt{lecture\_topic},
\texttt{dataset}, \texttt{model\_architecture}, \texttt{task\_type},
\texttt{metric}, \texttt{benchmark}, \texttt{framework},
\texttt{algorithm}, \texttt{paper\_citation}. \\

\midrule

\textbf{Visual}
& 6
&
\textit{Used by KG4VD for page-image grounded entities:}
\texttt{visual\_object}, \texttt{chart\_element},
\texttt{diagram\_component}, \texttt{table\_region},
\texttt{layout\_region}, \texttt{figure\_panel}. \\

\bottomrule
\end{tabular}
\caption{
Predefined entity-type schema used on DLVQA.
}
\label{tab:entity_types}
\vspace{-10pt}
\end{table*}

%% file: table/main_dlvqa_subsetwise.tex
\begin{table*}[t]
\centering
\resizebox{\textwidth}{!}{%
\begin{tabular}{llcccccccccccccccc}
\toprule
\multirow{3}{*}{\textbf{Method}}
& \multirow{3}{*}{\textbf{Metric}}
& \multicolumn{16}{c}{\textbf{Document-level VQA (DLVQA)}} \\
\cmidrule(lr){3-18}
&
& \multicolumn{4}{c}{\textbf{DLCV Slides}}
& \multicolumn{4}{c}{\textbf{Picture Books}}
& \multicolumn{4}{c}{\textbf{Environmental Reports}}
& \multicolumn{4}{c}{\textbf{World History Textbook}} \\
\cmidrule(lr){3-6}
\cmidrule(lr){7-10}
\cmidrule(lr){11-14}
\cmidrule(lr){15-18}
&
& \textbf{Ours Win} & \textbf{Ours Loss} & \textbf{Tie} & $\boldsymbol{\Delta}$
& \textbf{Ours Win} & \textbf{Ours Loss} & \textbf{Tie} & $\boldsymbol{\Delta}$
& \textbf{Ours Win} & \textbf{Ours Loss} & \textbf{Tie} & $\boldsymbol{\Delta}$
& \textbf{Ours Win} & \textbf{Ours Loss} & \textbf{Tie} & $\boldsymbol{\Delta}$ \\
\midrule

\multicolumn{18}{c}{\textit{\textbf{Without Retrieval}}} \\
\midrule
No Documents
& \textit{Comp} & 90.53 & 1.15 & 8.32 & 89.41 & 94.34 & 0.10 & 5.57 & 94.27 & 85.31 & 2.60 & 12.10 & 82.74 & 83.62 & 3.07 & 13.33 & 80.58 \\
& \textit{Div} & 94.81 & 0.87 & 4.32 & 93.97 & 98.03 & 0.40 & 1.57 & 97.66 & 90.39 & 1.52 & 8.10 & 88.91 & 88.97 & 1.73 & 9.33 & 87.26 \\
& \textit{Empo} & 89.52 & 1.69 & 8.74 & 87.86 & 93.04 & 0.75 & 6.16 & 92.32 & 84.70 & 2.97 & 12.27 & 81.76 & 83.14 & 3.39 & 13.43 & 79.78 \\

\midrule
\multicolumn{18}{c}{\textit{\textbf{Embedding RAG}}} \\
\midrule
NaiveRAG~\citep{guo2024lightrag}
& \textit{Comp} & 52.70 & 11.12 & 36.21 & 41.59 & 52.46 & 11.19 & 36.39 & 41.28 & 49.51 & 12.00 & 38.51 & 37.52 & 47.33 & 12.61 & 40.09 & 34.73 \\
& \textit{Div} & 66.59 & 6.86 & 26.56 & 59.73 & 66.39 & 6.89 & 26.74 & 59.50 & 63.89 & 7.25 & 28.86 & 56.64 & 62.05 & 7.52 & 30.44 & 54.53 \\
& \textit{Empo} & 56.88 & 8.59 & 34.51 & 48.27 & 56.66 & 8.65 & 34.68 & 47.99 & 53.94 & 9.37 & 36.67 & 44.54 & 51.92 & 9.91 & 38.14 & 42.00 \\

VisRAG~\citep{yu2024visrag}
& \textit{Comp} & 79.87 & 3.68 & 16.48 & 76.19 & 80.13 & 3.61 & 16.29 & 76.52 & 83.49 & 2.68 & 13.86 & 80.81 & 75.31 & 4.95 & 19.77 & 70.40 \\
& \textit{Div} & 90.87 & 1.33 & 7.78 & 89.49 & 91.09 & 1.30 & 7.59 & 89.74 & 93.93 & 0.88 & 5.16 & 93.00 & 87.03 & 1.89 & 11.07 & 85.09 \\
& \textit{Empo} & 83.90 & 2.39 & 13.69 & 81.49 & 84.14 & 2.33 & 13.51 & 81.80 & 87.25 & 1.50 & 11.24 & 85.73 & 79.71 & 3.50 & 16.76 & 76.18 \\

ColQwen~\citep{faysse2024colpali}
& \textit{Comp} & 49.99 & 11.94 & 38.07 & 38.08 & 51.12 & 11.63 & 37.26 & 39.52 & 48.17 & 12.45 & 39.39 & 35.76 & 45.84 & 13.10 & 41.08 & 32.76 \\
& \textit{Div} & 63.56 & 6.03 & 30.40 & 57.56 & 64.51 & 5.89 & 29.59 & 58.65 & 62.02 & 6.25 & 31.72 & 55.79 & 60.03 & 6.55 & 33.41 & 53.52 \\
& \textit{Empo} & 53.32 & 9.68 & 37.03 & 43.62 & 54.36 & 9.40 & 36.27 & 44.93 & 51.64 & 10.13 & 38.26 & 41.49 & 49.48 & 10.71 & 39.84 & 38.76 \\

GME~\citep{zhang2024gme}
& \textit{Comp} & 87.23 & 1.27 & 11.48 & 86.01 & 84.63 & 1.99 & 13.36 & 82.69 & 84.97 & 1.89 & 13.11 & 83.13 & 82.97 & 2.45 & 14.57 & 80.57 \\
& \textit{Div} & 94.63 & 0.15 & 5.23 & 94.49 & 92.43 & 0.47 & 7.11 & 91.97 & 92.72 & 0.43 & 6.86 & 92.30 & 91.02 & 0.67 & 8.32 & 90.36 \\
& \textit{Empo} & 88.55 & 1.62 & 9.84 & 86.94 & 86.15 & 2.26 & 11.60 & 83.90 & 86.47 & 2.17 & 11.37 & 84.30 & 84.63 & 2.67 & 12.71 & 81.98 \\

\midrule
\multicolumn{18}{c}{\textit{\textbf{Text-Only Graph-based RAG}}} \\
\midrule
GraphRAG~\citep{edge2024local}
& \textit{Comp} & 43.25 & 15.84 & 40.92 & 27.43 & 54.39 & 12.75 & 32.87 & 41.65 & 40.91 & 16.48 & 42.61 & 24.44 & 37.45 & 17.45 & 45.12 & 20.00 \\
& \textit{Div} & 50.27 & 18.17 & 31.59 & 32.08 & 59.69 & 16.80 & 23.54 & 42.87 & 48.29 & 18.46 & 33.28 & 29.81 & 45.35 & 18.89 & 35.79 & 26.44 \\
& \textit{Empo} & 45.61 & 15.38 & 38.98 & 30.21 & 55.89 & 12.64 & 31.45 & 43.23 & 43.45 & 15.96 & 40.57 & 27.47 & 40.25 & 16.82 & 42.92 & 23.41 \\

LightRAG~\citep{guo2024lightrag}
& \textit{Comp} & 69.08 & 3.35 & 27.58 & 65.73 & 70.89 & 2.84 & 26.27 & 68.05 & 61.36 & 5.48 & 33.16 & 55.88 & 56.87 & 6.73 & 36.39 & 50.14 \\
& \textit{Div} & 82.03 & 2.07 & 15.93 & 79.97 & 83.57 & 1.85 & 14.62 & 81.73 & 75.50 & 3.02 & 21.51 & 72.49 & 71.70 & 3.58 & 24.74 & 68.13 \\
& \textit{Empo} & 72.58 & 2.87 & 24.57 & 69.72 & 74.25 & 2.42 & 23.34 & 71.84 & 65.46 & 4.76 & 29.79 & 60.70 & 61.31 & 5.87 & 32.82 & 55.46 \\

HippoRAG2~\citep{gutierrez2025rag}
& \textit{Comp} & 87.36 & 1.90 & 10.78 & 85.44 & 83.22 & 3.05 & 13.78 & 80.15 & 82.79 & 3.16 & 14.09 & 79.60 & 77.55 & 4.61 & 17.87 & 72.93 \\
& \textit{Div} & 95.56 & 0.68 & 3.75 & 94.91 & 92.06 & 1.19 & 6.75 & 90.90 & 91.70 & 1.24 & 7.06 & 90.49 & 87.28 & 1.89 & 10.84 & 85.42 \\
& \textit{Empo} & 89.47 & 0.41 & 10.12 & 89.11 & 85.65 & 1.43 & 12.92 & 84.27 & 85.25 & 1.54 & 13.21 & 83.77 & 80.43 & 2.82 & 16.75 & 77.65 \\

\midrule
\multicolumn{18}{c}{\textit{\textbf{Multimodal Graph-based RAG}}} \\
\midrule
RAGAnything~\citep{guo2025rag}
& \textit{Comp} & 60.34 & 12.80 & 26.86 & 47.54 & 61.56 & 12.46 & 25.98 & 49.09 & 58.81 & 13.23 & 27.97 & 45.58 & 54.89 & 14.31 & 30.79 & 40.59 \\
& \textit{Div} & 70.07 & 7.77 & 22.16 & 62.30 & 71.10 & 7.62 & 21.28 & 63.47 & 68.77 & 7.96 & 23.27 & 60.81 & 65.46 & 8.45 & 26.09 & 57.02 \\
& \textit{Empo} & 64.73 & 9.95 & 25.33 & 54.78 & 65.85 & 9.65 & 24.50 & 56.20 & 63.31 & 10.32 & 26.36 & 52.99 & 59.71 & 11.28 & 29.01 & 48.43 \\

MegaRAG~\citep{hsiao2025megarag}
& \textit{Comp} & 53.43 & 15.77 & 30.80 & 37.66 & 53.21 & 15.83 & 30.96 & 37.37 & 52.00 & 16.17 & 31.84 & 35.83 & 51.76 & 16.23 & 32.00 & 35.54 \\
& \textit{Div} & 64.80 & 9.75 & 25.45 & 55.05 & 64.61 & 9.78 & 25.61 & 54.84 & 63.59 & 9.92 & 26.49 & 53.67 & 63.40 & 9.95 & 26.65 & 53.44 \\
& \textit{Empo} & 59.51 & 12.35 & 28.14 & 47.17 & 59.31 & 12.40 & 28.29 & 46.91 & 58.19 & 12.70 & 29.11 & 45.50 & 57.99 & 12.75 & 29.26 & 45.22 \\

\midrule
Avg.
& \textit{Comp} & 67.38 & 7.88 & 24.76 & 59.50 & 68.60 & 7.55 & 23.88 & 61.05 & 64.73 & 8.61 & 26.67 & 56.12 & 61.37 & 9.56 & 29.09 & 51.81 \\
& \textit{Div} & 77.31 & 5.37 & 17.32 & 71.96 & 78.34 & 5.22 & 16.44 & 73.13 & 75.07 & 5.69 & 19.23 & 69.39 & 72.24 & 6.12 & 21.65 & 66.12 \\
& \textit{Empo} & 70.41 & 6.49 & 23.10 & 63.92 & 71.53 & 6.19 & 22.27 & 65.34 & 67.97 & 7.14 & 24.88 & 60.82 & 64.85 & 7.98 & 27.15 & 56.88 \\

\bottomrule
\end{tabular}%
}
\caption{Subset-level reference-free pairwise KG4VD win-rate evaluation on DLVQA. \textit{Comp}, \textit{Div}, and \textit{Empo} denote comprehensiveness, diversity, and empowerment, respectively. $\Delta$ denotes \textit{Ours Win} minus \textit{Ours Loss}.} 
\label{tab:dlvqa_subset_reference_free_winrate}
\end{table*}

%% file: table/appendix_dlvqa_domainwise.tex
\begin{table*}[t]
\centering
\resizebox{\textwidth}{!}{%
\begin{tabular}{lcccccccccccc}
\toprule
\multirow{3}{*}{\textbf{Method}} 
& \multicolumn{12}{c}{\textbf{Document-level VQA (DLVQA)}} \\
\cmidrule(lr){2-13}
& \multicolumn{3}{c}{\textbf{DLCV Slides}}
& \multicolumn{3}{c}{\textbf{Picture Books}}
& \multicolumn{3}{c}{\textbf{Environmental Reports}}
& \multicolumn{3}{c}{\textbf{World History Textbook}} \\
\cmidrule(lr){2-4}
\cmidrule(lr){5-7}
\cmidrule(lr){8-10}
\cmidrule(lr){11-13}
& \textbf{Prec.} & \textbf{Rec.} & \textbf{F1}
& \textbf{Prec.} & \textbf{Rec.} & \textbf{F1}
& \textbf{Prec.} & \textbf{Rec.} & \textbf{F1}
& \textbf{Prec.} & \textbf{Rec.} & \textbf{F1} \\
\midrule

\multicolumn{13}{c}{\textit{\textbf{Without Retrieval}}} \\
\midrule
No Documents
& 54.46 & 23.86 & 33.18
& 25.06 & 22.06 & 23.46
& 78.76 & 15.76 & 26.26
& 63.64 & 15.64 & 25.11 \\

\midrule
\multicolumn{13}{c}{\textit{\textbf{Embedding RAG}}} \\
\midrule
NaiveRAG~\citep{guo2024lightrag}
& 72.26 & 30.25 & 42.65
& 67.86 & 21.55 & 32.71
& 90.46 & 20.65 & 33.62
& 76.46 & 19.95 & 31.64 \\

VisRAG~\citep{yu2024visrag}
& 75.80 & 37.80 & 50.44
& \textbf{79.56} & 23.08 & 35.78
& 79.56 & 23.08 & 35.78
& 78.60 & 27.60 & 40.85 \\

ColQwen~\citep{faysse2024colpali}
& \underline{81.59} & 34.70 & 48.69
& 67.39 & 28.10 & 39.66
& 97.29 & 25.10 & 39.90
& \textbf{84.77} & 24.30 & 37.77 \\

GME~\citep{zhang2024gme}
& 80.26 & 33.76 & 47.53
& \underline{73.96} & 33.86 & 46.45
& \textbf{98.36} & 22.86 & 37.10
& 83.14 & 22.84 & 35.84 \\

\midrule
\multicolumn{13}{c}{\textit{\textbf{Text-Only Graph-based RAG}}} \\
\midrule
GraphRAG~\citep{edge2024local}
& 75.56 & 47.11 & 58.04
& 40.66 & 28.21 & 33.31
& 96.16 & 34.11 & 50.36
& 79.34 & 35.49 & 49.04 \\

LightRAG~\citep{guo2024lightrag}
& 66.14 & 39.69 & 49.61
& 52.14 & 33.39 & 40.71
& 93.14 & \underline{37.79} & \underline{53.77}
& \underline{84.34} & \textbf{39.57} & \textbf{53.87} \\

HippoRAG2~\citep{gutierrez2025rag}
& 67.63 & 41.32 & 51.30
& 67.53 & 41.12 & 51.12
& 92.03 & 32.32 & 47.84
& 79.21 & 36.60 & 50.07 \\

\midrule
\multicolumn{13}{c}{\textit{\textbf{Multimodal Graph-based RAG}}} \\
\midrule
RAGAnything~\citep{guo2025rag}
& 73.72 & 46.98 & 57.39
& 55.12 & 41.58 & 47.40
& 85.42 & 35.18 & 49.84
& 76.30 & \underline{37.78} & \underline{50.54} \\

MegaRAG~\citep{hsiao2025megarag}
& 74.70 & \underline{50.61} & \underline{60.34}
& 57.80 & \textbf{47.11} & \underline{51.91}
& 92.50 & 34.71 & 50.48
& 71.80 & 32.21 & 44.47 \\

KG4VD (ours)
& \textbf{83.43} & \textbf{51.98} & \textbf{64.05}
& 68.13 & \underline{44.48} & \textbf{53.82}
& \underline{98.13} & \textbf{38.38} & \textbf{55.18}
& 80.03 & 34.96 & 48.66 \\

\bottomrule
\end{tabular}%
}
\caption{Subset-level DLVQA results under reference-based evaluation. Precision measures the proportion of generated statements supported by the reference evidence, recall measures the coverage of reference information, and F1 is their harmonic mean.}
\label{tab:dlvqa_subset_finesure_results}
\vspace{-10pt}
\end{table*}

%% file: table/appendix_judge_example.tex
\begin{table*}[t]
\centering\small
\renewcommand{\arraystretch}{1.25}
\begin{tabularx}{\textwidth}{p{0.085\textwidth} X}
\toprule
\multicolumn{2}{l}{\textbf{Question}} \\
\midrule
\multicolumn{2}{p{0.98\textwidth}}{
\textit{``Based on the CS231n lectures, what are the provided examples
of supervised learning, and how do the slides explicitly formulate the
input data and goals for unsupervised and self-supervised learning?
Be sure to include the specific models and representation evaluation
metrics discussed for self-supervised learning.''}
} \\
\midrule
\multicolumn{2}{l}{\textbf{Candidate answers}} \\
\midrule
\textsc{KG4VD} 
& ``Supervised learning utilizes labeled data $(x,y)$; examples include classification, regression, object detection, semantic segmentation, and image captioning. Unsupervised learning operates on unlabeled data $(x)$ to uncover hidden structures \textbf{[$\ldots$]}. Self-supervised learning generates pseudo labels from the data, $(x, \text{pseudo generated }y)$, and the framework often involves embedding networks that compare outputs from input images using a distance function.'' \\

\textsc{MegaRAG} 
& ``\textbf{[$\ldots$]} Specific models discussed for self-supervised learning include autoencoders, Contrastive Predictive Coding (CPC), and SimCLR \textbf{[$\ldots$]}. Evaluation metrics often involve clustering performance and visualization techniques like $t$-SNE, as well as the $L_2$ loss between original input and reconstructed data.'' \\

\textsc{RAG-Anything} 
& ``\textbf{[$\ldots$]} pretext tasks such as predicting image transformations, inpainting, rearrangement, and colorization are employed \textbf{[$\ldots$]}. Self-supervised methods such as SimCLR and MoCo utilize contrastive representation learning, often measured by metrics like mean Average Precision (\%mAP) and mean Intersection over Union (\%mIoU).'' \\

\midrule
\multicolumn{2}{l}{\textbf{Win-rate verdicts}} \\
\midrule

\multicolumn{2}{l}{\textsc{KG4VD} (Answer 1) vs.\ \textsc{MegaRAG} (Answer 2)}  \\
\addlinespace[-1pt]
\quad gpt
& \emph{Comp.} ``Covers more lecture specifics, including self-supervised models and evaluation methods; Answer 2 is more generic and incomplete.''
\newline\emph{Div.} ``Provides broader examples, learning paradigms, pretext tasks, and evaluation approaches.''
\newline\emph{Emp.} ``Gives concrete models and metrics, helping readers judge methods better.'' \\

\quad gemini
& \emph{Comp.} ``Answer 1 explicitly names specific self-supervised models and evaluation metrics, whereas Answer 2 omits these required details entirely.''
\newline\emph{Div.} ``Provides a broader range of technical insights, including specific pretext tasks and visualization techniques like $t$-SNE.''
\newline\emph{Emp.} ``Offers concrete examples and methodologies, enabling the reader to better understand the practical implementation of these learning paradigms.'' \\

\midrule
\multicolumn{2}{l}{\textsc{KG4VD} (Answer 1) vs.\ \textsc{RAG-Anything} (Answer 2)} \\
\addlinespace[-1pt]
\quad gpt
& \emph{Comp.} ``Covers more requested specifics: supervised examples, unsupervised/self-supervised goals, pretext tasks, and mentions SimCLR/MoCo plus metrics.''
\newline\emph{Div.} ``Includes a wider range of self-supervised methods and evaluation angles, while Answer 2 stays generic and narrow.''
\newline\emph{Emp.} ``Gives more concrete models, tasks, and metrics, helping readers judge and connect the lecture content better.'' \\

\quad gemini
& \emph{Comp.} ``Answer 1 explicitly addresses the specific models (SimCLR, MoCo) and metrics requested, whereas Answer 2 omits these details entirely.''
\newline\emph{Div.} ``Provides a broader range of technical examples, including specific pretext tasks and evaluation metrics, offering more depth.''
\newline\emph{Emp.} ``Provides actionable technical details and specific terminology, better equipping the reader to understand the actual implementation.'' \\

\midrule
\multicolumn{2}{p{0.98\textwidth}}{
\textbf{Outcome:} \textsc{KG4VD} wins all aggregated criterion-judge verdicts against both baselines, i.e., \(6/6\) against \textsc{MegaRAG} and \(6/6\) against \textsc{RAG-Anything}.
} \\
\bottomrule
\end{tabularx}
\caption{Pairwise win-rate case study on \texttt{DLCV\_q075}. 
Both judges prefer \textsc{KG4VD} over both baselines across Comprehensiveness, Diversity, and Empowerment. 
Candidate answers are excerpted with \textbf{[$\ldots$]}; judge explanations are quoted verbatim.}
\label{tab:judge_case_winrate}
\end{table*}

\begin{table*}[t]
\centering\small
\renewcommand{\arraystretch}{1.25}
\begin{tabularx}{\textwidth}{p{0.085\textwidth} c c X}
\toprule
\multicolumn{4}{l}{\textbf{Question}} \\
\midrule
\multicolumn{4}{p{0.98\textwidth}}{
\textit{``In the `CS231n: Deep Learning for Computer Vision' lectures, how is the Softmax classifier defined,
placed among other methods, and compared with SVM?''}
} \\
\midrule
\textbf{Method} & \textbf{faith\textsubscript{gpt}} & \textbf{faith\textsubscript{gem}} & \textbf{Candidate answer and per-sentence judge labels} \\
\midrule

\textsc{KG4VD}
& \textbf{1.00} & \textbf{1.00}
& 
\begin{minipage}[t]{\linewidth}
\textbf{\emph{Answer.}}
``The Softmax classifier interprets raw classifier scores as probabilities, converting them into a distribution that sums to one. For scores [cat~3.2, car~5.1, frog~$-1.7$], Softmax yields $[0.13, 0.87, 0.00]$. In contrast, SVM maximises the margin between classes with a loss function that emphasises the distance between the correct class and others.''

\smallskip
\noindent\rule{\linewidth}{0.3pt}
\smallskip

\textbf{\emph{Judge labels.}}
4/4 \texttt{no error} from both judges; representative reason: ``Aligns with the transcript's definition of multinomial logistic regression and matches the worked example on the cited page.''
\end{minipage}
\\

\midrule

\textsc{MegaRAG}
& 0.71 & 0.86
& 
\begin{minipage}[t]{\linewidth}
\textbf{\emph{Answer.}}
``\textbf{[$\ldots$]} The associated loss function is the cross-entropy loss, which maximises the probability of the correct class. \textbf{[$\ldots$]} The SVM loss penalises misclassifications based on the margin \textbf{[$\ldots$]}. This distinction influences the choice of method based on the classification problem's requirements, with Softmax being effective for scenarios needing class probabilities and SVM for \textit{robust decision boundaries}.''

\smallskip
\noindent\rule{\linewidth}{0.3pt}
\smallskip

\textbf{\emph{Judge labels.}}
5/7 \texttt{no error} from \textsc{gpt-5.4-mini} and 6/7 from \textsc{gemini-3.1-flash-lite}; flagged sentence: ``\textit{robust decision boundaries}.'' 
\textsc{gpt-5.4-mini} marks it as an \texttt{out-of-context error}, explaining that ``the transcript does not characterise SVM as providing robust decision boundaries,'' while \textsc{gemini-3.1-flash-lite} accepts it as a paraphrase.
\end{minipage}
\\

\midrule

\textsc{RAG-Anything}
& \textbf{0.00} & \textbf{0.25}
& 
\begin{minipage}[t]{\linewidth}
\textbf{\emph{Answer.}}
``\textbf{[$\ldots$]} It is a crucial method placed alongside others like Linear Classifiers and Support Vector Machines (SVM). While the Softmax classifier focuses on generating probability distributions, SVMs utilise a \textit{margin-based approach, emphasising the separation of classes in a more geometric context}. This comparison highlights the Softmax's probabilistic framework against the margin-driven objective of SVMs, showcasing their different methodologies.''

\smallskip
\noindent\rule{\linewidth}{0.3pt}
\smallskip

\textbf{\emph{Judge labels.}}
0/4 \texttt{no error} from \textsc{gpt-5.4-mini} and 1/4 from \textsc{gemini-3.1-flash-lite}; flagged categories include \texttt{out-of-context error} for ``geometric context'' and ``margin-driven objective,'' which do not appear in the cited transcript, and \texttt{entity error} for conflating ``Linear Classifiers'' with a different lecture.
\end{minipage}
\\

\midrule
\multicolumn{4}{p{0.98\textwidth}}{
\textbf{Outcome:} \textsc{KG4VD} obtains faithfulness \(=1.00\) from both judges and is the only method with zero flagged errors. \textsc{RAG-Anything} fails on every sentence under \textsc{gpt-5.4-mini} and on 3/4 sentences under \textsc{gemini-3.1-flash-lite}.
} \\
\bottomrule
\end{tabularx}
\caption{Faithfulness case study on \texttt{DLCV\_q036}. 
\textsc{KG4VD} is the only answer judged fully faithful by both judges. 
Candidate answers are excerpted with \textbf{[$\ldots$]}; labels show the judge category and brief reason for flagged sentences.}
\label{tab:judge_case_finesure}
\end{table*}

%% file: main.bib
@String(CVPR= {IEEE Conf. Comput. Vis. Pattern Recog.})

@String(ICCV= {Int. Conf. Comput. Vis.})

@String(ECCV= {Eur. Conf. Comput. Vis.})

@String(ICLR = {Int. Conf. Learn. Represent.})

@String(IJCAI = {IJCAI})

@String(AAAI = {AAAI})

@String(CVPR  = {CVPR})

@String(ICCV  = {ICCV})

@String(ECCV  = {ECCV})

@String(ICLR  = {ICLR})

@techreport{openai2025gpt5systemcard,
  title        = {GPT-5 System Card},
  author       = {{OpenAI}},
  institution  = {OpenAI},
  year         = {2025},
}

@article{beyer2024paligemma,
  title={Paligemma: A versatile 3b vlm for transfer},
  author={Beyer, Lucas and Steiner, Andreas and Pinto, Andr{\'e} Susano and Kolesnikov, Alexander and Wang, Xiao and Salz, Daniel and Neumann, Maxim and Alabdulmohsin, Ibrahim and Tschannen, Michael and Bugliarello, Emanuele and others},
  journal={arXiv preprint arXiv:2407.07726},
  year={2024}
}

@inproceedings{ma2024unifying,
  title={Unifying multimodal retrieval via document screenshot embedding},
  author={Ma, Xueguang and Lin, Sheng-Chieh and Li, Minghan and Chen, Wenhu and Lin, Jimmy},
  publisher={ In \textit{EMNLP}},
  year={2024}
}

@inproceedings{faysse2024colpali,
  title={Colpali: Efficient document retrieval with vision language models},
  author={Faysse, Manuel and Sibille, Hugues and Wu, Tony and Omrani, Bilel and Viaud, Gautier and Hudelot, C{\'e}line and Colombo, Pierre},
  publisher={ In \textit{ICLR}},
  year={2025}
}

@inproceedings{yu2024visrag,
  title={Visrag: Vision-based retrieval-augmented generation on multi-modality documents},
  author={Yu, Shi and Tang, Chaoyue and Xu, Bokai and Cui, Junbo and Ran, Junhao and Yan, Yukun and Liu, Zhenghao and Wang, Shuo and Han, Xu and Liu, Zhiyuan and others},
  publisher={ In \textit{ICLR}},
  year={2025}
}

@inproceedings{yang2018hotpotqa,
  title={HotpotQA: A Dataset for Diverse, Explainable Multi-hop Question Answering},
  author={Yang, Zhilin and Qi, Peng and Zhang, Saizheng and Bengio, Yoshua and Cohen, William and Salakhutdinov, Ruslan and Manning, Christopher D},
  publisher={ In \textit{EMNLP}},
  year={2018}
}

@article{trivedi2022musique,
  title   = {{MuSiQue}: Multihop Questions via Single-hop Question Composition},
  author  = {Trivedi, Harsh and Balasubramanian, Niranjan and Khot, Tushar and Sabharwal, Ashish},
  journal = {Transactions of the Association for Computational Linguistics},
  year    = {2022}
}

@inproceedings{
cheng2024multihop,
title={Multi-hop Question Answering under Temporal Knowledge Editing},
author={Keyuan Cheng and Gang Lin and Haoyang Fei and Yuxuan Zhai and Lu Yu and Muhammad Asif Ali and Lijie Hu and Di Wang},
publisher={ In \textit{COLM}},
year={2024}
}

@inproceedings{tanaka2023slidevqa,
  title={Slidevqa: A dataset for document visual question answering on multiple images},
  author={Tanaka, Ryota and Nishida, Kyosuke and Nishida, Kosuke and Hasegawa, Taku and Saito, Itsumi and Saito, Kuniko},
  publisher={ In \textit{AAAI}},
  year={2023}
}

@inproceedings{gao-etal-2022-graph,
    title = "Graph-augmented Learning to Rank for Querying Large-scale Knowledge Graph",
    author = "Gao, Hanning  and
      Wu, Lingfei  and
      Hu, Po  and
      Wei, Zhihua  and
      Xu, Fangli  and
      Long, Bo",
    year = "2022",
    publisher = " In \textit{AACL}"
}

@inproceedings{
li2025simple,
title={Simple is Effective: The Roles of Graphs and Large Language Models In Knowledge-Graph-Based Retrieval-Augmented Generation},
author={Mufei Li and Siqi Miao and Pan Li},
publisher={ In \textit{ICLR}},
year={2025}
}

@inproceedings{he2024g,
  title={G-retriever: Retrieval-augmented generation for textual graph understanding and question answering},
  author={He, Xiaoxin and Tian, Yijun and Sun, Yifei and Chawla, Nitesh and Laurent, Thomas and LeCun, Yann and Bresson, Xavier and Hooi, Bryan},
  publisher={ In \textit{NeurIPS}},
  year={2024}
}

@inproceedings{10.1007/978-3-030-21348-0_30,
author = {Liu, Ye and Li, Hui and Garcia-Duran, Alberto and Niepert, Mathias and Onoro-Rubio, Daniel and Rosenblum, David S.},
title = {MMKG: Multi-modal Knowledge Graphs},
year = {2019},
publisher = { In \textit{ESWC}},


}

@inproceedings{zhangmultimodal,
  title={Multimodal Analogical Reasoning over Knowledge Graphs},
  author={Zhang, Ningyu and Li, Lei and Chen, Xiang and Liang, Xiaozhuan and Deng, Shumin and Chen, Huajun},
  publisher={ In \textit{ICLR}},
  year={2023}
}

@article{zhao2023boosting,
  title   = {Boosting Entity-Aware Image Captioning with Multi-Modal Knowledge Graph},
  author  = {Zhao, Wentian and Wu, Xinxiao},
  journal = {IEEE Transactions on Multimedia},
  year    = {2023}
}

@inproceedings{xie2017image,
  title={Image-embodied knowledge representation learning},
  author={Xie, Ruobing and Liu, Zhiyuan and Luan, Huanbo and Sun, Maosong},
  publisher={ In \textit{IJCAI}},
  year={2017}
}

@inproceedings{sun2020multi,
  title={Multi-modal knowledge graphs for recommender systems},
  author={Sun, Rui and Cao, Xuezhi and Zhao, Yan and Wan, Junchen and Zhou, Kun and Zhang, Fuzheng and Wang, Zhongyuan and Zheng, Kai},
  publisher={ In \textit{CIKM}},
  year={2020}
}

@inproceedings{lee-etal-2024-multimodal,
    title = "Multimodal Reasoning with Multimodal Knowledge Graph",
    author = "Lee, Junlin  and
      Wang, Yequan  and
      Li, Jing  and
      Zhang, Min",
    editor = "Ku, Lun-Wei  and
      Martins, Andre  and
      Srikumar, Vivek",
    year = "2024",
    publisher = " In \textit{ACL}"
}

@article{edge2024local,
  title={From local to global: A graph rag approach to query-focused summarization},
  author={Edge, Darren and Trinh, Ha and Cheng, Newman and Bradley, Joshua and Chao, Alex and Mody, Apurva and Truitt, Steven and Larson, Jonathan},
  journal={arXiv preprint arXiv:2404.16130},
  year={2024}
}

@inproceedings{zhang2024gme,
      title={GME: Improving Universal Multimodal Retrieval by Multimodal LLMs}, 
      author={Zhang, Xin and Zhang, Yanzhao and Xie, Wen and Li, Mingxin and Dai, Ziqi and Long, Dingkun and Xie, Pengjun and Zhang, Meishan and Li, Wenjie and Zhang, Min},
      publisher={ In \textit{CVPR}},
      year={2025},
}

@inproceedings{guo2024lightrag,
  title={LightRAG: Simple and fast retrieval-augmented generation},
  author={Guo, Zirui and Xia, Lianghao and Yu, Yanhua and Ao, Tu and Huang, Chao},
  publisher={ In \textit{Findings of EMNLP}},
  year={2025}
}

@article{wang2024mineru,
  title={Mineru: An open-source solution for precise document content extraction},
  author={Wang, Bin and Xu, Chao and Zhao, Xiaomeng and Ouyang, Linke and Wu, Fan and Zhao, Zhiyuan and Xu, Rui and Liu, Kaiwen and Qu, Yuan and Shang, Fukai and others},
  journal={arXiv preprint arXiv:2409.18839},
  year={2024}
}

@inproceedings{gutierrez2025rag,
  title={From rag to memory: Non-parametric continual learning for large language models},
  author={Guti{\'e}rrez, Bernal Jim{\'e}nez and Shu, Yiheng and Qi, Weijian and Zhou, Sizhe and Su, Yu},
  publisher={ In \textit{ICML}},
  year={2025}
}

@inproceedings{liu2025aligning,
  title={Aligning vision to language: Text-free multimodal knowledge graph construction for enhanced llms reasoning},
  author={Liu, Junming and Meng, Siyuan and Gao, Yanting and Mao, Song and Cai, Pinlong and Yan, Guohang and Chen, Yirong and Bian, Zilin and Shi, Botian and Wang, Ding},
  publisher={ In \textit{ICCV}},
  year={2025}
}

@inproceedings{bu2025query,
  title={Query-Driven Multimodal GraphRAG: Dynamic Local Knowledge Graph Construction for Online Reasoning},
  author={Bu, Chenyang and Chang, Guojie and Chen, Zihao and Dang, CunYuan and Wu, Zhize and He, Yi and Wu, Xindong},
  publisher={In \textit{Findings of ACL}},
  year={2025}
}

@inproceedings{
ma2025thinkongraph,
title={Think-on-Graph 2.0: Deep and Faithful Large Language Model Reasoning with Knowledge-guided Retrieval Augmented Generation},
author={Shengjie Ma and Chengjin Xu and Xuhui Jiang and Muzhi Li and Huaren Qu and Cehao Yang and Jiaxin Mao and Jian Guo},
publisher={In \textit{ICLR}},
year={2025}
}

@article{comanici2025gemini,
  title={Gemini 2.5: Pushing the frontier with advanced reasoning, multimodality, long context, and next generation agentic capabilities},
  author={Comanici, Gheorghe and Bieber, Eric and Schaekermann, Mike and Pasupat, Ice and Sachdeva, Noveen and Dhillon, Inderjit and Blistein, Marcel and Ram, Ori and Zhang, Dan and Rosen, Evan and others},
  journal={arXiv preprint arXiv:2507.06261},
  year={2025}
}

@article{bai2025qwen2,
  title={Qwen2.5-vl technical report},
  author={Bai, Shuai and Chen, Keqin and Liu, Xuejing and Wang, Jialin and Ge, Wenbin and Song, Sibo and Dang, Kai and Wang, Peng and Wang, Shijie and Tang, Jun and others},
  journal={arXiv preprint arXiv:2502.13923},
  year={2025}
}

@misc{meta2025llama4,
  title        = {Introducing LLaMA 4: Advancing Multimodal Intelligence},
  author       = {Meta AI},
  year         = {2025},
}

@article{team2025gemma,
  title={Gemma 3 technical report},
  author={Team, Gemma and Kamath, Aishwarya and Ferret, Johan and Pathak, Shreya and Vieillard, Nino and Merhej, Ramona and Perrin, Sarah and Matejovicova, Tatiana and Ram{\'e}, Alexandre and Rivi{\`e}re, Morgane and others},
  journal={arXiv preprint arXiv:2503.19786},
  year={2025}
}

@article{kuang2025natural,
  title={Natural language understanding and inference with mllm in visual question answering: A survey},
  author={Kuang, Jiayi and Shen, Ying and Xie, Jingyou and Luo, Haohao and Xu, Zhe and Li, Ronghao and Li, Yinghui and Cheng, Xianfeng and Lin, Xika and Han, Yu},
  journal={ACM Computing Surveys},
  volume={57},
  number={8},
  pages={1--36},
  year={2025},
  publisher={ACM New York, NY}
}

@article{mcinnes2017hdbscan,
  title={hdbscan: Hierarchical density based clustering.},
  author={McInnes, Leland and Healy, John and Astels, Steve and others},
  journal={J. Open Source Softw.},
  year={2017}
}

@inproceedings{wang2025taming,
 title = {Taming the Untamed: Graph-Based Knowledge Retrieval and Reasoning for MLLMs to Conquer the Unknown},
 author = {Wang, Bowen and Jiang, Zhouqiang and Susumu, Yasuaki and Miwa, Shotaro and Chen, Tianwei and Nakashima, Yuta},
 publisher = {In \textit{ICCV}},
 year = {2025}
}

@inproceedings{wei2023uniir,
  title={Uniir: Training and benchmarking universal multimodal information retrievers},
  author={Wei, Cong and Chen, Yang and Chen, Haonan and Hu, Hexiang and Zhang, Ge and Fu, Jie and Ritter, Alan and Chen, Wenhu},
  publisher={In \textit{ECCV}},
  year={2024}
}

@inproceedings{10.5555/3737916.3740957,
author = {Ma, Yubo and Zang, Yuhang and Chen, Liangyu and Chen, Meiqi and Jiao, Yizhu and Li, Xinze and Lu, Xinyuan and Liu, Ziyu and Ma, Yan and Dong, Xiaoyi and Zhang, Pan and Pan, Liangming and Jiang, Yu-Gang and Wang, Jiaqi and Cao, Yixin and Sun, Aixin},
title = {MMLONGBENCH-DOC: benchmarking long-context document understanding with visualizations},
year = {2024},
publisher = {In \textit{NeurIPS}},
}

@inproceedings{
madaan2023selfrefine,
title={Self-Refine: Iterative Refinement with Self-Feedback},
author={Aman Madaan and Niket Tandon and Prakhar Gupta and Skyler Hallinan and Luyu Gao and Sarah Wiegreffe and Uri Alon and Nouha Dziri and Shrimai Prabhumoye and Yiming Yang and Shashank Gupta and Bodhisattwa Prasad Majumder and Katherine Hermann and Sean Welleck and Amir Yazdanbakhsh and Peter Clark},
booktitle={NeurIPS},
year={2023}
}

@article{renze2024self,
  title={Self-reflection in llm agents: Effects on problem-solving performance},
  author={Renze, Matthew and Guven, Erhan},
  journal={arXiv:2405.06682},
  year={2024}
}

@article{guo2025rag,
  title={Rag-anything: All-in-one rag framework},
  author={Guo, Zirui and Ren, Xubin and Xu, Lingrui and Zhang, Jiahao and Huang, Chao},
  journal={arXiv preprint arXiv:2510.12323},
  year={2025}
}

@inproceedings{hsiao2025megarag,
  title={MegaRAG: Multimodal Knowledge Graph-Based Retrieval Augmented Generation},
  author={Hsiao, Chi-Hsiang and Wang, Yi-Cheng and Lin, Tzung-Sheng and Yeh, Yi-Ren and Chen, Chu-Song},
  booktitle={ACL},
  year={2026}
}

@inproceedings{song-etal-2024-finesure,
    title = "{F}ine{S}ur{E}: Fine-grained Summarization Evaluation using {LLM}s",
    author = "Song, Hwanjun  and
      Su, Hang  and
      Shalyminov, Igor  and
      Cai, Jason  and
      Mansour, Saab",
    booktitle = "ACL",
    year = "2024",
}

@inproceedings{shao-etal-2024-assisting,
    title = "Assisting in Writing {W}ikipedia-like Articles From Scratch with Large Language Models",
    author = "Shao, Yijia  and
      Jiang, Yucheng  and
      Kanell, Theodore  and
      Xu, Peter  and
      Khattab, Omar  and
      Lam, Monica",
    booktitle = "NAACL",
    year = "2024",
}

@inproceedings{10.1145/775152.775191,
author = {Jeh, Glen and Widom, Jennifer},
title = {Scaling personalized web search},
year = {2003},
publisher = {ACM},
booktitle = {WWW},
}

@article{yang2023set,
  title={Set-of-mark prompting unleashes extraordinary visual grounding in gpt-4v},
  author={Yang, Jianwei and Zhang, Hao and Li, Feng and Zou, Xueyan and Li, Chunyuan and Gao, Jianfeng},
  journal={arXiv preprint arXiv:2310.11441},
  year={2023}
}

@inproceedings{fang2026cogniverse,
  title={Cogniverse: Revolutionizing multi-modal retrieval-augmented generation with cognitive reflection and geometric reasoning},
  author={Fang, Xiang and Fang, Wanlong and Wang, Changshuo},
  booktitle={CVPR},
  year={2026}
}

@InProceedings{Park_2026_CVPR,
    author    = {Park, Hyeongcheol and Seo, Jiyoung and Mun, Jaewon and Park, Hogun and Byeon, Wonmin and Kim, Sung June and Im, Hyeonsoo and Lee, JeungSub and Kim, Sangpil},
    title     = {M{\textasciicircum}3KG-RAG: Multi-hop Multimodal Knowledge Graph-enhanced Retrieval-Augmented Generation},
    booktitle = {CVPR},
    year      = {2026},
}

@inproceedings{10.1145/3770854.3783944,
    author = {Ren, Xubin and Xu, Lingrui and Xia, Long and Wang, Shuaiqiang and Yin, Dawei and Huang, Chao},
    title = {VideoRAG: Retrieval-Augmented Generation with Extreme Long-Context Videos},
    year = {2026},
    booktitle = {KDD},
}

@inproceedings{van2023document,
  title={Document understanding dataset and evaluation (dude)},
  author={Van Landeghem, Jordy and Tito, Rub{\`e}n and Borchmann, {\L}ukasz and Pietruszka, Micha{\l} and Joziak, Pawel and Powalski, Rafal and Jurkiewicz, Dawid and Coustaty, Micka{\"e}l and Anckaert, Bertrand and Valveny, Ernest and others},
  publisher = {In \textit{ICCV}},
  year={2023}
}

@inproceedings{masry2022chartqa,
  title={Chartqa: A benchmark for question answering about charts with visual and logical reasoning},
  author={Masry, Ahmed and Do, Xuan Long and Tan, Jia Qing and Joty, Shafiq and Hoque, Enamul},
    publisher = {In \textit{Findings of ACL}},
  year={2022}
}

@article{islam2023financebench,
  title={Financebench: A new benchmark for financial question answering},
  author={Islam, Pranab and Kannappan, Anand and Kiela, Douwe and Qian, Rebecca and Scherrer, Nino and Vidgen, Bertie},
  journal={arXiv preprint arXiv:2311.11944},
  year={2023}
}

@inproceedings{tanaka2025vdocrag,
  title={Vdocrag: Retrieval-augmented generation over visually-rich documents},
  author={Tanaka, Ryota and Iki, Taichi and Hasegawa, Taku and Nishida, Kyosuke and Saito, Kuniko and Suzuki, Jun},
  booktitle={CVPR},
  year={2025},
}

@inproceedings{cho2024m3docrag,
  title={M3docrag: Multi-modal retrieval is what you need for multi-page multi-document understanding},
  author={Cho, Jaemin and Mahata, Debanjan and Irsoy, Ozan and He, Yujie and Bansal, Mohit},
  booktitle={ICCV},
  year={2025}
}
